\documentclass[aps,prd,reprint,superscriptaddress,nofootinbib,longbibliography,floatfix]{revtex4-2}

\usepackage{tikz}
\usetikzlibrary{arrows.meta,positioning,decorations.pathmorphing}

\usepackage{amsmath}
\usepackage{amssymb}
\usepackage{bm}
\usepackage{graphicx}
\usepackage{booktabs}
\usepackage{xcolor}
\usepackage{hyperref}
\usepackage{microtype}
\usepackage{tikz}
\usetikzlibrary{arrows.meta,positioning}

\hypersetup{colorlinks=true,linkcolor=blue,citecolor=blue,urlcolor=blue}

\newcommand{\sigman}{\sigma_{N}}

\newcommand{\vmin}{v_{\min}}
\newcommand{\mchi}{m_{\chi}}
\newcommand{\er}{E_{R}}
\newcommand{\GeV}{\mathrm{GeV}}
\newcommand{\TeV}{\mathrm{TeV}}
\newcommand{\keV}{\mathrm{keV}}

\begin{document}

\title{Dark Matter at the Kinematic Edge: Interpreting the 248 keV LZ Nuclear-Recoil Candidate}

\author{Mattia Di Mauro}\email{dimauro.mattia@gmail.com}
\affiliation{Istituto Nazionale di Fisica Nucleare, Sezione di Torino, Via P. Giuria 1, 10125 Torino, Italy}

\date{\today}

\begin{abstract}
The LUX-ZEPLIN (LZ) Collaboration recently reported one event consistent with a $248\,\keV$ nuclear recoil in a $2.84$ tonne-year exposure, with a maximum local significance of $3.4\sigma$ and a global significance of $2.6\sigma$. We investigate whether the dark matter (DM)--nucleon interactions favored by this high-energy event can arise from particle DM models that simultaneously reproduce the observed relic abundance and satisfy indirect-detection constraints. Using the published LZ efficiency and operator significances, we show that elastic spin-independent (SI) scattering poorly explains an isolated high-energy recoil because its spectrum is concentrated at lower energies, whereas elastic spin-dependent (SD) $\mathcal{O}_4$ scattering remains viable. Endothermic scattering instead naturally suppresses the low-energy rate and shifts the recoil spectrum toward the observed energy.
A thermal pseudo-Dirac fermion with an off-diagonal vector interaction provides a simple realization of this mechanism. For $m_\chi\simeq1\,\TeV$, the relic-density requirement predicts $\sigma_N\simeq6.5\times10^{-43}\,\mathrm{cm}^2$, while a splitting $\delta\simeq297\,\keV$ shifts the recoil spectrum into the LZ event region. Present-day indirect-detection signals can be strongly suppressed because freeze-out proceeds mainly through coannihilation, while the excited state is depleted at late times. A thermal Higgsino provides a more predictive realization: its relic abundance fixes the mass near $1.1\,\TeV$, while a splitting $\delta\simeq377\,\keV$ is required to reproduce the event. This interpretation is testable through the associated gamma-ray line signal. Overall, combining direct detection, relic density, and indirect detection significantly restricts the viable interpretations of the LZ event and provides concrete targets for future searches.
\end{abstract}

\maketitle

\section{Introduction}

A broad range of astrophysical and cosmological observations provides compelling evidence for dark matter (DM), including galactic rotation curves, gravitational lensing and galaxy-cluster dynamics, and measurements of the cosmic microwave background and large-scale structure \cite{Rubin1980,Clowe2006,Planck2018}. These observations indicate that DM constitutes about $85\%$ of the matter content of the Universe. Nevertheless, its microscopic nature remains unknown and no unambiguous non-gravitational signal of a DM particle has yet been established.

Weakly interacting massive particles (WIMPs) have long represented one of the benchmark candidates for particle DM. In the standard thermal scenario, interactions of approximately weak strength can naturally produce the observed relic abundance through thermal freeze-out, the so-called WIMP miracle. This has motivated extensive direct-detection, indirect-detection, and collider searches. Increasingly stringent constraints have, however, excluded large regions of the simplest WIMP parameter space, with viable thermal solutions often surviving near resonances or for interaction structures that suppress one or more experimental probes \cite{Arcadi2018,LZ2025WIMP,XENONnT2025,DiMauroXie2026,KongDiMauro2026,KoechlerDiMauro2025}. Secluded DM provides an important alternative within the thermal paradigm: DM annihilation into lighter hidden-sector particles can determine the relic abundance while the coupling connecting the dark and Standard Model (SM) sectors remains much smaller \cite{PospelovRitzVoloshin2008,DiMauroWang2026,DiMauroNeutrino2026}. More generally, these scenarios motivate thermal DM models whose laboratory phenomenology differs substantially from the simplest elastic spin-independent (SI) or spin-dependent (SD) WIMP picture.

In this context, the recent extended-energy analysis of the LUX-ZEPLIN (LZ) experiment is particularly interesting \cite{LZ2026Extended}. LZ extends the nuclear-recoil window to approximately $270\,\keV$ and reports one event consistent with
$E_R=248\pm23\,(\mathrm{stat})\pm23\,(\mathrm{sys})\,\keV$
in a $2.84$ tonne-year exposure. The maximum local significance among the interaction hypotheses tested is $3.4\sigma$, while the significance after accounting for the look-elsewhere effect is $2.6\sigma$. Although this does not constitute evidence for DM, the unusually large recoil energy and the small expected background make it interesting to investigate which DM interactions can naturally produce such an event.

Importantly, LZ tests a broad set of interaction structures rather than a single DM hypothesis. The analysis considers the covariant WIMP--nucleon interactions $\mathcal{L}_1-\mathcal{L}_{20}$, as well as the conventional SI and SD operators $\mathcal{O}_1$ and $\mathcal{O}_4$ and their inelastic extensions, for both isoscalar and isovector couplings \cite{Fan2010,Fitzpatrick2013,Anand2014,LZ2026Extended}. The maximum local significance of $3.4\sigma$ is not unique to one interaction. Among the covariant interactions, $\mathcal{L}_{10}^{s}$, $\mathcal{L}_{10}^{v}$, and $\mathcal{L}_{16}^{s}$ reach $3.4\sigma$ for heavy WIMPs. Among the inelastic models, $\mathcal{O}_1^{v}$ and $\mathcal{O}_4^{s,v}$ also reach $3.4\sigma$ for appropriate DM masses and splittings, while the inelastic isoscalar SI interaction $\mathcal{O}_1^{s}$ reaches $3.3\sigma$. The latter is particularly interesting because it provides a simple realization of endothermic scattering and resembles the interaction expected for a pseudo-Dirac Higgsino \cite{Graham2025}.

The high recoil energy strongly constrains the possible interpretation. Conventional elastic SI scattering predicts a rapidly falling recoil spectrum, due to both the Galactic velocity distribution and the nuclear form factor. Normalizing such a spectrum to produce an event near $248\,\keV$ therefore generally predicts many more events at lower energies, making standard elastic SI scattering a poor explanation of an isolated high-energy event. This conclusion does not apply to all elastic interactions: the SD operator $\mathcal{O}_4$, for example, produces a different nuclear response and remains phenomenologically viable in the LZ analysis. Inelastic scattering provides another natural mechanism for generating a harder spectrum \cite{TuckerSmith2001,TuckerSmith2005,Barello2014,Bramante2016}. If the incoming state $\chi_1$ scatters into a slightly heavier state $\chi_2$, a mass splitting of a few hundred keV suppresses low-energy recoils and can shift a sizeable fraction of the rate toward the high-energy region probed by LZ. We study these kinematic effects quantitatively below.

The purpose of this work is to determine whether the interactions capable of explaining the LZ event can be embedded in physically consistent particle-DM models. Reproducing a recoil spectrum is not sufficient: the same model must also account for the cosmological DM abundance and remain compatible with indirect-detection and other laboratory constraints. We therefore construct explicit particle realizations motivated by the interaction structures favored in the LZ analysis and connect their direct-detection phenomenology to their cosmological history and present-day annihilation signals.

We first consider a pseudo-Dirac fermion with an off-diagonal vector interaction. Small Majorana masses split an underlying Dirac state into two nearby Majorana eigenstates, and in the heavy-mediator limit the interaction reduces to the inelastic isoscalar $\mathcal{O}_1$ structure studied by LZ. The same underlying coupling controls direct detection, $\chi_1 N\rightarrow\chi_2 N$, and early-Universe coannihilation, $\chi_1\chi_2\rightarrow q\bar q$, allowing the LZ signal region to be directly compared with the thermal relic-density requirement. We then consider a nearly pure thermal Higgsino as a more predictive realization of inelastic scattering. Its relic abundance selects a mass near $1.1\,\TeV$, while its electroweak interaction fixes the scattering strength and the LZ event primarily constrains the neutral-state mass splitting \cite{KrallReece2018,Graham2025}. We also discuss particle realizations associated with the SD $\mathcal{O}_4$ interaction and with the covariant operators displaying the largest LZ significances.

Finally, we confront these interpretations with indirect-detection constraints, with particular emphasis on gamma-ray line searches. This comparison is especially relevant for the Higgsino, for which the electroweak interactions responsible for the thermal relic abundance also predict potentially observable present-day annihilation signals. Combining the LZ recoil information with relic-density and indirect-detection constraints therefore provides a substantially more restrictive test of a DM interpretation than the direct-detection spectrum alone.

Our analysis is a phenomenological recoil-energy recast based on the public LZ detector information. The LZ Collaboration instead performs an unbinned profile-likelihood analysis in the calibrated $\{S1c,\log_{10}S2c\}$ plane, including the full NEST detector response and simultaneous fits to science and veto samples \cite{LZ2026Extended}. We use the published efficiency, nuclear-recoil response, and operator results to identify the relevant mass, splitting, and interaction scales, while propagating the dominant detector and astrophysical uncertainties. A full experimental parameter inference would ultimately require the official LZ event-level likelihood.

\section{The LZ high-energy nuclear-recoil candidate}

\subsection{Exposure, event properties, and detector response}

The new LZ analysis uses $220$ live days collected from 27 March 2023 to 1 April 2024. The fiducial liquid-xenon mass is $4.71\pm0.08$ tonnes, corresponding to the quoted $2.84$ tonne-year exposure \cite{LZ2026Extended}. The WIMP-search region of interest (ROI) is defined by $3<S1c<600$ phd, $S2>645$ phd, and $10^{2.75}<S2c<10^{4.15}$ phd\footnote{The quantities $S1c$ and $S2c$ denote the corresponding position-corrected signal sizes, while ``phd'' stands for photons detected.}. The prompt scintillation signal $S1$ is produced directly at the interaction vertex in the liquid xenon, whereas the delayed $S2$ signal arises from ionization electrons that drift upward, are extracted into the gas phase, and generate electroluminescence. The nuclear-recoil signal efficiency averages approximately $96\%$ between $14$ and $250\,\keV$, and LZ quotes $50\%$ efficiency at $5.4\,\keV$ and $269.9\,\keV$ \cite{LZ2026Extended}. The Supplemental Material also provides an energy-dependent gray uncertainty envelope, assessed with $^3$H, $^{220}$Rn, and AmLi calibration data. We use this published energy dependence directly in the recast, as described below.

The event of interest is characterized by
\begin{align}
S1c &= 540.1\ \mathrm{phd},\\
S2c &= 9268\ \mathrm{phd},\\
\er &= 248\pm23\,(\mathrm{stat})\pm23\,(\mathrm{sys})\ \keV.
\end{align}
It was recorded at 21:22:39 UTC on 16 June 2023 and lies $1.5\sigma$ below the median nuclear-recoil (NR) band but $6.7\sigma$ below the electron-recoil (ER) median at the same $S1c$ \cite{LZ2026Extended}. Its reconstructed position is well inside the fiducial volume. The Collaboration discusses several rare background mechanisms and finds no obvious detector anomaly temporally or spatially associated with the event. A $^{57}\mathrm{Co}$ calibration source had been removed $25$ minutes earlier on the opposite side of the detector; the previous AmBe nuclear-recoil calibration was on 8 June, and no anomalous event populations were identified around the candidate \cite{LZ2026Extended}.

For the numerical recast we digitize the final black ``+ ROI'' efficiency curve and the gray uncertainty envelope directly from the vector graphics in LZ Supplemental Fig.~S2 \cite{LZ2026Extended}. We interpolate these curves as functions of the true nuclear-recoil energy and use the central curve in every event-rate integral in this paper. The digitized central efficiency at the candidate energy is
\begin{equation}
\epsilon(248\,\keV)=0.946,
\end{equation}
while the lower and upper edges of the published gray envelope give $0.884$ and $0.980$, respectively. At $269.9\,\keV$ the digitized central curve gives $0.498$, providing a direct cross-check of the published $50\%$ high-energy point. The digitization is not an official LZ numerical acceptance table; it is a reconstruction of the curve displayed in the published figure. Further details of the digitization and its validation are collected in Appendix~\ref{app:efficiency}.

\begin{figure}[t]
\includegraphics[width=\columnwidth]{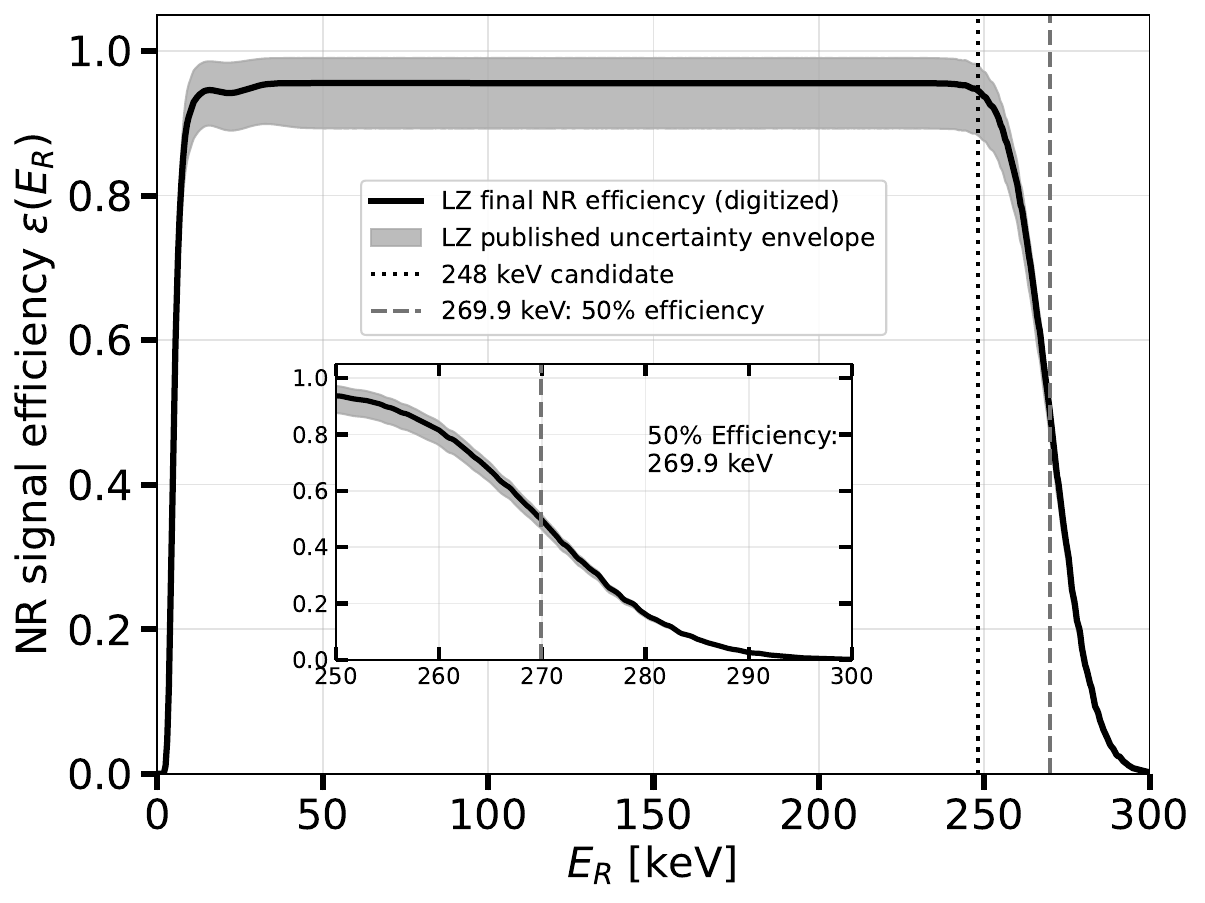}
\caption{Final LZ nuclear-recoil efficiency used in the public recast, vector-digitized from Supplemental Fig.~S2 of Ref.~\cite{LZ2026Extended}. The black curve is the final efficiency after the trigger, event-selection, and ROI requirements, and the gray region is the calibration-derived uncertainty envelope shown by LZ. The dotted vertical line marks the $248\,\keV$ candidate and the dashed line marks the published $50\%$ high-energy point at $269.9\,\keV$. The inset reproduces the useful high-recoil zoom over $250$--$300\,\keV$.}
\label{fig:efficiency}
\end{figure}

Fig.~\ref{fig:efficiency} shows why the efficiency treatment matters for this analysis. The candidate itself is still accepted with high probability, $\epsilon\simeq0.95$, but it lies immediately before the rapid high-energy turnoff. The inset makes the upper-edge behavior especially clear: the efficiency falls from about $0.94$ at $250\,\keV$ to $0.50$ at $269.9\,\keV$, $0.16$ at $280\,\keV$, and only a few percent by $290\,\keV$. Consequently, predicted spectra extending toward or beyond the upper ROI boundary must be folded with the measured energy-dependent efficiency rather than with a constant plateau.

We propagate the gray LZ envelope as an explicit detector-efficiency systematic by repeating the rate calculation with its lower and upper edges. Since LZ does not assign a confidence level to this plotted envelope, we treat the two edges as systematic variations rather than as a statistical interval. The systematic uncertainty on the LZ nuclear-recoil efficiency propagates into an uncertainty on the signal normalization as a function of recoil energy and, consequently, into the value of the mass splitting $\delta$ required to reproduce the observed event. At fixed halo parameters, however, this effect is relatively small. For the $1\,\TeV$ pseudo-Dirac benchmark, varying the efficiency within the gray uncertainty band reported by LZ shifts the preferred splitting from the central value $\delta=297.05\,\keV$ to the range $295.59$--$297.80\,\keV$. For the $1.1\,\TeV$ Higgsino benchmark, the corresponding shift is from $\delta=377.07\,\keV$ to $376.79$--$377.21\,\keV$. Thus, the detector-efficiency systematic induces only a modest uncertainty on the inferred splitting. As discussed below, a significantly larger uncertainty arises from the poorly constrained high-velocity tail of the Galactic DM distribution, which directly affects the population of particles energetic enough to produce these large nuclear recoils.

\subsection{What LZ actually tested: operator basis and inelastic benchmarks}
\label{sec:lzoperators}

It is important to distinguish the nonrelativistic operator probed by the nucleus from the Lorentz structure of an ultraviolet model. In the standard nonrelativistic effective field theory (NREFT), Galilean invariance allows the scattering amplitude to be organized in operators built from $i\bm q/m_N$, the transverse relative velocity $\bm v^{\perp}$, the DM spin $\bm S_\chi$, and the nucleon spin $\bm S_N$ \cite{Fitzpatrick2013,Anand2014}. The two operators most directly relevant to the inelastic LZ analysis are
\begin{align}
\mathcal{O}_1 &= \mathbf{1}_{\chi}\mathbf{1}_N,\\
\mathcal{O}_4 &= \bm S_\chi\cdot\bm S_N.
\end{align}
Thus, $\mathcal{O}_1$ is the standard SI operator, whereas $\mathcal{O}_4$ describes the conventional SD interaction between the DM and nucleon spins.

For a given nonrelativistic operator $\mathcal{O}_i$, we denote by $c_i^p$ and $c_i^n$ the effective coupling coefficients of DM to protons and neutrons, respectively, such that schematically
\begin{equation}
\mathcal{L}_{\rm NR}\supset
c_i^p\,\mathcal{O}_i^{(p)}
+
c_i^n\,\mathcal{O}_i^{(n)} .
\end{equation}
The superscripts $s$ and $v$ used by LZ denote the corresponding isoscalar and isovector combinations,
\begin{equation}
c_i^s=\frac{c_i^p+c_i^n}{2},
\qquad
c_i^v=\frac{c_i^p-c_i^n}{2}.
\end{equation}
A purely isoscalar interaction therefore corresponds to equal proton and neutron couplings, $c_i^p=c_i^n$, whereas a purely isovector interaction corresponds to opposite-sign couplings, $c_i^p=-c_i^n$.

Inelasticity does not change the nonrelativistic operator itself. Instead, it modifies the energy-conservation relation and therefore the minimum incoming DM velocity $v_{\min}(E_R)$ required to produce a recoil of energy $E_R$ \cite{Barello2014}. The standard elastic SI and SD hypotheses correspond respectively to $\mathcal{O}_1$ and $\mathcal{O}_4$ with $\delta=0$, while $\delta>0$ defines their endothermic inelastic extensions.

This makes clear why the extended-energy LZ analysis should not be interpreted as replacing the standard SI or SD searches. LZ has already performed dedicated low-energy WIMP searches optimized for the conventional elastic spectra. The purpose of the extended-energy analysis is instead to probe interactions for which a larger fraction of the recoil spectrum lies above $\sim100\,\keV$ and therefore becomes accessible only when the ROI is extended to approximately $270\,\keV$.

In addition to the inelastic $\mathcal{O}_1^{s,v}$ and $\mathcal{O}_4^{s,v}$ hypotheses, LZ studies the covariant interactions $\mathcal{L}_1$--$\mathcal{L}_{20}$. Their nonrelativistic reductions generate different dependences on momentum transfer, velocity, DM spin, nucleon spin, and nuclear response. The LZ paper uses the inelastic $\mathcal{O}_1^s$ and the elastic $\mathcal{L}_{10}^s$ interaction as illustrative examples. The former is an inelastic SI interaction and is described by LZ as Higgsino-like, whereas $\mathcal{L}_{10}^s$ corresponds to a magnetic-moment interaction producing a broad, double-peaked recoil spectrum in the extended ROI \cite{LZ2026Extended}.

The distinction between these descriptions is important for model building. A nonrelativistic operator such as $\mathcal{O}_1$ does not uniquely determine the underlying relativistic interaction. In this work we choose an off-diagonal vector-vector interaction as a concrete realization because it naturally arises for pseudo-Dirac DM and simultaneously provides an unsuppressed coannihilation channel in the early Universe. Alternative UV completions associated with $\mathcal{O}_4$ and with the covariant interactions tested by LZ are discussed separately below.

\subsection{Why elastic SI and SD behave differently}
\label{sec:elasticdiagnostic}

Both standard elastic SI and SD scattering obey the same kinematic relation $v_{\min}\propto\sqrt{E_R}$ at $\delta=0$, so neither develops the finite-energy minimum characteristic of endothermic scattering. The $248\,\keV$ candidate is nevertheless kinematically accessible to sufficiently heavy elastic DM. For a xenon nucleus of mass $m_A$, the maximum recoil energy produced by a DM particle with speed $v$ is
\begin{equation}
E_{R,\rm el}^{\max}=\frac{2\mu_A^2v^2}{m_A},
\end{equation}
where $\mu_A$ is the DM--nucleus reduced mass. Using the Standard Halo Model (SHM) parameters adopted below, with $v_{\rm esc}=544\,\mathrm{km/s}$
and a mean laboratory speed of approximately $250\,\mathrm{km/s}$,
a $248\,\keV$ elastic recoil is kinematically possible for
$m_\chi\gtrsim75\,\GeV$. The difficulty with the standard elastic SI interpretation is therefore not the ability to produce such a recoil, but the shape of the predicted recoil spectrum.

This becomes clear from the momentum transfer associated with the candidate. For a nuclear recoil of energy $E_R$,
\begin{equation}
q=\sqrt{2m_AE_R},
\end{equation}
which, taking $A=131$, gives
\begin{equation}
q(248\,\keV)\simeq0.246\,\GeV\simeq1.25\,\mathrm{fm}^{-1}.
\end{equation}
This is a large momentum transfer on nuclear scales. In a Helm description of the SI nuclear response, the coherence factor for $A=131$ is already strongly suppressed at this energy, $F^2\simeq2.6\times10^{-4}$. The natural-xenon isotope mixture smooths the diffraction structure, but the physical conclusion is unchanged: the familiar coherent enhancement of SI scattering is severely reduced near the energy of the candidate. A complete calculation requires the isotope-dependent nuclear responses used by
LZ \cite{Fitzpatrick2013,Anand2014,LZ2026Extended}.

Consequently, although an elastic SI WIMP can kinematically produce a $248\,\keV$ recoil, the rate at this energy lies in the strongly suppressed tail of its spectrum. Normalizing the interaction strength
to yield one event in this region therefore predicts a much larger population of lower-energy recoils. As shown quantitatively in Sec.~\ref{sec:elasticSInogo}, this makes the conventional elastic $\mathcal{O}_1$ interpretation incompatible with the observed recoil distribution.

That no-go does not automatically apply to standard elastic SD scattering. The $\mathcal{O}_4$ response is controlled by the spin structure functions of the spin-carrying xenon isotopes rather than by the coherent SI nuclear response. LZ Supplemental Table S8 gives a $2.7\sigma$ local preference for elastic $\mathcal{O}_4^{s}$ and $\mathcal{O}_4^{v}$ at $m_\chi=1\,\TeV$, while elastic $\mathcal{O}_1^s$ gives $0.0\sigma$. Thus elastic SD remains a genuine phenomenological alternative and deserves a dedicated analysis using the appropriate DMFormfactor nuclear response together with a consistent relic-density calculation.

LZ scans inelastic $\mathcal{O}_1^{s,v}$ and $\mathcal{O}_4^{s,v}$, together with a larger set of covariant interactions $\mathcal{L}_1$--$\mathcal{L}_{20}$. The paper uses $\mathcal{O}_1^s$ and $\mathcal{L}_{10}^s$ as illustrative spectra. The former is the inelastic SI case and is explicitly described as Higgsino-like; the latter is a magnetic-moment interaction that produces a broad double-peaked spectrum in the high-energy region \cite{LZ2026Extended}. The distinction matters for model building: the identity operator $\mathcal{O}_1$ can descend from several relativistic interactions. We choose an off-diagonal vector-vector interaction below because it naturally emerges from a pseudo-Dirac fermion and simultaneously provides an unsuppressed $s$-wave coannihilation channel at freeze-out.

For a $1\,\TeV$ WIMP, the LZ results provide a useful operator-level diagnostic. The local significance of $\mathcal{O}_1^s$ rises from $0.8\sigma$ at $\delta=100\,\keV$ to $2.7\sigma$, $3.0\sigma$, and $3.3\sigma$ at $\delta=200$, $300$, and $350\,\keV$, respectively. The isovector $\mathcal{O}_1^v$ case reaches $3.4\sigma$ at $\delta=300$--$350\,\keV$, and the $\mathcal{O}_4$ cases also reach $3.4\sigma$ for large splittings. Similarly, $\mathcal{L}_{10}^{s,v}$ reaches $3.4\sigma$ around the TeV scale. Therefore the candidate does not identify a unique interaction. What it robustly selects across many successful templates is a spectrum with appreciable support at high recoil energy.

\begin{figure}[t]
\includegraphics[width=\columnwidth]{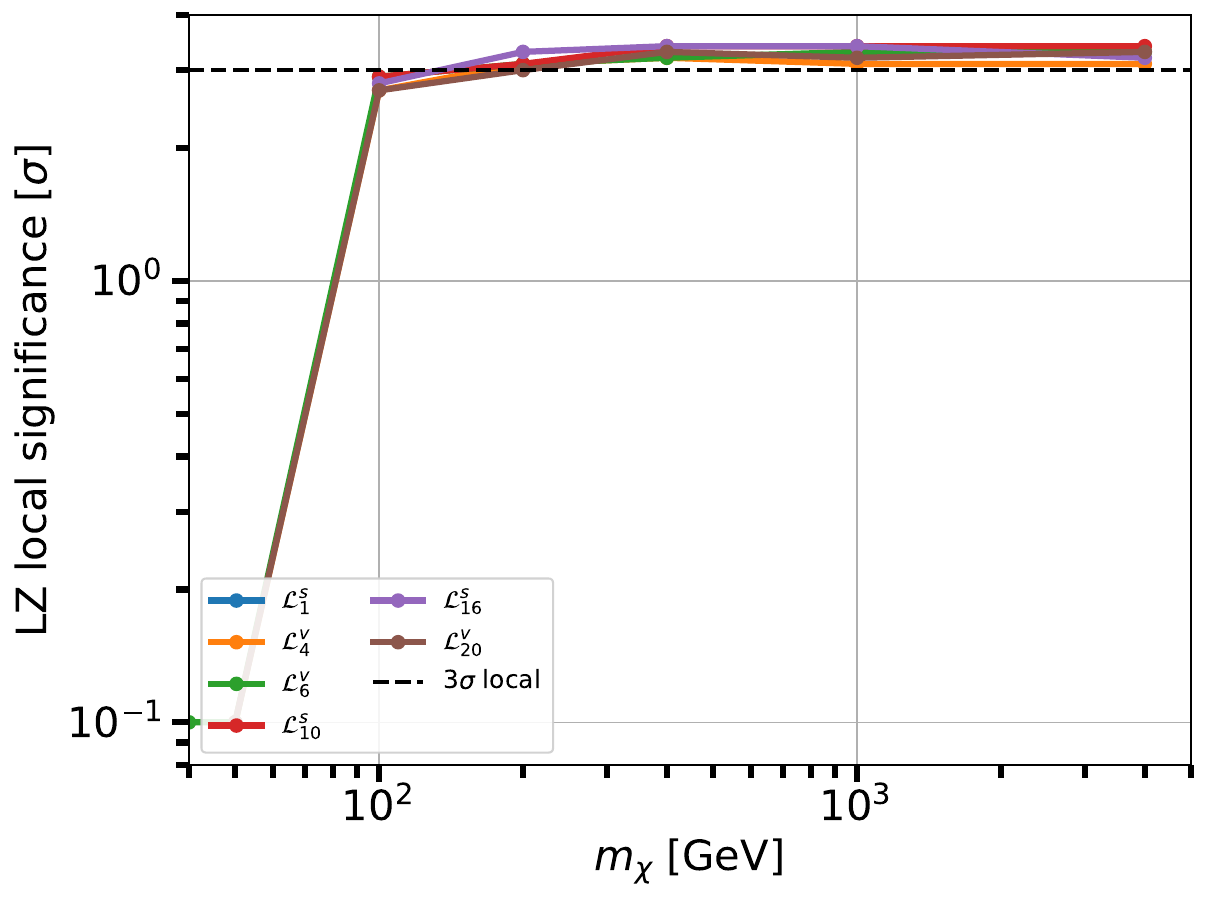}
\caption{Representative local significances from LZ Supplemental Table S7 for covariant interactions $\mathcal{L}_i^{s,v}$. The plotted numbers are transcribed from the LZ table and are not recomputed in this work. The rapid rise above $m_\chi\sim100$--$200\,\GeV$ illustrates that the isolated high-energy candidate can be described by several Lorentz structures; $\mathcal{L}_{10}$ is only one illustrative magnetic-moment example used by LZ.}
\label{fig:lzlagrangians}
\end{figure}

The same conclusion follows from the tabulation of the local significance for the complete covariant set $\mathcal{L}_1$--$\mathcal{L}_{20}$. The often-quoted $\mathcal{L}_{10}^s$ curve in the main LZ paper is therefore an illustrative example rather than a statement that a magnetic moment is uniquely favored. At $m_\chi=1\,\TeV$, $\mathcal{L}_{10}^s$ and $\mathcal{L}_{16}^s$ reach $3.4\sigma$, while many other structures lie in the $3.0$--$3.3\sigma$ range. Fig.~\ref{fig:lzlagrangians} shows representative entries from the exact LZ table. This broad degeneracy is one reason to start our particle interpretation from the simplest UV-motivated inelastic realization and then discuss alternative operator completions separately.
In particular, Fig.~\ref{fig:lzlagrangians} emphasizes the mass dependence rather than only the best-fit TeV point. Interactions that are essentially insensitive to the event at tens of GeV can reach local significances near $3\sigma$ once $m_\chi\gtrsim200\,\GeV$, because sufficiently heavy incident particles can populate the extended recoil window. The similarity of several curves also shows that the high-energy event is primarily selecting spectral hardness, not one unique relativistic bilinear.

\begin{figure}[t]
\includegraphics[width=\columnwidth]{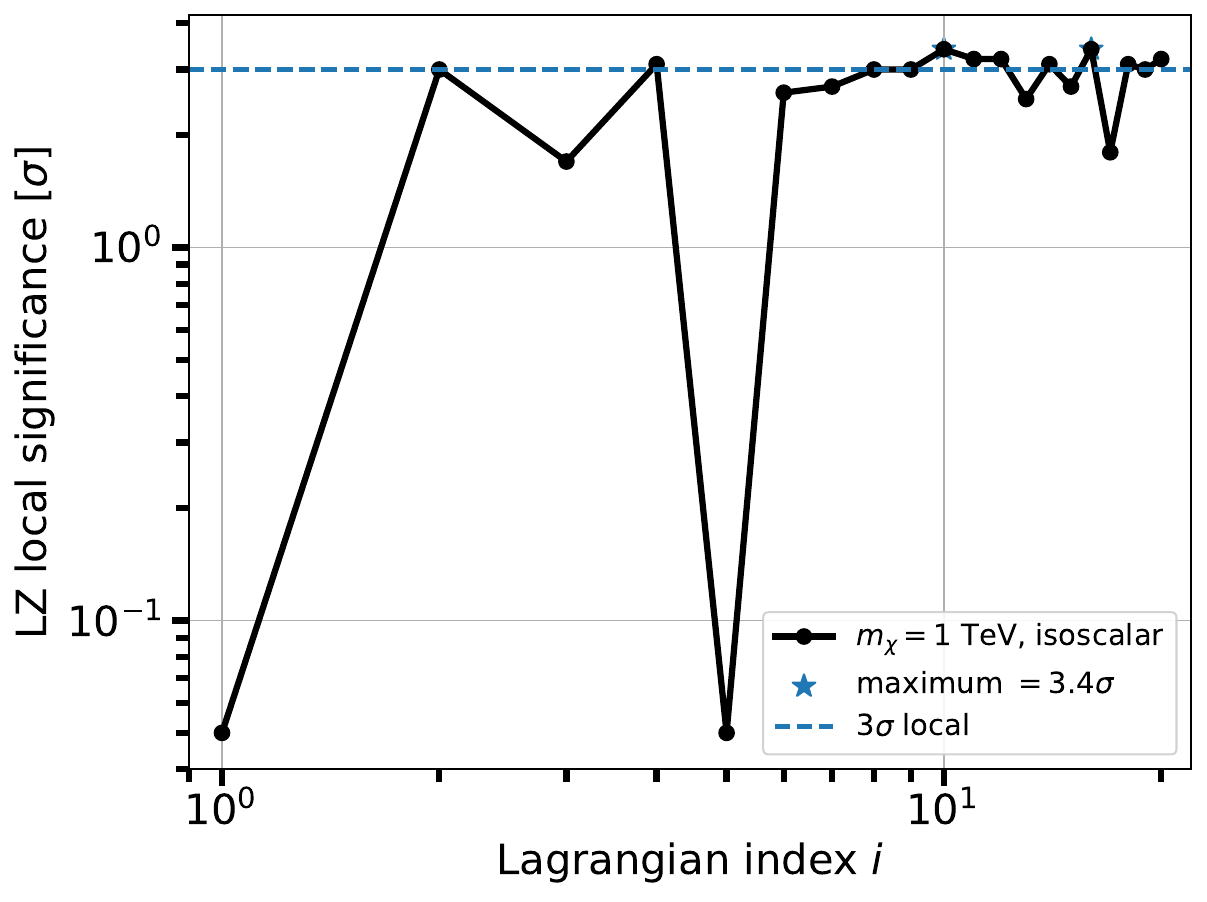}
\caption{This figure shows the local significance as a function of the covariant-Lagrangian index $i$ for the isoscalar interactions $\mathcal{L}_i^s$ and $m_\chi=1\,\TeV$. The exact tabulated zeros are displayed at a plotting floor of $0.05\sigma$ only because logarithmic axes are used. The largest value is $3.4\sigma$, reached by both $\mathcal{L}_{10}^s$ and $\mathcal{L}_{16}^s$.}
\label{fig:lisoscalar}
\end{figure}

\begin{figure}[t]
\includegraphics[width=\columnwidth]{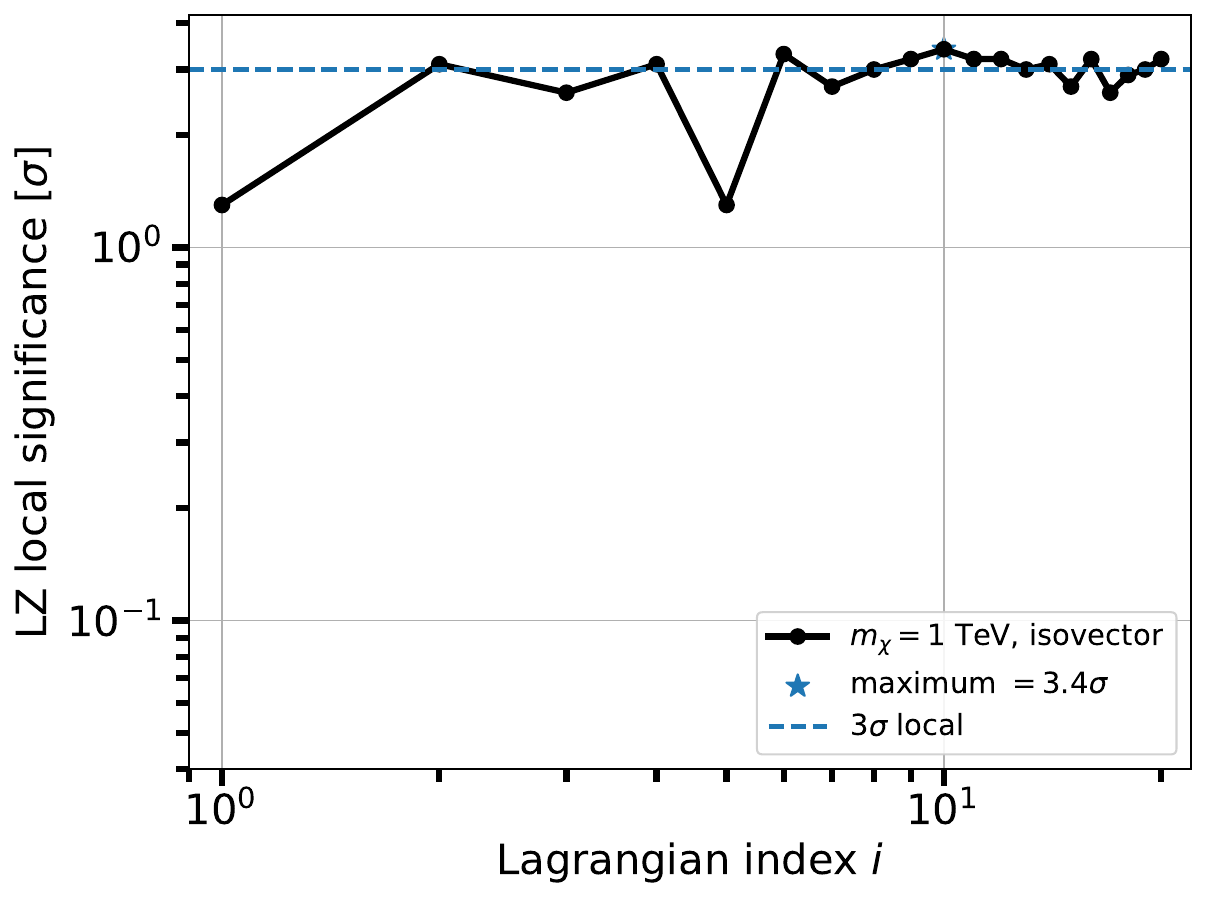}
\caption{As Fig.~\ref{fig:lisoscalar}, but for the isovector interactions $\mathcal{L}_i^v$ at $m_\chi=1\,\TeV$. The largest tabulated value is $3.4\sigma$, obtained for $\mathcal{L}_{10}^v$. The broad plateau near $3\sigma$ demonstrates that the single high-energy event does not identify a unique Lorentz structure.}
\label{fig:lisovector}
\end{figure}

To expose the operator degeneracy at fixed mass,
Figs.~\ref{fig:lisoscalar} and~\ref{fig:lisovector} display all twenty covariant interactions at $m_\chi=1\,\TeV$. The isoscalar panel has a tied maximum at $\mathcal{L}_{10}^s$ and $\mathcal{L}_{16}^s$, whereas the isovector panel peaks at $\mathcal{L}_{10}^v$. In both cases many interactions cluster between roughly $3.0$ and $3.3\sigma$, so the location of the maximum should not be interpreted as a measurement of a unique ultraviolet
interaction.

A useful counterexample is provided by $\mathcal{L}_5$, whose local significance remains very small compared with many of the interactions favored by the high-energy event. This behavior is closely related to that of $\mathcal{L}_1$. In the nonrelativistic reduction, both
$\mathcal{L}_1$ and $\mathcal{L}_5$ generate the same leading operator $\mathcal{O}_1=\mathbf{1}_\chi\mathbf{1}_N$, corresponding to the standard SI interaction, differing only by an overall normalization of the effective coupling \cite{Anand2014}. Since this normalization is a
free parameter in the fit, the two interactions predict essentially the same recoil-spectrum shape and therefore nearly identical statistical behavior in the LZ analysis. Their small local significance indicates that the standard $\mathcal{O}_1$-like spectral shape places too little relative weight in the high-recoil-energy region around the
$248\,\keV$ candidate, in contrast to interactions such as
$\mathcal{L}_{10}$ and $\mathcal{L}_{16}$ that generate substantially harder recoil spectra.

The fixed-mass comparison also clarifies why $\mathcal{L}_{10}^s$ is useful as an illustrative model without being uniquely selected by the data. Its $3.4\sigma$ value is part of a broad set of high-significance Lorentz structures, while the low-significance operators provide useful counterexamples whose recoil spectra do not place enough weight near the observed event.

\begin{figure}[t]
\includegraphics[width=\columnwidth]{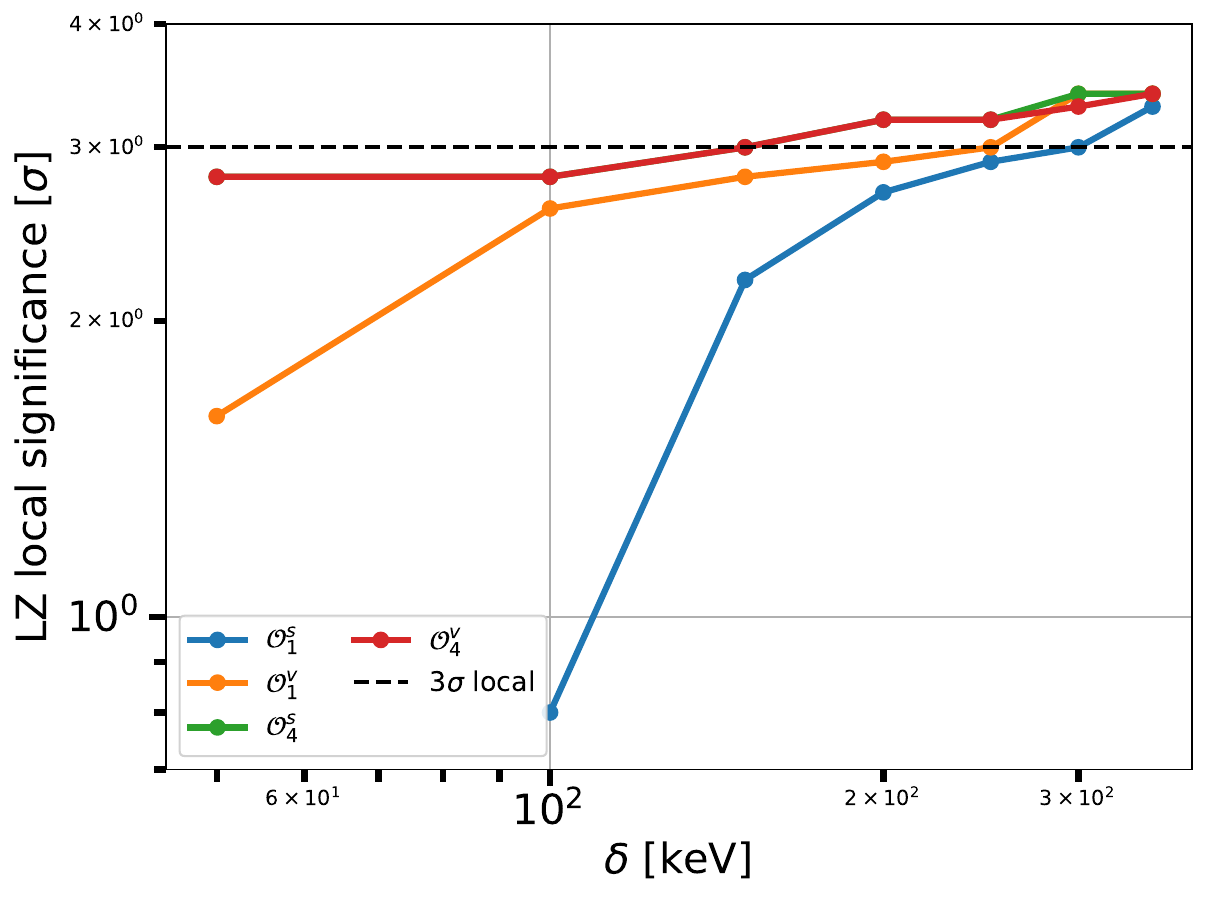}
\caption{Local significances reported by LZ for the $m_\chi=1\,\TeV$ inelastic $\mathcal{O}_1^{s,v}$ and $\mathcal{O}_4^{s,v}$ hypotheses, transcribed from Table S8 of the LZ Supplemental Material. The plot emphasizes that several operator choices become comparably successful once the splitting pushes the recoil spectrum into the high-energy region.}
\label{fig:lzsignificance}
\end{figure}

The complementary nonrelativistic scan is shown in Fig.~\ref{fig:lzsignificance}. Here the role of inelasticity can be isolated directly. For $m_\chi=1\,\TeV$, the isoscalar $\mathcal{O}_1^s$ hypothesis has essentially no preference in the elastic limit, while its local significance increases to $2.7\sigma$ at $\delta=200\,\keV$, $3.0\sigma$ at $\delta=300\,\keV$, and $3.3\sigma$ at $\delta=350\,\keV$. The isovector $\mathcal{O}_1^v$ interaction reaches $3.4\sigma$ already at $\delta=300\,\keV$. By contrast, $\mathcal{O}_4^{s,v}$ is already viable for $\delta=0$, with a local significance of $2.7\sigma$, and reaches about $3.3$--$3.4\sigma$ for splittings of $\delta\simeq300$--$350\,\keV$. This is the central reason we distinguish the strong elastic-SI no-go from the much weaker statement that can be made about elastic SD scattering.

\section{From recoil energy to DM parameters}

\subsection{Inelastic kinematics}
\label{sec:inelastickinematics}

The LZ results discussed above indicate that interactions producing a hard recoil spectrum can describe the $248\,\keV$ candidate more successfully than conventional elastic SI scattering. Endothermic inelastic scattering provides a particularly simple kinematic mechanism for generating such a spectrum. In this case, part of the incoming DM kinetic energy is required to produce a heavier final DM state, which suppresses low-energy scattering and can shift the recoil spectrum toward larger energies. Before computing the event rate for specific particle models, it is therefore useful to determine which combinations of DM mass and mass splitting can kinematically produce the observed recoil.

We consider the endothermic transition
\begin{equation}
\chi_1 + A \rightarrow \chi_2 + A,
\end{equation}
with
\begin{equation}
\delta=m_{\chi_2}-m_{\chi_1}>0.
\end{equation}

The modification of the recoil kinematics follows directly from nonrelativistic energy and momentum conservation. Denoting by $\bm q$ the momentum transferred to a nucleus initially at rest, momentum conservation gives $\bm p'=\bm p-\bm q$. Neglecting corrections of order $\delta/m_\chi$ in the kinetic-energy terms, energy conservation for the endothermic process reads
\begin{equation}
\frac{p^2}{2m_\chi}=\frac{(\bm p-\bm q)^2}{2m_\chi}+\frac{q^2}{2m_A}+\delta .
\end{equation}
Using $\bm p=m_\chi\bm v$ and defining the DM--nucleus reduced mass as
\begin{equation}
\mu_A=\frac{m_\chi m_A}{m_\chi+m_A},
\end{equation}
the energy-conservation condition can be rewritten as
\begin{equation}
\bm v\cdot\bm q=\frac{q^2}{2\mu_A}+\delta .
\end{equation}
For a fixed recoil energy, the minimum incident speed is obtained when $\bm v$ is parallel to $\bm q$. Since $q=\sqrt{2m_AE_R}$, the minimum DM speed required to produce a nuclear recoil of energy $E_R$ is \cite{TuckerSmith2001,Barello2014}
\begin{equation}
v_{\min}(E_R)=\frac{m_AE_R/\mu_A+\delta}{\sqrt{2m_AE_R}}.
\label{eq:vmin}
\end{equation}
A derivation from energy and momentum conservation, together with the allowed recoil interval at fixed incoming speed, is given in Appendix~\ref{app:inelastic_kinematics}.

For elastic scattering, $\delta=0$, and $v_{\min}$ increases monotonically with $E_R$. For endothermic scattering, instead, the additional energy required to produce $\chi_2$ introduces a term proportional to $\delta/\sqrt{E_R}$, and $v_{\min}(E_R)$ develops a minimum at
\begin{equation}
\begin{aligned}
E_R^{\star}&=\delta\frac{\mu_A}{m_A},\\
v_{\min}^{\star}&=\sqrt{\frac{2\delta}{\mu_A}}.
\end{aligned}
\label{eq:estar}
\end{equation}
The quantity $E_R^\star$ is therefore the recoil energy at which the required incoming DM speed is smallest for a given $(m_\chi,\delta)$. Since the number of available halo particles generally decreases rapidly toward large velocities, this kinematic minimum identifies the recoil-energy region with the largest available DM phase space, although the actual recoil spectrum also depends on the interaction and nuclear response.

If the observed LZ recoil lies close to this characteristic energy, $E_R\simeq E_R^\star$, the required splitting can be estimated as
\begin{equation}
\delta\simeq E_R\frac{m_A}{\mu_A}.
\label{eq:deltaest}
\end{equation}
Although the DM-mass dependence is not explicit in this expression, it is contained in the reduced mass $\mu_A$. Substituting its definition gives
\begin{equation}
\delta\simeq E_R\left(1+\frac{m_A}{m_\chi}\right).
\label{eq:deltaestexplicit}
\end{equation}
This form makes the dependence on $m_\chi$ transparent. In the heavy-DM limit, $m_\chi\gg m_A$, one has $\mu_A\simeq m_A$ and therefore $\delta\simeq E_R$. For lighter DM, $\mu_A<m_A$, so a larger mass splitting is required to place the minimum of $v_{\min}$ at the same recoil energy.

For $A=131$ and $E_R=248\,\keV$, Eq.~\eqref{eq:deltaest} gives the values reported in Table~\ref{tab:kinematics}.

\begin{table}[b]
\caption{Kinematic ridge obtained by requiring the $248\,\keV$ recoil to coincide with the minimum of $v_{\min}(E_R)$ for $A=131$. The dependence on $m_\chi$ enters through the DM--nucleus reduced mass $\mu_A$. The last column gives the corresponding minimum incoming speed $v_{\min}^{\star}$. Values exceeding the approximate maximum DM speed in the detector frame are kinematically inaccessible in the baseline SHM.}
\label{tab:kinematics}
\begin{ruledtabular}
\begin{tabular}{ccc}
$m_\chi$ [GeV] & $\delta$ [keV] & $v_{\min}(248\,\keV)$ [km/s]\\
\hline
200  & 399 & 973\\
300  & 349 & 850\\
400  & 324 & 789\\
1000 & 278 & 678\\
1100 & 276 & 671\\
4000 & 256 & 623\\
\end{tabular}
\end{ruledtabular}
\end{table}

Table~\ref{tab:kinematics} illustrates two related effects. First, the splitting required to place the kinematic minimum at $248\,\keV$ decreases with increasing $m_\chi$ and asymptotically approaches $\delta=248\,\keV$ in the heavy-DM limit. Second, the corresponding minimum velocity also decreases as $m_\chi$ increases. Consequently, for low DM masses the same high-energy recoil requires particles increasingly close to, or even above, the maximum speed available in the Galactic halo. For example, a splitting $\delta=300\,\keV$ requires approximately $m_\chi\gtrsim286\,\GeV$ if the largest available detector-frame speed is approximated by $v_{\rm esc}+\bar v_E\simeq794\,\mathrm{km/s}$. A $\delta=370\,\keV$ transition instead requires $m_\chi\gtrsim0.78\,\TeV$ under the same assumptions. These are purely kinematic restrictions and do not depend on the interaction normalization or detector efficiency.

\begin{figure}[t]
\includegraphics[width=\columnwidth]{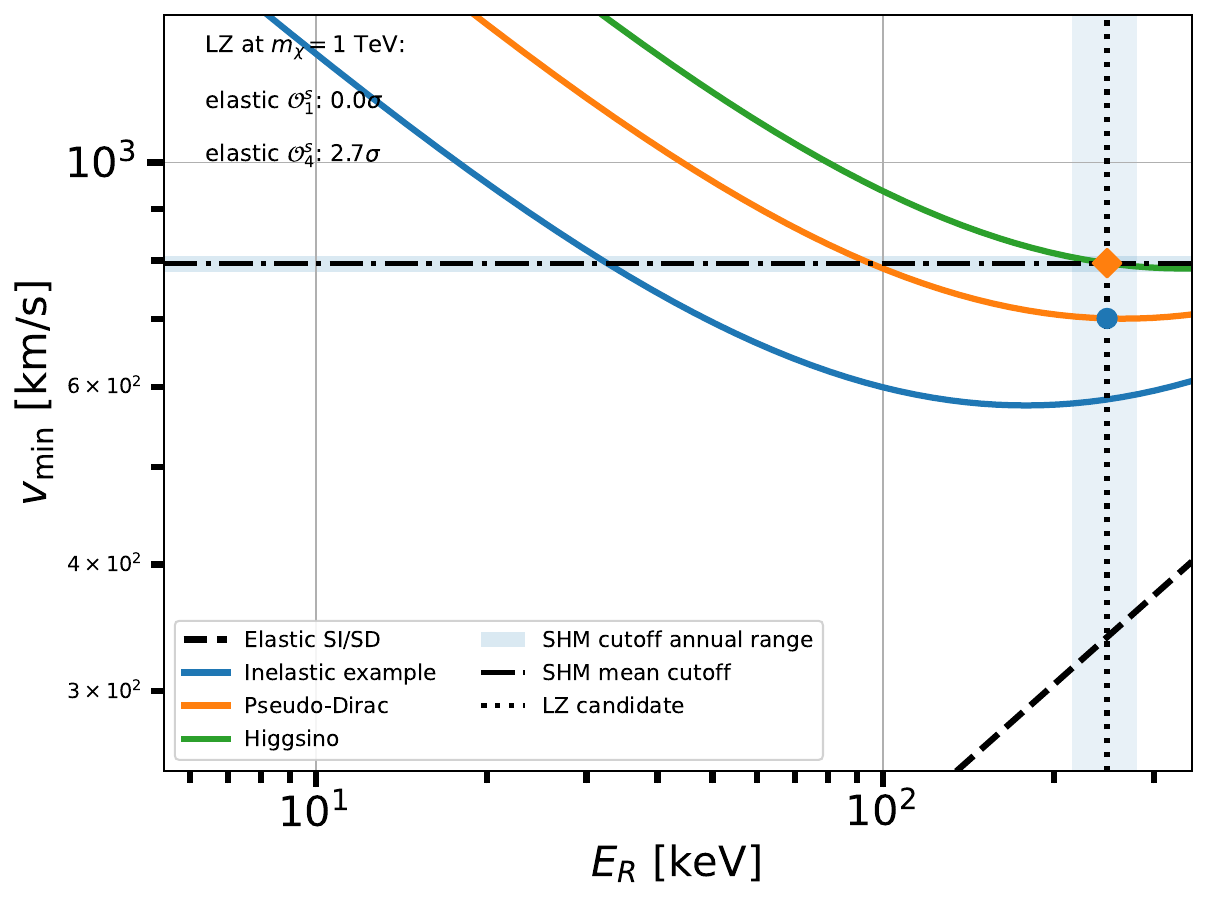}
\caption{Minimum incoming DM speed as a function of recoil energy for scattering on xenon. The dashed elastic curve is evaluated for $m_\chi=1\,\TeV$ and $\delta=0$ and therefore describes the common kinematics of standard SI and SD scattering. The inelastic example illustrates the effect of a nonzero splitting. The pseudo-Dirac and Higgsino curves use the benchmark parameters obtained from the relic-density and LZ-rate analyses discussed below. The horizontal dot-dashed line denotes the SHM detector-frame cutoff evaluated with the mean Earth velocity, $v_{\rm esc}+\bar v_E\simeq794\,\mathrm{km/s}$, while the shaded horizontal band shows its annual variation induced by the Earth's motion. The vertical dotted line marks the $248\,\keV$ candidate and the vertical shaded region its total $1\sigma$ energy uncertainty. At the candidate energy, the pseudo-Dirac benchmark requires $v_{\min}\simeq701\,\mathrm{km/s}$, whereas the Higgsino requires $v_{\min}\simeq795\,\mathrm{km/s}$ and therefore lies at the SHM mean kinematic cutoff.}
\label{fig:vmin}
\end{figure}

Fig.~\ref{fig:vmin} shows directly how inelasticity modifies the velocity required to produce a given nuclear recoil. The horizontal line labeled ``SHM mean cutoff'' corresponds to $v_{\rm esc}+\bar v_E\simeq794\,\mathrm{km/s}$, obtained from the baseline values $v_{\rm esc}=544\,\mathrm{km/s}$ and $\bar v_E=250.2\,\mathrm{km/s}$. This quantity is not the average DM speed in the Galaxy. Rather, it approximates the largest DM speed that can be observed in the detector frame when the mean Earth velocity relative to the halo is used. The horizontal shaded band shows how this detector-frame cutoff varies during the year as the Earth's velocity changes by approximately $\Delta v_E\simeq15\,\mathrm{km/s}$.

The pseudo-Dirac and Higgsino curves in Fig.~\ref{fig:vmin} are obtained by inserting into Eq.~\eqref{eq:vmin} the benchmark values of $(m_\chi,\delta)$ derived later from the model-specific relic-density and LZ-rate calculations. For the thermal pseudo-Dirac model we use $m_\chi=1\,\TeV$ and $\delta=297.05\,\keV$. The interaction strength is first fixed by the relic-density requirement, and the splitting is then determined by requiring the predicted LZ rate to reproduce one event. For the thermal Higgsino we use $m_\chi=1.1\,\TeV$ and $\delta=377.07\,\keV$: the DM mass is selected by the thermal relic-density condition, the scattering strength is fixed by the electroweak interaction, and the splitting is determined from the LZ rate. Once these benchmark parameters have been specified, the curves shown in Fig.~\ref{fig:vmin} are purely kinematic; the scattering cross section and detector efficiency do not enter directly into the calculation of $v_{\min}(E_R)$.

The elastic curve is particularly useful for separating kinematics from the recoil-spectrum shape. For a TeV-scale WIMP, the $248\,\keV$ recoil is not kinematically difficult to produce: one finds $v_{\min}^{\rm el}\simeq340\,\mathrm{km/s}$, well below the SHM detector-frame cutoff. The difficulty of the conventional elastic SI interpretation therefore arises from its recoil-spectrum shape and from the strong loss of nuclear coherence at high recoil energy, rather than from kinematics. Elastic SD scattering has the same favorable elastic kinematics but a different nuclear response and, as discussed above, remains a viable interpretation of the event.

For $\delta>0$, the situation changes qualitatively because $v_{\min}(E_R)$ develops a minimum at finite recoil energy. The pseudo-Dirac benchmark requires $v_{\min}(248\,\keV)\simeq701\,\mathrm{km/s}$, which probes the high-velocity tail but remains appreciably below the SHM cutoff. The Higgsino benchmark instead requires $v_{\min}(248\,\keV)\simeq795\,\mathrm{km/s}$, placing the event at the kinematic limit for the mean Earth velocity. This explains why the Higgsino interpretation is substantially more sensitive than the pseudo-Dirac benchmark to the poorly constrained high-speed tail of the Galactic DM distribution and to annual modulation.

The same kinematic information can be displayed more globally in the $(m_\chi,\delta)$ plane. For any given $m_\chi$, one can first ask which value of $\delta$ places the minimum of $v_{\min}(E_R)$ exactly at the observed recoil energy. Requiring $E_R^\star=248\,\keV$ gives
\begin{equation}
E_R^\star=\delta\frac{\mu_A}{m_A}=248\,\keV,
\end{equation}
or equivalently Eq.~\eqref{eq:deltaestexplicit}. We refer to the corresponding locus in the $(m_\chi,\delta)$ plane as the $248\,\keV$ kinematic ridge. The ridge is not a kinematic boundary and should not be interpreted as the preferred model parameter space. It simply identifies the combinations of $(m_\chi,\delta)$ for which a $248\,\keV$ recoil occurs exactly at the minimum of $v_{\min}(E_R)$.

A different quantity is the largest endothermic splitting that can be excited at all for a given DM mass. From Eq.~\eqref{eq:estar}, the smallest velocity required for an inelastic transition is $v_{\min}^{\star}=\sqrt{2\delta/\mu_A}$. Requiring this minimum velocity not to exceed the largest available detector-frame speed, $v_{\min}^{\star}\leq v_{\max}$, gives
\begin{equation}
\delta_{\max}(m_\chi)=\frac{1}{2}\mu_A v_{\max}^2.
\label{eq:deltamax}
\end{equation}
This relation defines a true kinematic ceiling: for $\delta>\delta_{\max}$, no recoil energy can satisfy the inelastic scattering condition, irrespective of the interaction strength.

\begin{figure}[t]
\includegraphics[width=\columnwidth]{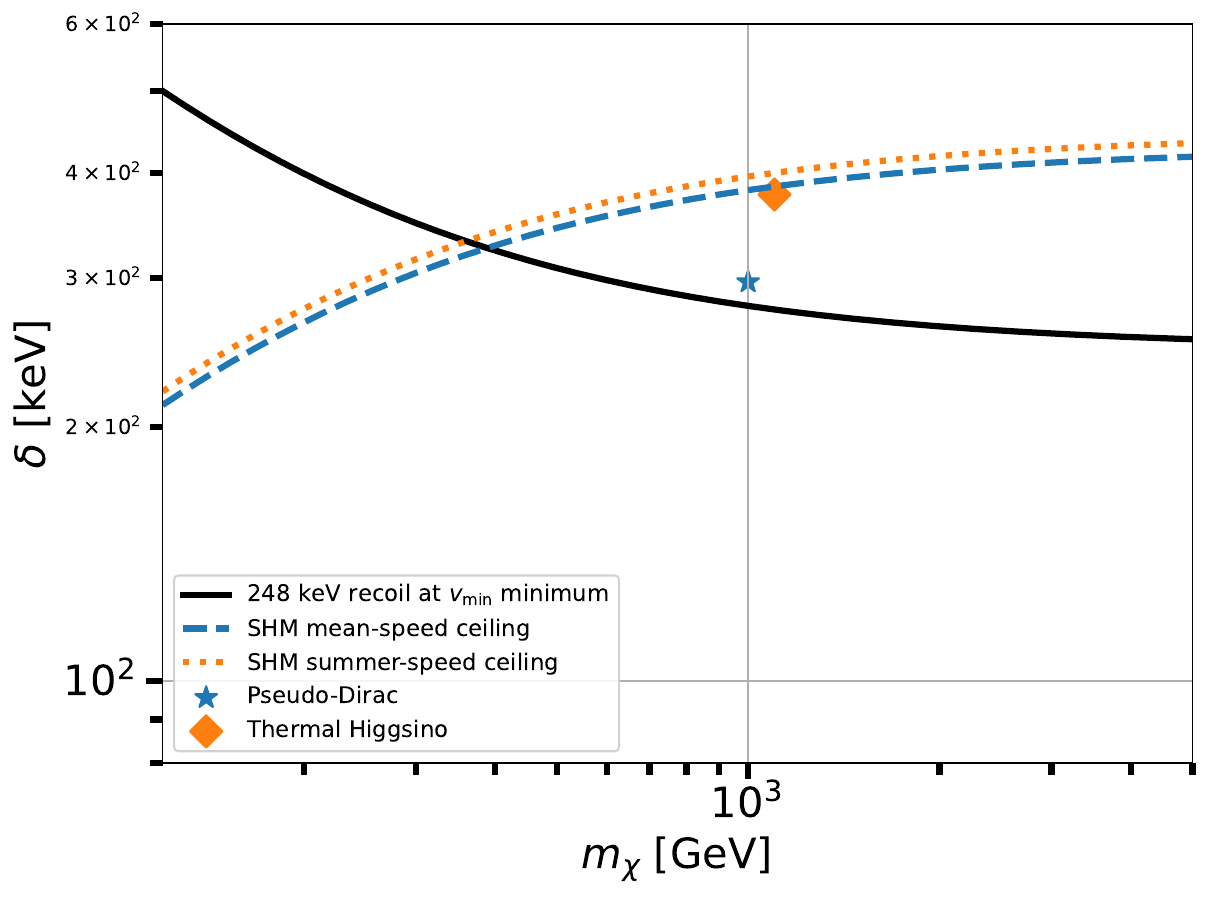}
\caption{Kinematic structure in the $m_\chi$--$\delta$ plane for scattering on $A=131$. The solid curve, labeled ``$248\,\keV$ recoil at $v_{\min}$ minimum'', is the kinematic ridge defined by $E_R^\star=248\,\keV$; it is not a boundary, but the locus for which the observed recoil coincides with the minimum of $v_{\min}(E_R)$. The dashed curve is the kinematic ceiling obtained using the mean detector-frame maximum speed, $v_{\max}^{\rm mean}=v_{\rm esc}+\bar v_E$, while the dotted curve uses the larger summer-like value $v_{\max}^{\rm summer}=v_{\rm esc}+\bar v_E+\Delta v_E$. The region above either ceiling is kinematically inaccessible for the corresponding choice of $v_{\max}$. The pseudo-Dirac and thermal Higgsino points show the benchmark parameters obtained independently from the relic-density and LZ-rate analyses; they are therefore not required to lie on the $248\,\keV$ kinematic ridge. The Higgsino lies much closer to the kinematic ceiling and is consequently more sensitive to the fastest particles in the local halo.}
\label{fig:kinematicmap}
\end{figure}

We consider two choices for $v_{\max}$. The first uses the mean Earth velocity, $v_{\max}^{\rm mean}=v_{\rm esc}+\bar v_E$, and defines the SHM mean-speed ceiling shown by the dashed curve in Fig.~\ref{fig:kinematicmap}. The second uses a summer-like value, $v_{\max}^{\rm summer}=v_{\rm esc}+\bar v_E+\Delta v_E$, corresponding approximately to the time of year when the Earth moves fastest relative to the halo; this gives the slightly higher dotted curve. For $m_\chi=1.1\,\TeV$ and xenon, these choices give $\delta_{\max}\simeq385\,\keV$ and approximately $400\,\keV$, respectively. A splitting near $377\,\keV$ is therefore kinematically allowed, but it lies sufficiently close to the ceiling that only the fastest local DM particles can contribute efficiently.

Fig.~\ref{fig:kinematicmap} makes the different roles of the three curves transparent. The solid curve is the $248\,\keV$ kinematic ridge. At low $m_\chi$, the splitting required to place the minimum of $v_{\min}$ at $248\,\keV$ rises rapidly because the DM--nucleus reduced mass becomes smaller. In the heavy-DM limit, $\mu_A\rightarrow m_A$, and the ridge asymptotically approaches $\delta\simeq248\,\keV$. Importantly, points away from this curve can still produce a $248\,\keV$ recoil; the recoil simply does not coincide with the minimum of $v_{\min}(E_R)$.

The dashed and dotted curves instead represent genuine kinematic boundaries. For a point above a given ceiling, even the fastest DM particle allowed by that choice of halo velocity cannot overcome the endothermic mass splitting, so scattering is impossible at any recoil energy. Points below the ceiling are kinematically accessible, although this alone does not imply a large event rate: the rate still depends on the number of halo particles with the required velocity, the interaction strength, the nuclear response, and the detector efficiency. The pseudo-Dirac benchmark lies sufficiently below the ceiling that a broader part of the high-speed halo can contribute. The thermal Higgsino lies much closer to the boundary, so its predicted LZ rate depends more strongly on the extreme high-velocity tail and on the seasonal variation of the detector-frame velocity distribution.

\subsection{Recoil rate and the SHM}

For an isoscalar SI contact interaction, the differential rate per detector mass can be written as \cite{LewinSmith1996,Baxter2021}
\begin{equation}
\frac{dR}{d\er}
=\frac{\rho_\chi}{\mchi}\frac{\sigman}{2\mu_N^2}
\sum_i \xi_i A_i^2 F_i^2(\er)\,\eta\!\left[\vmin(\er)\right],
\label{eq:ratecompact}
\end{equation}
where $\xi_i$ denotes the isotope weighting per unit target mass, $F_i$ is the nuclear form factor, $\mu_N$ is the DM--nucleon reduced mass, and
\begin{equation}
\eta(v_{\min})=\int_{v>v_{\min}}d^3v\,\frac{f_{\rm lab}(\bm{v})}{v}
\end{equation}
is the mean inverse speed. Our implementation evaluates each xenon isotope explicitly, uses natural isotopic abundances, and uses a Helm form factor.

We adopt the direct-detection SHM conventions recommended in Ref.~\cite{Baxter2021} with:
\begin{align}
\rho_\chi &= 0.3\ \GeV\,\mathrm{cm}^{-3},\\
v_0 &= 238\ \mathrm{km/s},\\
v_{\rm esc} &= 544\ \mathrm{km/s}.
\end{align}
The Solar peculiar velocity gives $|\bm v_0+\bm v_\odot|\simeq250.2\,\mathrm{km/s}$. For annual averaging we use a one-harmonic approximation
\begin{equation}
v_E(t)=250.2\,\mathrm{km/s}+14.9\,\mathrm{km/s}\cos\!\left[\omega(t-t_0)\right],
\end{equation}
with $t_0$ near 2 June. This approximation is adequate for exposing the extreme-tail sensitivity, but a precision analysis should use the exact LZ live-time distribution and full vector Earth velocity.

The detected number of events is
\begin{equation}
N_{\rm sig}=MT\int d\er\,\epsilon(\er)\frac{dR}{d\er},
\label{eq:nsig}
\end{equation}
with $MT=2.84$ tonne-year, efficiency $\epsilon(E_R)$ taken from Fig.~\ref{fig:efficiency}. For a simple central rate estimator we take one observed event minus the quoted high-$S1$ integrated background guide $0.0106$, giving $N_{\rm sig}\simeq0.989$. The $0.0106$ count number comes from a particular high-$S1$ projection in the LZ analysis and is not the complete event-level background likelihood; it is used here only to show that replacing one by $0.989$ has negligible effect on the inferred cross-section scale. A one-bin profile-likelihood exercise would give a very broad $90\%$ interval, roughly $0.095<N_{\rm sig}<3.64$, illustrating the statistical uncertainty from a single event. Details of the rate normalization and the LZ cross-section convention are summarized in Appendix~\ref{app:rate}, while the one-event statistical guide is derived in Appendix~\ref{app:poisson}.

LZ also provides the normalization needed to compare $\mathcal{O}_1^s$ directly with a conventional per-nucleon cross section. With $v_{\rm EW}=246.2\,\GeV$, the Supplemental Material writes
\begin{equation}
\left(c_1^s v_{\rm EW}^2\right)^2
=\sigma_N^{\rm SI}\frac{\pi v_{\rm EW}^4}{\mu_N^2},
\label{eq:lzo1conversion}
\end{equation}
so that
\begin{equation}
\sigma_N^{\rm SI}=\frac{\mu_N^2}{\pi}\left(c_1^s\right)^2.
\label{eq:lzo1conversion2}
\end{equation}
This is exactly the normalization that will emerge from our universal-vector contact model after identifying $c_1^s=3g_\chi g_q/m_V^2$. For a weak neutral-current vector charge, as relevant to the Higgsino, LZ notes that the xenon cross-section limit is obtained by multiplying the scalar-normalized result by approximately $3.2$. This factor reflects the neutron-dominated weak charge rather than a change in the inelastic kinematics.

\subsection{Why ordinary elastic SI scattering fails quantitatively}
\label{sec:elasticSInogo}

Before fitting an inelastic model, it is useful to perform a normalization-independent test of the recoil-spectrum shape. For an elastic $1\,\TeV$ SI WIMP, our detected spectrum predicts that only
\begin{equation}
\frac{N(E_R>200\,\keV)}{N(5.4<E_R<300\,\keV)}\simeq1.37\times10^{-3}
\label{eq:elastictailfraction}
\end{equation}
of accepted signal events occur above $200\,\keV$. Thus, if a single detected event were drawn from the elastic-SI signal spectrum, the probability for it to occur above $200\,\keV$ would be only about $0.14\%$. The candidate is therefore not merely a somewhat energetic recoil; it lies in the per-mille tail of the standard elastic-SI spectrum even for TeV-scale DM.

A more candidate-specific estimate can be obtained using the measured recoil energy and its uncertainty. Combining the statistical and systematic uncertainties in quadrature gives $\Delta E_R\simeq32.5\,\keV$, corresponding approximately to the interval $215.5<E_R<280.5\,\keV$. In our elastic-SI spectrum, only
\begin{equation}
P_{\rm el}^{1\sigma}\equiv\frac{N(215.5<E_R<280.5\,\keV)}{N(5.4<E_R<300\,\keV)}\simeq6.6\times10^{-4}
\end{equation}
of accepted signal events fall in this interval. Therefore, conditional on observing one elastic-SI signal event in the accepted recoil range, the probability for it to appear within approximately one total energy uncertainty of the observed LZ candidate is only about $0.066\%$. We stress that this quantity is a spectral probability within our simplified recoil-energy recast, not an experimental $p$-value and not a replacement for the full LZ profile-likelihood analysis.

For a second, complementary test we determine the elastic cross section required to produce one detected event between $215$ and $300\,\keV$, a window centered on the candidate and extending through the high-energy efficiency turnoff. The public recast gives
\begin{equation}
\sigma_N^{\rm el}\simeq2.7\times10^{-44}\,\mathrm{cm}^2.
\end{equation}
With the same normalization, the expected number of signal events between $5.4$ and $55\,\keV$ is
\begin{equation}
N_{5.4-55}\simeq1.42\times10^{3}.
\label{eq:elasticcatastrophe}
\end{equation}
The incompatibility is therefore not caused by the inability of a TeV-scale elastic WIMP to produce a $248\,\keV$ recoil; as discussed above, such a recoil is kinematically allowed. Rather, an interaction strength large enough to populate the extreme high-energy tail inevitably produces an enormous low-energy SI signal.

The same conclusion can be formulated as a normalization-independent spectral probability. If the elastic-SI spectrum is normalized such that the expected number of events in the $215$--$300\,\keV$ interval is $\lambda_{\rm high}=1$, the same spectral shape predicts $\lambda_{\rm low}\simeq1416$ events between $5.4$ and $55\,\keV$. The physically relevant quantity is therefore the ratio of the expected populations in the two energy intervals. Conditional on detecting one signal event in their union, the probability for an elastic-SI event to occur in the high-energy interval rather than in the low-energy interval is
\begin{equation}
P_{\rm high}^{\rm el}
=
\frac{\lambda_{\rm high}}
{\lambda_{\rm low}+\lambda_{\rm high}}
\simeq
\frac{1}{1417}
\simeq7.1\times10^{-4}.
\end{equation}
Thus, within the elastic-SI signal hypothesis, an event in the high-energy window is expected only about once for every $1.4\times10^3$ events in the low-energy interval. This quantity is a conditional spectral probability rather than an experimental $p$-value: the real LZ data contain low-energy background events and must be analyzed with the full multidimensional likelihood. Nevertheless, it makes clear that explaining the $248\,\keV$ candidate with conventional elastic SI scattering would require a very large accompanying low-energy DM population that is not observed.

The situation is very different for the inelastic benchmarks. Table~\ref{tab:spectralcounts} reports the expected number of detected signal events in representative recoil-energy intervals for the elastic-SI normalization described above and for the thermal pseudo-Dirac and Higgsino benchmarks introduced below. The latter two benchmarks are normalized by their underlying particle models: their interaction strengths are fixed by the relic-density or electroweak requirements, while the mass splittings are selected by requiring approximately one LZ signal event in the present exposure.

\begin{table}[t]
\caption{Expected detected signal counts in the $2.84$ tonne-year LZ exposure for the spectra shown in Fig.~\ref{fig:spectra}. The elastic SI spectrum is normalized to one event in $215<E_R<300\,\keV$. The pseudo-Dirac and Higgsino benchmarks use their relic-density-motivated interaction strengths and the splittings determined from the LZ rate. The last two columns give the fraction of accepted events above $200\,\keV$ and within the approximate $1\sigma$ energy interval of the candidate, $215.5<E_R<280.5\,\keV$.}
\label{tab:spectralcounts}
\begin{ruledtabular}
\begin{tabular}{lcccc}
Model & $N_{5.4-55}$ & $N_{55-200}$ & $f_{>200}$ & $f_{1\sigma}$\\
\hline
Elastic SI & $1.42\times10^{3}$ & $55.4$ & $1.37\times10^{-3}$ & $6.6\times10^{-4}$\\
Pseudo-Dirac & $\simeq0$ & $0.73$ & $0.262$ & $0.139$\\
Higgsino & $\simeq0$ & $0.070$ & $0.929$ & $0.733$\\
\end{tabular}
\end{ruledtabular}
\end{table}

For the pseudo-Dirac benchmark, the total expected signal in the accepted $5.4$--$300\,\keV$ range is approximately $0.99$ events. The endothermic threshold makes the $5.4$--$55\,\keV$ contribution kinematically inaccessible in the truncated SHM used in our calculation. About $0.73$ events are expected between $55$ and $200\,\keV$ and about $0.26$ events above $200\,\keV$, corresponding to a high-energy fraction of approximately $26\%$. The probability for a signal event to fall within the approximate $1\sigma$ energy interval surrounding the LZ candidate is about $14\%$. Thus, unlike elastic SI scattering, the observed high recoil is not an exceptionally rare tail event in the pseudo-Dirac model, although the spectrum remains relatively broad.

The Higgsino spectrum is considerably more concentrated at high recoil energy. The benchmark again predicts approximately $0.99$ accepted signal events in total, but only about $0.070$ events below $200\,\keV$ and approximately $0.92$ events above $200\,\keV$. Hence about $93\%$ of accepted Higgsino events occur above $200\,\keV$, and approximately $73\%$ fall within the $215.5$--$280.5\,\keV$ interval surrounding the observed candidate. This strong concentration at high recoil energy follows from the larger splitting, $\delta\simeq377\,\keV$, which pushes the scattering close to the Galactic kinematic boundary.

\begin{figure}[t]
\includegraphics[width=\columnwidth]{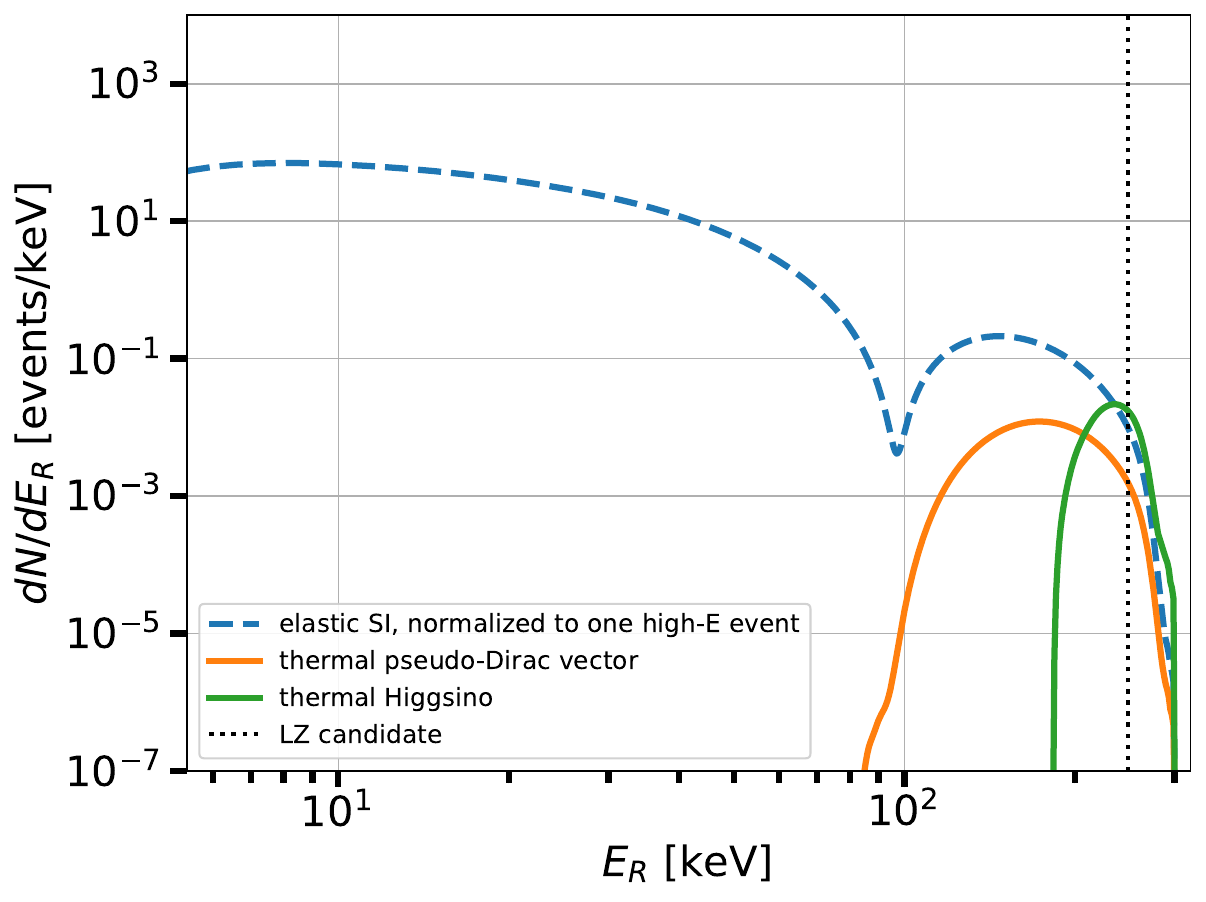}
\caption{Detected recoil spectra in the $2.84$ tonne-year LZ exposure after folding the theoretical recoil rates with the digitized LZ nuclear-recoil efficiency. The dashed elastic-SI spectrum is normalized to one expected event in the $215$--$300\,\keV$ interval. This artificial normalization makes the spectral inconsistency transparent: producing one event near the observed high-energy candidate requires approximately $1.42\times10^3$ additional signal events between $5.4$ and $55\,\keV$. The pseudo-Dirac and Higgsino curves instead use the interaction strengths and mass splittings obtained from the relic-density and LZ-rate analyses. Their total expected yields are approximately one event, but their spectral distributions differ significantly: the pseudo-Dirac signal is broad, with about $26\%$ of the accepted rate above $200\,\keV$, whereas approximately $93\%$ of the Higgsino signal lies above $200\,\keV$. The vertical dotted line marks the $248\,\keV$ LZ candidate. The oscillatory structure of the elastic-SI high-energy tail originates from the xenon nuclear form factor rather than from detector acceptance.}
\label{fig:spectra}
\end{figure}

Fig.~\ref{fig:spectra} provides a direct visual comparison of the three hypotheses. The vertical axis is the expected number of detected events per unit recoil energy, after folding the theoretical recoil spectrum with the LZ efficiency and the $2.84$ tonne-year exposure. The normalization of the elastic-SI curve is deliberately chosen to yield one event in the high-energy window and should therefore not be interpreted as a best-fit elastic-SI model. Its very large low-energy amplitude demonstrates the central problem: the $248\,\keV$ event can be reached only through the extreme tail of a spectrum whose dominant contribution lies at much lower recoil energies.

The pseudo-Dirac spectrum behaves qualitatively differently. The inelastic threshold eliminates the low-energy part of the spectrum and produces a broad distribution extending from approximately the $100\,\keV$ region toward the upper end of the LZ ROI. Its maximum lies below the observed candidate, so it would be inaccurate to state that the pseudo-Dirac spectrum is sharply peaked at $248\,\keV$. Rather, the important point is that the candidate lies within a substantial high-energy tail: approximately one quarter of the accepted pseudo-Dirac signal is predicted above $200\,\keV$. Additional events in this model would therefore be expected over a relatively broad recoil-energy range.

The Higgsino spectrum is much narrower and is shifted further toward high recoil energies because of its larger inelastic splitting. Most of its accepted signal is predicted above $200\,\keV$, and the observed event lies close to the region containing the majority of the expected rate. This provides a much more natural spectral explanation of an isolated high-energy recoil than elastic SI scattering. At the same time, as shown in Fig.~\ref{fig:vmin}, this spectral concentration is obtained by pushing the scattering close to the maximum velocities available in the local halo. The Higgsino interpretation is therefore spectrally more concentrated around the LZ event but astrophysically more sensitive to the extreme velocity tail.

The comparison in Fig.~\ref{fig:spectra} therefore highlights two distinct ways of producing the observed high-energy candidate. The pseudo-Dirac benchmark substantially suppresses the low-energy spectrum while retaining a relatively broad range of allowed recoils, whereas the Higgsino benchmark concentrates almost the entire accepted rate at high recoil energies. Future events would distinguish these possibilities: a population extending down toward $100$--$200\,\keV$ would be more characteristic of the pseudo-Dirac benchmark, while a concentration of events near the upper part of the LZ recoil window would favor the more strongly endothermic Higgsino-like scenario.

\section{A minimal pseudo-Dirac vector model}
\label{sec:minpseudo}

\subsection{Physical origin of the pseudo-Dirac interaction}
\label{sec:pseudodiracorigin}

The inelastic $\mathcal{O}_1$ interaction considered above can arise naturally if DM is not a single isolated state but instead originates from a Dirac fermion that is weakly split into two nearly degenerate Majorana states by small Majorana mass terms \cite{TuckerSmith2001,DeSimone2010}. This pseudo-Dirac construction is particularly attractive for the LZ interpretation because the small mass splitting required by the recoil kinematics does not have to be introduced by hand between two unrelated particles; rather, it results from a small breaking of the symmetry that would otherwise protect an exactly Dirac DM state. Moreover, a vector interaction that would be elastic for the original Dirac fermion becomes predominantly off diagonal in the Majorana mass basis, so that it naturally mediates transitions between the two states, $\chi_1\rightarrow\chi_2$, and thereby generates the endothermic scattering required to reproduce the high-energy recoil spectrum.

We start from two left-handed Weyl fermions, $\xi$ and $\eta$, carrying opposite dark charges. In the absence of Majorana masses they combine into the four-component Dirac field
\begin{equation}
\Psi=
\begin{pmatrix}
\xi_\alpha\\
\eta^{\dagger\dot{\alpha}}
\end{pmatrix},
\end{equation}
with Dirac mass $m_D$. The field $\Psi$ therefore describes a single Dirac particle when the dark fermion number is exactly conserved. A small violation of this symmetry allows Majorana masses for the two Weyl components, leading to
\begin{equation}
\mathcal{L}_{\rm mass}
=-m_D\,\xi\eta
-\frac{1}{2}m_L\,\xi\xi
-\frac{1}{2}m_R\,\eta\eta
+\mathrm{h.c.},
\label{eq:pseudodiracmass}
\end{equation}
where $|m_L|,|m_R|\ll m_D$. In the $(\xi,\eta)$ basis, the corresponding mass matrix is
\begin{equation}
\mathcal{M}=
\begin{pmatrix}
m_L & m_D\\
m_D & m_R
\end{pmatrix}.
\label{eq:pseudodiracmatrix}
\end{equation}

When $m_L=m_R=0$, the two Weyl fields $\xi$ and $\eta$ combine into an exactly Dirac fermion of mass $m_D$. Switching on the small Majorana masses breaks the conserved Dirac fermion number and splits the Dirac state into two nearly degenerate Majorana mass eigenstates, which we denote by $\chi_1$ and $\chi_2$. To leading order in $m_{L,R}/m_D$, their left-handed components are
\begin{align}
\chi_{1L} &\simeq \frac{i}{\sqrt{2}}\left(\xi-\eta\right),\\
\chi_{2L} &\simeq \frac{1}{\sqrt{2}}\left(\xi+\eta\right),
\end{align}
where the factor of $i$ is a convenient phase convention that ensures positive physical masses. The corresponding four-component Majorana fields are obtained by adding the charge-conjugate right-handed components. Their masses are
\begin{align}
m_{\chi_1}&\simeq m_D-\frac{m_L+m_R}{2},\\
m_{\chi_2}&\simeq m_D+\frac{m_L+m_R}{2},
\end{align}
so that
\begin{equation}
\delta\equiv m_{\chi_2}-m_{\chi_1}\simeq m_L+m_R.
\end{equation}

The original Dirac field $\Psi$ can consequently be expressed in terms of the two Majorana mass eigenstates. Up to convention-dependent phases and corrections suppressed by $m_{L,R}/m_D$, one may write schematically
\begin{equation}
\Psi\simeq\frac{1}{\sqrt{2}}\left(\chi_2+i\chi_1\right).
\end{equation}
Thus the $\Psi$ particle is replaced by two almost degenerate Majorana particles, $\chi_1$ and $\chi_2$. A $300\,\keV$ splitting for a TeV-scale DM particle corresponds to a relative symmetry-breaking scale of only $\delta/m_\chi\sim3\times10^{-7}$. Such a small splitting is technically natural because in the limit $m_L,m_R\rightarrow0$ the symmetry associated with conserved Dirac fermion number is restored.

The origin of the inelastic interaction becomes particularly transparent if the original Dirac field couples vectorially to a neutral mediator $V_\mu$,
\begin{equation}
\mathcal{L}\supset g_\chi V_\mu\bar\Psi\gamma^\mu\Psi
+g_qV_\mu\sum_q\bar q\gamma^\mu q.
\label{eq:diracvector}
\end{equation}
Here $\Psi$ is the original four-component Dirac field defined above, before the small Majorana masses are diagonalized. Substituting the decomposition given above into the Dirac vector current shows why the interaction becomes inelastic. A diagonal vector current of a Majorana fermion vanishes identically,
\begin{equation}
\bar\chi_i\gamma^\mu\chi_i=0,
\end{equation}
and therefore the leading vector current that survives connects the two different Majorana mass eigenstates,
\begin{equation}
\bar\Psi\gamma^\mu\Psi
\longrightarrow
i\bar\chi_2\gamma^\mu\chi_1,
\label{eq:offdiagonalcurrent}
\end{equation}
up to the phase convention used in defining $\chi_1$ and $\chi_2$ and corrections suppressed by the small Majorana deformation. For Majorana fields the bilinear $\bar\chi_2\gamma^\mu\chi_1$ is anti-Hermitian, so the combination $i\bar\chi_2\gamma^\mu\chi_1$ is Hermitian. The important physical point is that the mediator couples predominantly between the two states rather than separately to $\chi_1$ or $\chi_2$. Consequently, an incoming ground-state particle $\chi_1$ scatters through
\begin{equation}
\chi_1+A\rightarrow\chi_2+A,
\end{equation}
and must supply the additional energy $\delta$ required to produce the heavier state. The endothermic kinematics discussed in Sec.~\ref{sec:inelastickinematics} therefore follows directly from the pseudo-Dirac structure rather than being imposed as an independent assumption.

At momentum transfers much smaller than $m_V$, the mediator can be integrated out, giving
\begin{equation}
\mathcal{L}_{\rm eff}=
i\frac{g_\chi g_q}{m_V^2}
\left(\bar\chi_2\gamma^\mu\chi_1\right)
\sum_q\bar q\gamma_\mu q.
\label{eq:effvector}
\end{equation}
In the nonrelativistic limit, the time component of the vector current gives the leading contribution. For universal couplings to the light quarks, the matrix element of the quark vector current simply counts the three valence quarks in a nucleon. The effective proton and neutron couplings are therefore equal,
\begin{equation}
c_1^p=c_1^n\equiv c_1^s=\frac{3g_\chi g_q}{m_V^2},
\label{eq:mapo1}
\end{equation}
so that Eq.~\eqref{eq:effvector} reduces to the inelastic isoscalar operator $\mathcal{O}_1^s$ used by LZ. Combining Eq.~\eqref{eq:mapo1} with the LZ normalization in Eq.~\eqref{eq:lzo1conversion2} gives
\begin{equation}
\sigma_N=\frac{\mu_N^2}{\pi}
\left(\frac{3g_\chi g_q}{m_V^2}\right)^2.
\label{eq:sigmageneric}
\end{equation}
This explicit matching is one reason to use the vector pseudo-Dirac model as our baseline realization. Other relativistic interactions can also reduce to $\mathcal{O}_1$ in the nonrelativistic limit, but the vector pseudo-Dirac construction simultaneously explains why the scattering is inelastic and provides a direct connection between the interaction probed by LZ and the coannihilation process relevant in the early Universe.

The model defined by Eqs.~\eqref{eq:pseudodiracmass}--\eqref{eq:effvector} should be regarded as a minimal phenomenological realization rather than a complete ultraviolet theory. In simplified models with a spin-1 mediator, embedding the interaction into a consistent gauge theory generally requires specifying the symmetry-breaking sector that generates $m_V$, the origin of the DM mass terms, and the additional states required by gauge invariance and perturbative consistency \cite{Kahlhoefer2016}. If $V_\mu$ gauges a current involving SM fermions, anomaly cancellation can in addition require new fermionic degrees of freedom, whose masses and interactions may themselves lead to additional phenomenology \cite{FileviezPerez2010,Duerr2014,Ekstedt2016}. In the pseudo-Dirac case, the complete theory must also specify the mechanism that generates the small Majorana masses and determine the cosmological evolution and eventual depletion of the excited state $\chi_2$ \cite{CarrilloGonzalez2021}. These ingredients do not alter the basic direct-detection mechanism described above, but they become important when collider, cosmological, and indirect-detection constraints are considered.

\subsection{Freeze-out and an analytic direct--relic relation}

An important feature of the pseudo-Dirac construction is that the same small mass splitting plays very different roles at chemical freeze-out and in the present Galactic halo, thereby providing a particularly simple connection between the cosmological DM abundance and direct detection. For $m_\chi=1\,\TeV$, a typical freeze-out temperature is $T_f\simeq m_\chi/20\simeq50\,\GeV$, so that for $\delta\simeq300\,\keV$
\begin{equation}
\frac{\delta}{T_f}\sim6\times10^{-6}.
\end{equation}
The two Majorana states $\chi_1$ and $\chi_2$ therefore have nearly identical equilibrium abundances at freeze-out and behave, to excellent accuracy, as the two components of the original quasi-Dirac fermion, allowing efficient $\chi_1\chi_2$ coannihilation. In the Galactic halo, by contrast, the kinetic energy available in a DM--nucleus collision is only of order $\mu_A v^2/2\sim0.1$--$0.4\,\mathrm{MeV}$ for xenon and the velocities relevant here. A splitting of a few hundred keV is therefore completely negligible compared with the thermal energy at freeze-out, but becomes an order-one kinematic threshold for present-day direct detection. This hierarchy of scales is central to the phenomenology: it allows the model to retain an approximately standard thermal history while strongly suppressing low-energy nuclear recoils and shifting the detectable signal toward the high-recoil-energy region probed by the LZ event.

For the vector interaction of Eq.~\eqref{eq:effvector}, the dominant annihilation process at freeze-out is the coannihilation
\begin{equation}
\chi_1\chi_2\rightarrow q\bar q .
\end{equation}
Since $\delta\ll T_f$, the equilibrium abundances of $\chi_1$ and $\chi_2$ are equal. Neglecting subleading annihilation channels, the standard coannihilation formalism \cite{GriestSeckel1991} then gives
\begin{equation}
\langle\sigma v\rangle_{\rm eff}\simeq\frac{1}{2}\langle\sigma v\rangle_{12},
\end{equation}
where the factor $1/2$ follows from the statistical weights of the $12$ and $21$ initial states.

In the heavy-mediator limit, $m_V^2\gg4m_\chi^2$, and assuming universal vector couplings to the six kinematically accessible quark flavors, the leading $s$-wave contribution is approximately
\begin{align}
\langle\sigma v\rangle_{12}&\simeq
\frac{18g_\chi^2g_q^2m_\chi^2}{\pi m_V^4},\\
\langle\sigma v\rangle_{\rm eff}&\simeq
\frac{9g_\chi^2g_q^2m_\chi^2}{\pi m_V^4}.
\label{eq:svgeneric}
\end{align}
The same combination of couplings and mediator mass controls the direct-detection cross section,
\begin{equation}
\sigma_N=
\frac{\mu_N^2}{\pi}
\left(\frac{3g_\chi g_q}{m_V^2}\right)^2
=
\frac{9\mu_N^2g_\chi^2g_q^2}{\pi m_V^4}.
\end{equation}
Comparing the two expressions immediately gives, in natural units $\hbar=c_{\rm light}=1$,
\begin{equation}
\langle\sigma v\rangle_{\rm eff}
\simeq
\frac{m_\chi^2}{\mu_N^2}\,\sigma_N .
\label{eq:masterrelationnatural}
\end{equation}
Here the relative velocity entering $\langle\sigma v\rangle$ is dimensionless in natural units. Consequently, both $\sigma_N$ and $\langle\sigma v\rangle_{\rm eff}$ have dimensions of inverse energy squared in this convention.

When $\sigma_N$ is expressed in $\mathrm{cm}^2$ and the annihilation rate in $\mathrm{cm^3\,s^{-1}}$, one must restore the conversion between a dimensionless velocity in natural units and a velocity in conventional units. The relation then becomes
\begin{equation}
\left[
\frac{\langle\sigma v\rangle_{\rm eff}}
{\mathrm{cm^3\,s^{-1}}}
\right]
\simeq
\frac{m_\chi^2}{\mu_N^2}
\left[
\frac{\sigma_N}{\mathrm{cm^2}}
\right]
\left(2.998\times10^{10}\,\mathrm{cm\,s^{-1}}\right).
\label{eq:masterrelation}
\end{equation}
The numerical factor $2.998\times10^{10}\,\mathrm{cm\,s^{-1}}$ is therefore only a unit-conversion factor corresponding to the speed of light; it is not an additional model parameter. The coefficient in Eq.~\eqref{eq:masterrelation} is derived explicitly in Appendix~\ref{app:direct_relic}.

Eq.~\eqref{eq:masterrelation} is one of the main advantages of this minimal setup. In the heavy-mediator regime, the same combination $g_\chi g_q/m_V^2$ controls both the inelastic DM--nucleon scattering rate and the coannihilation rate relevant for freeze-out. Once the DM mass is specified, the thermal relic requirement therefore predicts the corresponding direct-detection cross section, up to corrections from the finite mediator mass, the resonance region, and the detailed freeze-out calculation.

For a standard radiation-dominated cosmological history, the observed abundance $\Omega_{\rm DM}h^2\simeq0.12$ \cite{Planck2018} is reproduced by an $s$-wave effective annihilation cross section of order a few $10^{-26}\,\mathrm{cm^3\,s^{-1}}$ \cite{Bringmann2012Thermal}. As a representative reference value we adopt
\begin{equation}
\langle\sigma v\rangle_{\rm eff}\simeq2.2\times10^{-26}\,\mathrm{cm^3\,s^{-1}}.
\end{equation}
Using Eq.~\eqref{eq:masterrelation}, for $m_\chi=1\,\TeV$ this corresponds to
\begin{equation}
\sigma_N^{\rm thermal}\simeq6.5\times10^{-43}\,\mathrm{cm^2}.
\label{eq:sigmathermal}
\end{equation}
The significance of this result is that the interaction strength required by thermal freeze-out is naturally of the same order as that needed to produce an observable high-energy LZ event once the inelastic splitting suppresses the otherwise dominant low-energy recoil population.

It is useful to characterize the strength of the contact interaction by defining the effective scale
\begin{equation}
\Lambda\equiv\frac{m_V}{\sqrt{g_\chi g_q}},
\label{eq:Lambda}
\end{equation}
so that the coefficient of the four-fermion operator in Eq.~\eqref{eq:effvector} can be written simply as
\begin{equation}
\frac{g_\chi g_q}{m_V^2}=\frac{1}{\Lambda^2}.
\end{equation}
The thermal benchmark in Eq.~\eqref{eq:sigmathermal} corresponds to
\begin{equation}
\Lambda\simeq6.2\,\TeV.
\end{equation}
Thus, $\Lambda$ should not be interpreted as an additional particle mass: it is the effective suppression scale of the contact interaction and combines the physical mediator mass with its couplings to DM and quarks. For example, if $g_\chi g_q\simeq1$, then $\Lambda\simeq m_V$; for smaller couplings, the same value of $\Lambda$ corresponds to a lighter mediator. Consequently, the value of $\Lambda$ alone does not guarantee the validity of the heavy-mediator approximation. Direct detection requires only $m_V^2\gg |q^2|$, with $|q|\lesssim\mathcal{O}(0.1$--$0.3)\,\GeV$ in the present analysis, whereas the contact approximation for freeze-out is more restrictive and requires $m_V^2\gg s\simeq4m_\chi^2$.

For the benchmark $\Lambda\simeq6.2\,\TeV$ and $m_\chi=1\,\TeV$, the physical mediator mass is
\begin{equation}
m_V\simeq6.2\,\TeV\sqrt{g_\chi g_q}.
\end{equation}
Thus $g_\chi g_q\simeq1$ corresponds to a mediator well above the DM pair threshold, whereas $g_\chi g_q\simeq0.10$ places the mediator close to the $s$-channel resonance $m_V\simeq2m_\chi$. The effective scale $\Lambda$ therefore fixes the direct-detection interaction strength but does not by itself determine whether the freeze-out process lies in the contact or resonant regime.

A particularly important exception occurs close to the $s$-channel resonance,
\begin{equation}
m_V\simeq m_{\chi_1}+m_{\chi_2}\simeq2m_\chi.
\end{equation}
In this region the contact approximation used in Eq.~\eqref{eq:svgeneric} breaks down. The annihilation cross section contains the full mediator propagator,
\begin{equation}
\sigma_{12}v\propto
\frac{g_\chi^2g_q^2}
{(s-m_V^2)^2+m_V^2\Gamma_V^2},
\end{equation}
and must be thermally averaged over the DM velocity distribution. Close to the resonance, $\langle\sigma_{\rm eff}v\rangle$ can vary strongly with temperature, so the relic abundance is controlled by the full freeze-out integral rather than by a nearly constant annihilation cross section. Consequently, the commonly adopted reference value $\langle\sigma v\rangle_{\rm eff}\simeq2.2\times10^{-26}\,\mathrm{cm^3\,s^{-1}}$ should not be imposed as the value of the microscopic cross section at a single representative velocity. The simple direct--relic relation in Eq.~\eqref{eq:masterrelation} is therefore valid only in the off-resonance heavy-mediator regime.

This observation also opens an additional phenomenological possibility. Near the resonance, the observed relic abundance can be obtained with smaller values of $g_\chi g_q$ than in the contact-limit thermal benchmark, thereby reducing the direct-detection cross section. Conversely, reproducing the LZ event would then select a restricted combination of the mediator mass, width, couplings, and inelastic splitting rather than the essentially one-dimensional relation of Eq.~\eqref{eq:masterrelation}. A dedicated resonant analysis would therefore require solving the Boltzmann equation with the full velocity-dependent propagator rather than using the analytic contact-limit estimate.

Away from the resonance, and for mediator masses sufficiently above the DM pair-production threshold, the contact result provides a transparent first approximation. Propagator, finite-width, quark-mass, and freeze-out corrections can still modify the numerical relation at the tens-of-percent level, while collider constraints depend strongly on the mediator mass, its branching fractions, and the ultraviolet completion. We therefore use Eq.~\eqref{eq:masterrelation} as a benchmark correlation rather than as a model-independent prediction.

\subsection{Matching the LZ rate}

Having established the kinematic region in which endothermic scattering can populate the high-recoil-energy LZ window, we now ask whether the interaction strength required by the thermal relic abundance can simultaneously reproduce the observed event rate. In the minimal pseudo-Dirac model, this test is particularly predictive because the same combination of couplings and mediator mass controls both the early-Universe coannihilation rate and the inelastic DM--nucleon scattering cross section. We therefore first fix the cross section to the thermal value derived in Eq.~\eqref{eq:sigmathermal} and vary the inelastic splitting $\delta$, which determines how much of the Galactic velocity distribution can contribute to the LZ recoil window.

For $m_\chi=1\,\TeV$, integrating Eq.~\eqref{eq:nsig} over the LZ recoil-energy range, including the digitized energy-dependent efficiency and the annual-averaged SHM velocity distribution, gives approximately one expected signal event for
\begin{equation}
\begin{gathered}
m_\chi=1.0\,\TeV,\qquad \delta\simeq297\,\keV,\\
\sigma_N\simeq6.5\times10^{-43}\,\mathrm{cm}^2.
\end{gathered}
\label{eq:genericbenchmark}
\end{equation}
At the observed recoil energy this benchmark requires
\begin{equation}
v_{\min}(248\,\keV)\simeq701\,\mathrm{km/s},
\end{equation}
while the minimum of $v_{\min}(E_R)$ occurs at $E_R^\star\simeq265\,\keV$. The observed candidate therefore lies close to the kinematically favored recoil-energy region of the benchmark rather than in a strongly suppressed tail.

\begin{figure}[t]
\includegraphics[width=\columnwidth]{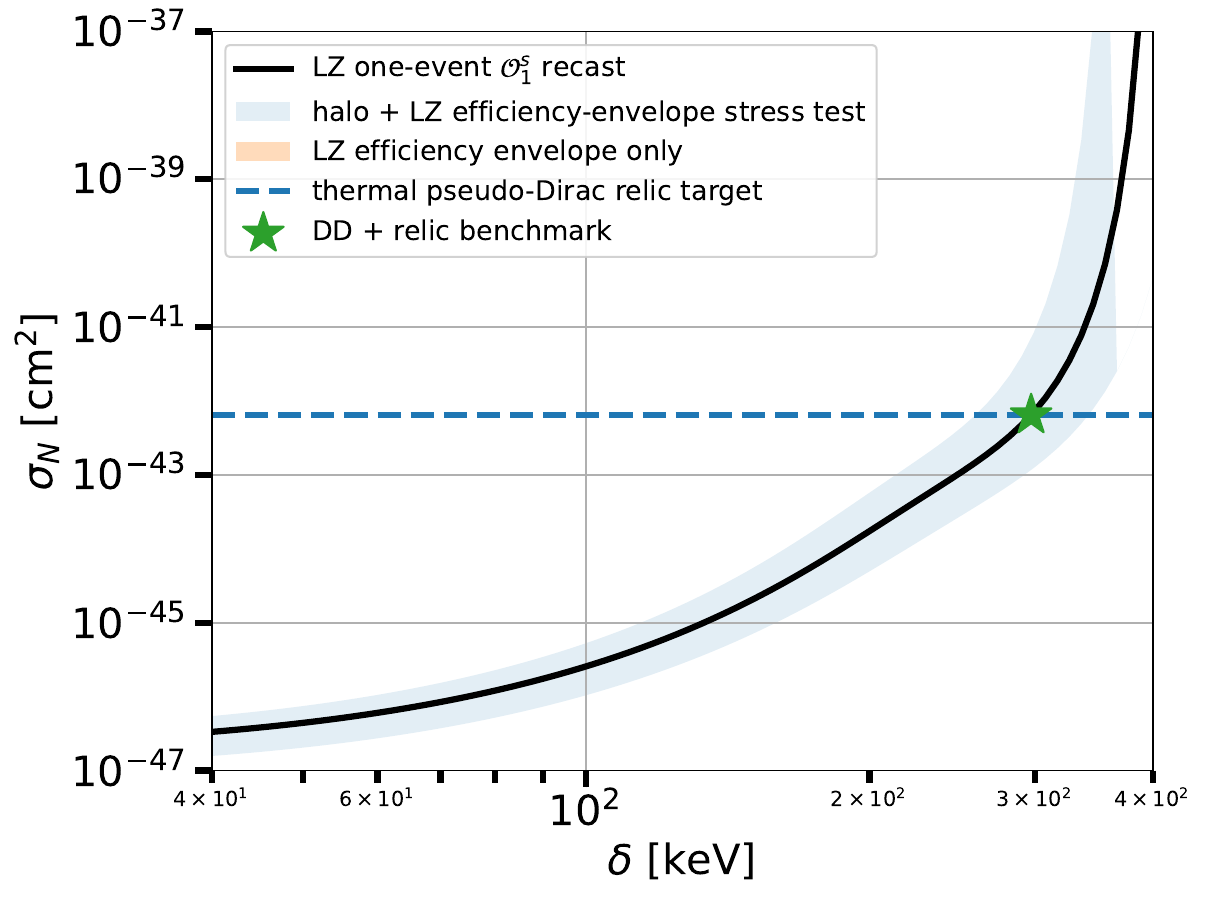}
\caption{Per-nucleon cross section required to produce approximately one LZ signal event as a function of the inelastic splitting for a $1\,\TeV$ isoscalar $\mathcal{O}_1$ interaction. The solid black curve is the central recoil-level recast obtained with the nominal SHM parameters and the vector-digitized central LZ efficiency. The narrow shaded band propagates only the uncertainty in the LZ efficiency, obtained from the lower and upper boundaries of the envelope reported in Fig.~S2 of Ref.~\cite{LZ2026Extended}. The broader shaded band additionally includes conservative variations of the local DM density and Galactic velocity parameters and should be interpreted as an astrophysical-plus-detector stress test rather than as a statistical confidence interval. The horizontal dashed line is the direct-detection cross section implied by the thermal relic-density requirement of the pseudo-Dirac model. The star marks the intersection of the thermal relic target with the central one-event LZ curve, defining the benchmark of Eq.~\eqref{eq:genericbenchmark}.}
\label{fig:sigmadelta}
\end{figure}

Fig.~\ref{fig:sigmadelta} shows how the cross section required to produce approximately one LZ signal event changes with $\delta$ at fixed $m_\chi=1\,\TeV$. For each value of the splitting, the solid black curve gives the value of $\sigma_N$ for which the predicted accepted signal yield is approximately one event. At small and moderate $\delta$, the required cross section rises gradually because the inelastic threshold progressively removes lower-velocity DM particles from the scattering phase space. Once $\delta$ approaches the largest splitting accessible to the Galactic halo, however, only particles in the extreme high-velocity tail can scatter and the required cross section rises very rapidly.
The horizontal dashed line in Fig.~\ref{fig:sigmadelta} has a different origin from the black curve. It is fixed by the early-Universe relic-density calculation and is approximately independent of $\delta$ as long as $\delta\ll T_f$, so that $\chi_1$ and $\chi_2$ remain effectively degenerate during freeze-out. The intersection between the two curves selects the central pseudo-Dirac benchmark, $\delta\simeq297\,\keV$. Thus Fig.~\ref{fig:sigmadelta} directly displays the consistency condition between cosmology and direct detection: points on the black curve reproduce the LZ event normalization, points on the horizontal line reproduce the reference thermal relic abundance, and their intersection satisfies both requirements simultaneously.
The two shaded regions quantify the stability of this conclusion. Varying only the LZ efficiency moves the one-event solution from $\delta\simeq295.6$ to $297.8\,\keV$, so the efficiency-only band is extremely narrow and lies close to the central black curve. By contrast, including the halo and local-density stress-test variations broadens the solution to approximately $\delta\simeq259$--$342\,\keV$. The dominant uncertainty is therefore associated with the poorly constrained high-speed tail of the local DM distribution rather than with the detector efficiency. The analogous halo-plus-efficiency range for the Higgsino benchmark is approximately $337$--$427\,\keV$, reflecting its greater proximity to the kinematic ceiling discussed in Sec.~\ref{sec:inelastickinematics}. The halo variations used here, including $\rho_{\rm DM}=0.2$--$0.6\,\GeV\,\mathrm{cm}^{-3}$, $v_0=220$--$250\,\mathrm{km/s}$ and $v_{\rm esc}=500$--$600\,\mathrm{km/s}$, are intended as conservative stress-test corners and not as a statistically defined halo posterior.

\begin{figure}[t]
\includegraphics[width=\columnwidth]{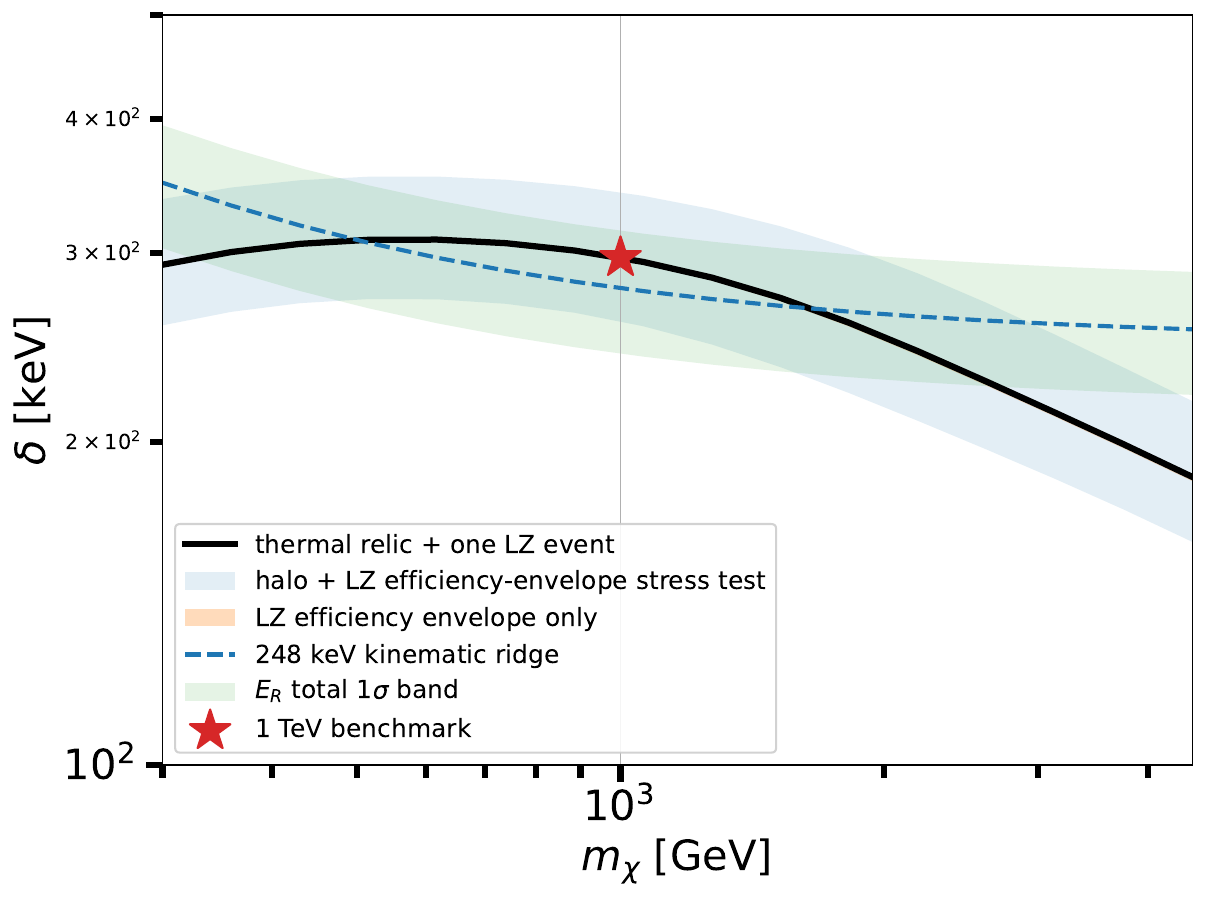}
\caption{Combined relic-density and direct-detection relation in the $(m_\chi,\delta)$ plane for the minimal universal-vector pseudo-Dirac model. The solid black curve contains the parameter combinations for which the thermal relic relation fixes the interaction strength and the corresponding LZ rate is approximately one event, using the nominal SHM parameters and central LZ efficiency. The narrow shaded band propagates the LZ efficiency uncertainty alone, whereas the broader band additionally includes the conservative halo and local-density stress-test variations. The dashed curve is the $248\,\keV$ kinematic ridge defined by $E_R^\star=248\,\keV$, namely the locus for which the observed recoil coincides with the minimum of $v_{\min}(E_R)$. The surrounding band propagates the total experimental uncertainty on the reconstructed recoil energy and therefore represents an uncertainty on the kinematic ridge rather than on the predicted event rate. The star marks the $m_\chi=1\,\TeV$ benchmark obtained from the simultaneous relic-density and one-event conditions.}
\label{fig:globalddrelic}
\end{figure}

A complementary view is obtained by allowing the DM mass to vary. For each value of $m_\chi$, the thermal relic relation in Eq.~\eqref{eq:masterrelation} fixes the corresponding scattering cross section, and we then determine the splitting $\delta$ for which the expected LZ signal yield is approximately one event. Fig.~\ref{fig:globalddrelic} therefore projects the simultaneous thermal-relic and direct-detection conditions directly into the $(m_\chi,\delta)$ plane.
The solid black curve in Fig.~\ref{fig:globalddrelic} is the central result of the minimal pseudo-Dirac model: every point on this locus simultaneously reproduces the reference thermal relic abundance and approximately one accepted LZ signal event. Within the mass range explicitly explored, $300\,\GeV\lesssim m_\chi\lesssim4.5\,\TeV$, such a solution exists continuously. Along the central curve, the required splitting is approximately $293\,\keV$ at $m_\chi=300\,\GeV$, remains of order $300\,\keV$ through the several-hundred-GeV to TeV region, and decreases to about $186\,\keV$ at $m_\chi=4.5\,\TeV$. The relatively mild variation around the TeV scale results from the interplay between the mass dependence of the thermal scattering cross section and the inelastic kinematics required to populate the high-recoil-energy LZ window.
The dashed curve in Fig.~\ref{fig:globalddrelic} has a purely kinematic interpretation. It is the $248\,\keV$ ridge introduced in Sec.~\ref{sec:inelastickinematics}, for which the observed recoil occurs exactly at the minimum of $v_{\min}(E_R)$. The relic-plus-LZ curve is not required to coincide with this ridge because the predicted event rate depends on the complete recoil spectrum, the halo velocity integral, the nuclear response, and the detector efficiency rather than only on the position of the minimum of $v_{\min}$. Nevertheless, the two loci remain relatively close over much of the parameter region, showing that the rate calculation naturally favors splittings for which the observed recoil lies near the kinematically preferred part of the spectrum.
The different shaded regions in Fig.~\ref{fig:globalddrelic} should also be distinguished carefully. The bands surrounding the solid black curve arise from uncertainties in the detector efficiency and halo model and therefore quantify how the rate-matching relation shifts under these variations. The band surrounding the dashed ridge instead propagates the total recoil-energy uncertainty,
\begin{equation}
E_R=248\pm\sqrt{23^2+23^2}\,\keV,
\end{equation}
through the relation $E_R^\star=\delta\mu_A/m_A$ and therefore describes only the uncertainty in the location of the kinematic ridge.
The endpoints of the displayed mass interval should not be interpreted as physical exclusion boundaries. Fig.~\ref{fig:globalddrelic} shows that the thermal relic-density condition and the LZ event can be simultaneously satisfied throughout the scanned range, but it does not by itself determine a finite allowed interval in $m_\chi$. Additional constraints can further restrict this parameter space, including the validity of the heavy-mediator approximation, perturbativity, the mediator mass and width, collider searches, indirect detection, and the cosmological evolution of the excited state. A fully allowed mass range therefore requires specifying the ultraviolet completion and imposing these additional constraints consistently.

\begin{figure}[t]
\includegraphics[width=\columnwidth]{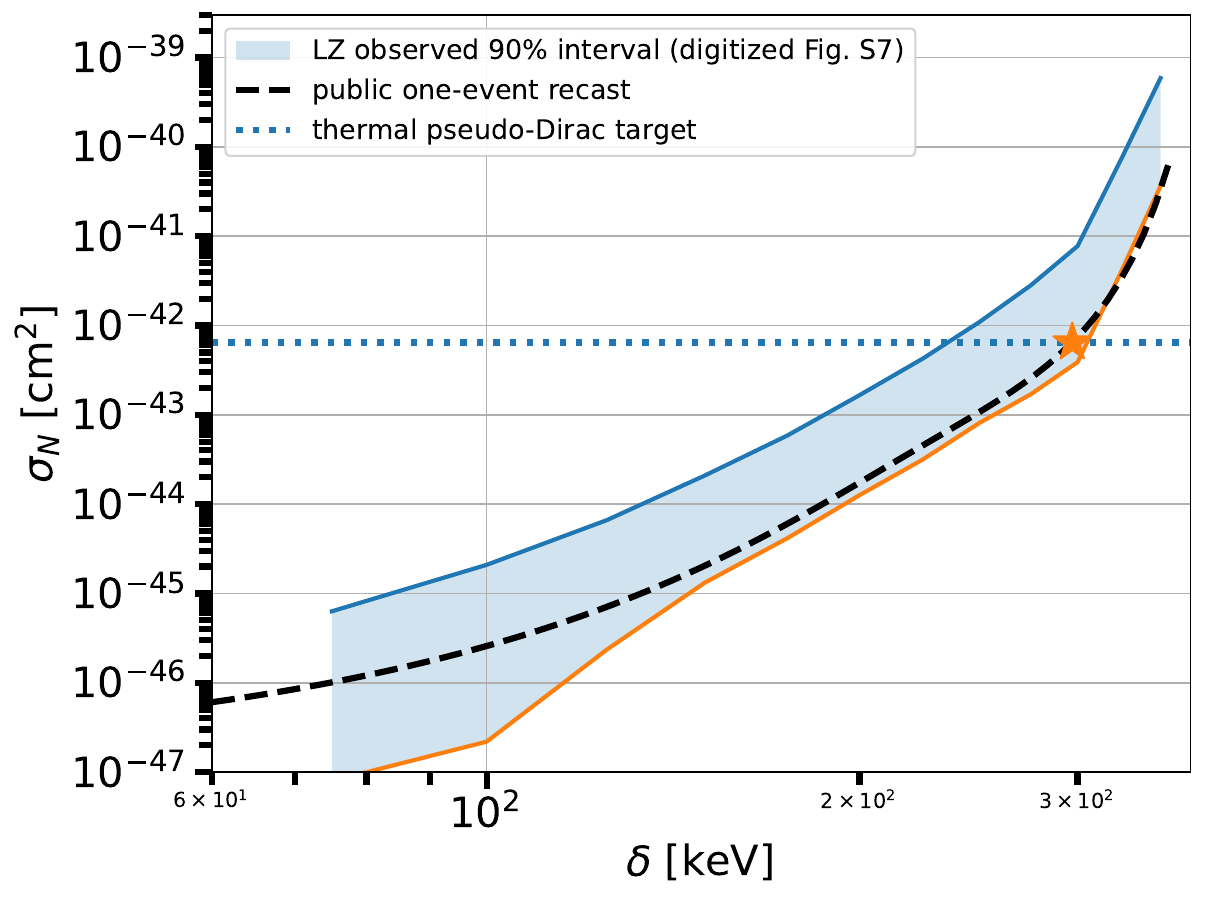}
\caption{Comparison between the simplified public recast and the LZ inelastic $\mathcal{O}_1^s$ result at $m_\chi=1\,\TeV$. The shaded region is an approximate digitization of the observed two-sided $90\%$ interval reported in LZ Fig.~S7, using the scalar/isoscalar cross-section normalization defined in the Supplemental Material. The dashed black curve is our recoil-level cross section required to produce $N_{\rm sig}=0.989$ accepted signal events for each value of $\delta$. The horizontal dotted line is the thermal pseudo-Dirac cross-section target obtained from the relic-density relation, and the star marks its intersection with the one-event recast. The digitized LZ interval is used only for validation of the normalization and overall $\delta$ dependence and is not included in the fit.}
\label{fig:lzo1comparison}
\end{figure}

Finally, we compare the normalization of the simplified recoil-level recast with the result published directly by LZ. The Supplemental Material provides the observed two-sided $90\%$ interval for inelastic $\mathcal{O}_1^s$ scattering as a function of $\delta$ at $m_\chi=1\,\TeV$. Fig.~\ref{fig:lzo1comparison} compares an approximate digitization of this interval with our one-event recast. The purpose of this comparison is only to test whether the simplified calculation reproduces the correct cross-section scale and the characteristic dependence on $\delta$; it is not intended to reproduce the full LZ likelihood.

The three ingredients in Fig.~\ref{fig:lzo1comparison} have distinct meanings. The shaded region is the experimentally inferred LZ interval and incorporates the Collaboration's full profile-likelihood analysis, including the detector response, background model, and nuclear-response calculation. Its rapid rise at large $\delta$ reflects the loss of kinematically accessible halo particles as the endothermic threshold approaches the maximum available DM speed. The dashed black curve is instead an event-normalization relation: at every value of $\delta$, $\sigma_N$ is adjusted until the simplified recoil-level calculation predicts approximately one accepted event. It is therefore not a confidence interval.

The horizontal dotted line represents the interaction strength required by the reference thermal relic abundance. Its intersection with the dashed curve reproduces the benchmark $\delta\simeq297\,\keV$ and $\sigma_N\simeq6.5\times10^{-43}\,\mathrm{cm}^2$ already identified in Fig.~\ref{fig:sigmadelta}. The fact that this point lies within the broad digitized LZ interval, and that the simplified one-event curve follows the same steep increase with $\delta$, provides a useful validation of the normalization and overall kinematic behavior of the recast. It should not, however, be interpreted as evidence that the recoil-level calculation reproduces the official LZ confidence construction, which would require the full event-level likelihood and detector response.

\subsection{Indirect detection in the generic model}

The pseudo-Dirac construction has an important consequence for indirect detection. The interaction responsible for the relic abundance is not necessarily the same annihilation process that remains active in the present Universe. As discussed above, the vector current is predominantly off diagonal in the Majorana mass basis,
\begin{equation}
V_\mu\,\bar\Psi\gamma^\mu\Psi
\;\longrightarrow\;
iV_\mu\bar\chi_2\gamma^\mu\chi_1.
\end{equation}
Consequently, the mediator couples one $\chi_1$ and one $\chi_2$, rather than two identical ground-state particles. At freeze-out this does not suppress the annihilation rate because $\delta\ll T_f$ and the two states have almost equal equilibrium abundances. The dominant process is therefore
\begin{equation}
\chi_1\chi_2\rightarrow V^\ast\rightarrow q\bar q,
\end{equation}
which provides the thermal coannihilation rate used in the relic-density calculation.

The situation can be very different at late times. After freeze-out, the heavier state $\chi_2$ may be depleted, while the cosmological DM abundance becomes dominated by the lighter and stable state $\chi_1$. If the present-day halo contains essentially only $\chi_1$, the tree-level $s$-channel process
\begin{equation}
\chi_1\chi_1\rightarrow V^\ast\rightarrow q\bar q
\end{equation}
is absent for the purely off-diagonal vector interaction, because a single $V$ vertex always converts $\chi_1$ into $\chi_2$. Equivalently, the diagonal Majorana vector current $\bar\chi_1\gamma^\mu\chi_1$ vanishes. The annihilation channel that was efficient during freeze-out, $\chi_1\chi_2\rightarrow q\bar q$, therefore becomes strongly suppressed today if very few $\chi_2$ particles survive. This provides a natural separation between the thermal relic-density requirement and present-day indirect-detection constraints.

This suppression can be made more quantitative by introducing the present-day excited-state fraction
\begin{equation}
f_2\equiv\frac{n_2}{n_1+n_2}.
\end{equation}
The annihilation rate per unit volume associated with the coannihilation channel is proportional to
\begin{equation}
\Gamma_{\rm ann}\propto n_1n_2\,\langle\sigma v\rangle_{12}.
\end{equation}
At freeze-out, $n_1\simeq n_2$ and therefore $f_2\simeq1/2$, so this channel is unsuppressed. In the present halo, if $f_2\ll1$, one has $n_2\simeq f_2 n_{\rm DM}$ and $n_1\simeq n_{\rm DM}$, so the indirect-detection rate becomes approximately proportional to $f_2\langle\sigma v\rangle_{12}$. Relative to the quasi-Dirac population at freeze-out, the present-day annihilation signal is therefore parametrically suppressed by the small residual abundance of the excited state. Thus, even though $\langle\sigma v\rangle_{12}$ can be of thermal strength, the observable gamma-ray or cosmic-ray signal today can be much smaller.

An additional annihilation channel can arise if the mediator is lighter than the DM particle. In particular, processes such as
\begin{equation}
\chi_1\chi_1\rightarrow VV
\end{equation}
can proceed through $t$- and $u$-channel exchange of $\chi_2$ when $m_\chi>m_V$. Such a channel would remain active even if the halo contains only $\chi_1$ and could therefore restore strong indirect-detection constraints. For this reason, in the simple benchmark considered here we focus on the regime $m_V>m_\chi$, for which annihilation into two on-shell mediators is kinematically forbidden. Off-shell, loop-induced, or higher-order annihilation processes can still be present, but they are model dependent and are generally suppressed relative to the freeze-out coannihilation channel.

The crucial assumption is therefore not merely the off-diagonal interaction, but also an efficient depletion of $\chi_2$ after chemical freeze-out. The excited-state abundance cannot in general be assumed to vanish automatically. After the total DM abundance freezes out, processes such as
\begin{equation}
\chi_2\chi_2\rightarrow\chi_1\chi_1
\end{equation}
can continue to convert the heavier state into the lighter one. Depending on the ultraviolet completion, $\chi_2$ may also decay into $\chi_1$ plus lighter dark-sector or SM states, or it may remain in chemical contact with the thermal bath through inelastic scattering. These processes can strongly reduce the excited-state fraction before recombination and before structure formation \cite{CarrilloGonzalez2021,Brahma2023,Chao2023Excited}. Their efficiency depends on the mediator mass, the coupling ratio $g_\chi/g_q$, the splitting $\delta$, and on any additional light states present in the complete theory.

Conversely, if a non-negligible population of $\chi_2$ survives, the coannihilation process $\chi_1\chi_2\rightarrow q\bar q$ becomes active again at late times. Energy injection around recombination can then be constrained by the CMB, while annihilation in Galactic halos can produce gamma rays, antiprotons, or other cosmic-ray signals \cite{Slatyer2016}. For this reason, the statement that the generic pseudo-Dirac model evades indirect detection is conditional: it requires a sufficiently small late-time excited-state fraction.

In the proof-of-principle analysis presented here, we therefore impose efficient depletion of $\chi_2$ as a consistency condition rather than deriving it from the simplified interaction alone. A complete calculation should follow the coupled Boltzmann equations for the two abundances,
\begin{equation}
Y_1=\frac{n_1}{s},
\qquad
Y_2=\frac{n_2}{s},
\end{equation}
including annihilation, coannihilation, conversion, and decay processes, and determine the residual excited-state fraction at recombination and in the present halo. This calculation can depend on parameters that are not fixed by the direct-detection rate alone, in particular the separate values of $g_\chi$ and $g_q$ rather than only their product $g_\chi g_q$.

The separation between the freeze-out and present-day populations is therefore one of the main phenomenological advantages of the pseudo-Dirac interpretation. During freeze-out, $\delta\ll T_f$ and both states are populated, allowing the thermal coannihilation rate required for $\Omega_{\rm DM}h^2\simeq0.12$. At late times, the same interaction can produce a much smaller annihilation signal if $\chi_2$ has been efficiently removed. This makes it possible, at least in principle, to reconcile the interaction strength suggested by the LZ event with the thermal relic abundance while avoiding the indirect-detection signal that would normally accompany an $s$-wave WIMP with $\langle\sigma v\rangle\sim10^{-26}\,\mathrm{cm^3\,s^{-1}}$.

A complete ultraviolet realization must therefore satisfy several conditions simultaneously: it must preserve the approximately off-diagonal interaction responsible for inelastic direct detection, reproduce the required thermal coannihilation rate, provide an efficient mechanism for depleting $\chi_2$ before the epochs constrained by the CMB and present-day indirect searches, forbid or sufficiently suppress alternative late-time annihilation channels such as $\chi_1\chi_1\rightarrow VV$, and remain compatible with collider, flavor, and other laboratory constraints. Establishing all of these conditions requires a dedicated ultraviolet model and a coupled cosmological analysis, which goes beyond the minimal phenomenological benchmark considered here.

\section{Thermal Higgsino completion}

We report here the basic aspects of the thermal Higgsino model and report in Appendix \ref{sec:higgsinotheory} a more detailed discussion.

\subsection{Why the Higgsino is a natural inelastic SI completion}

The generic pseudo-Dirac model discussed in Sec.~\ref{sec:minpseudo} shows that an off-diagonal vector interaction can naturally generate endothermic $\mathcal{O}_1$ scattering while maintaining an approximately standard thermal history. A particularly compelling realization of this mechanism is provided by a nearly pure Higgsino. In this case the pseudo-Dirac structure, the mediator responsible for direct detection, and the thermal annihilation channels are all already contained in the electroweak sector of supersymmetry. The model therefore contains considerably less freedom than the generic vector-mediator construction: the relevant neutral-current interaction is fixed by the SM weak coupling, while the neutral-state mass splitting is generated by mixing with the heavy Majorana gauginos \cite{NagataShirai2015,KrallReece2018,Graham2025}. This makes the Higgsino one of the simplest ultraviolet-motivated examples of inelastic DM and, in particular, a natural realization of the high-energy nuclear-recoil phenomenology probed by LZ.

In the limit in which the bino and wino masses, $M_1$ and $M_2$, are much larger than the Higgsino mass parameter $\mu$, the two neutral Higgsino Weyl fields $\widetilde H_u^0$ and $\widetilde H_d^0$ approximately form a Dirac fermion. In the formal limit $M_1,M_2\rightarrow\infty$, an approximate Higgsino-number symmetry is restored and the two neutral states become degenerate. For finite gaugino masses, electroweak-symmetry breaking induces small bino--Higgsino and wino--Higgsino mixings. Because the bino and wino are Majorana fermions, these mixings break the approximate Higgsino-number symmetry and split the neutral Dirac state into two nearby Majorana mass eigenstates, which we denote by $\widetilde H_1$ and $\widetilde H_2$ \cite{NagataShirai2015,KrallReece2018}. To leading order in the inverse gaugino masses, the neutral-state splitting is
\begin{equation}
\delta\equiv m_{\widetilde H_2}-m_{\widetilde H_1}
\simeq
m_Z^2
\left(
\frac{\sin^2\theta_W}{M_1}
+
\frac{\cos^2\theta_W}{M_2}
\right),
\label{eq:higgsinosplit}
\end{equation}
up to sign conventions and subleading corrections that depend on $\tan\beta$, the relative signs of $\mu$, $M_1$, and $M_2$, and threshold effects in the full supersymmetric spectrum. The important point for the present analysis is that splittings in the $100$--$1000\,\keV$ range arise naturally when the electroweak gauginos are very heavy. The nearly pure Higgsino is therefore precisely a pseudo-Dirac system of the type discussed in Sec.~\ref{sec:pseudodiracorigin}.

The direct-detection interaction is particularly simple. Before the neutral Higgsino Dirac state is split, it couples to the $Z$ boson through the electroweak neutral current. After diagonalization into the two Majorana states, the diagonal vector currents vanish, while the $Z$ interaction becomes dominantly off diagonal,
\begin{equation}
iZ_\mu\,
\overline{\widetilde H}_2\gamma^\mu\widetilde H_1
\end{equation}
\cite{NagataShirai2015,KrallReece2018}. Consequently, the leading tree-level $Z$-mediated nuclear process is not ordinary elastic scattering of the lightest neutral state, but the endothermic transition
\begin{equation}
\widetilde H_1+A\rightarrow\widetilde H_2+A.
\end{equation}
The Higgsino therefore realizes the same basic mechanism as the generic pseudo-Dirac model: the lightest state must acquire the energy $\delta$ required to produce its slightly heavier Majorana partner.

At momentum transfers relevant for direct detection, the $Z$ boson can be integrated out and the dominant nonrelativistic interaction is the identity operator $\mathcal{O}_1$. The Higgsino is therefore closely related to the inelastic $\mathcal{O}_1$ scenario considered by LZ. There is, however, an important difference from the universal-vector model introduced above. The $Z$ does not couple equally to protons and neutrons. Its coherent vector coupling is proportional to the nuclear weak charge
\begin{equation}
Q_W=N-(1-4\sin^2\theta_W)Z,
\end{equation}
so that the interaction is strongly neutron dominated because $1-4\sin^2\theta_W\simeq0.075$. Thus the Higgsino has the same recoil kinematics and leading nonrelativistic operator as inelastic $\mathcal{O}_1$ scattering, but a different proton--neutron coupling structure. This is the precise sense in which the inelastic $\mathcal{O}_1$ model discussed by LZ resembles a Higgsino scenario.

For a xenon isotope, the corresponding zero-momentum weak-charge cross section can be written as
\begin{equation}
\sigma_A=
\frac{G_F^2\mu_A^2}{2\pi}
\left[
N-(1-4\sin^2\theta_W)Z
\right]^2.
\end{equation}
It is useful to define the associated neutron-normalized cross section,
\begin{equation}
\sigma_n^{\widetilde H}
=
\frac{G_F^2\mu_N^2}{2\pi}
\simeq
7.4\times10^{-39}\,\mathrm{cm}^2,
\label{eq:higgsinosigma}
\end{equation}
for $m_\chi\simeq1.1\,\TeV$, where $\mu_N$ is the DM--nucleon reduced mass. This value is not a free parameter of the model: it follows directly from the SM weak interaction.\footnote{Several treatments in the literature use the smaller $G_F^2\mu^2/(8\pi)$ normalization for the Higgsino nuclear cross section, e.g., Refs.~\cite{Graham2025,KrallReece2018,NagataShirai2015}. The correct vector-like Higgsino result is larger by a factor of four. The discrepancy originates from assigning an additional factor $1/2$ to the off-diagonal neutral-Higgsino transition current; its derivation is given in Appendix~G and further discussed in Ref.~\cite{PospelovRamani2026}.} The Supplemental Material of the LZ analysis provides a useful independent normalization check, showing that converting from the scalar/isoscalar $\mathcal{O}_1^s$ convention to the weak vector charge changes the xenon cross-section normalization by a factor of approximately $3.2$ \cite{LZ2026Extended}. In our calculation we do not apply a single average conversion factor; instead, we evaluate the weak charge separately for each xenon isotope.

The Higgsino is also considerably more predictive than the generic pseudo-Dirac model from the cosmological point of view. A nearly pure Higgsino belongs to an electroweak multiplet containing two neutral Majorana states and a nearly degenerate charged state. During freeze-out, annihilation and coannihilation among these states into electroweak gauge bosons determine the relic abundance. In a standard radiation-dominated cosmology, a nearly pure thermal Higgsino reproduces the observed DM density for a mass close to $1.1\,\TeV$ \cite{KrallReece2018,Rodd2024,Graham2025}; this characteristic thermal mass has long made the Higgsino one of the canonical electroweak WIMP candidates. Recent indirect-detection studies likewise use the $m_\chi\simeq1.1\,\TeV$ thermal Higgsino as a sharply defined benchmark \cite{Rodd2024}.

The resulting phenomenology is therefore unusually constrained. In the generic vector pseudo-Dirac model, $m_\chi$, $g_\chi g_q/m_V^2$, and $\delta$ can in principle be varied independently before the relic-density and direct-detection requirements are imposed. For the nearly pure Higgsino, instead, the thermal relic abundance selects
\begin{equation}
m_\chi\simeq1.1\,\TeV,
\end{equation}
the $Z$ interaction fixes the leading inelastic scattering strength in Eq.~\eqref{eq:higgsinosigma}, and the remaining quantity most directly selected by the LZ event is the neutral-state splitting $\delta$. The high-energy LZ candidate can therefore be interpreted, within this framework, primarily as a probe of how close the Higgsino splitting lies to the maximum endothermic threshold accessible to the local Galactic velocity distribution.

This predictivity is also what makes the Higgsino interpretation particularly testable. The same electroweak structure that fixes its thermal relic abundance implies substantial annihilation into electroweak gauge bosons in the present Universe and generates characteristic gamma-ray continuum and line signals. Thermal Higgsino DM near $1.1\,\TeV$ is consequently an important target for current and future indirect searches with H.E.S.S., CTAO, and related gamma-ray experiments \cite{KrallReece2018,Rodd2024}. Unlike the generic pseudo-Dirac model, where late-time annihilation can be strongly suppressed if the excited state is depleted, the Higgsino therefore connects the LZ interpretation directly to an experimentally testable indirect-detection prediction.

\subsection{The splitting required by LZ}

The nearly pure Higgsino scenario is much more predictive than the generic pseudo-Dirac model because both the DM mass and the leading inelastic scattering strength are essentially fixed. For a standard thermal history, the relic abundance selects $m_\chi\simeq1.1\,\TeV$, while the $Z$-mediated weak interaction fixes the scattering cross section in Eq.~\eqref{eq:higgsinosigma}. The principal quantity that remains to be determined by the LZ event is therefore the neutral-state mass splitting $\delta$.

Using $m_\chi=1.1\,\TeV$, the isotope-dependent weak nuclear charge, natural xenon abundances, the vector-digitized final LZ efficiency, and the annually averaged SHM velocity distribution, we determine the splitting for which the predicted signal yield is approximately one event in the present LZ exposure. This gives
\begin{equation}
\begin{gathered}
m_\chi=1.1\,\TeV,\qquad \delta\simeq377\,\keV,\\
\sigma_n^{\widetilde H}\simeq7.4\times10^{-39}\,\mathrm{cm}^2.
\end{gathered}
\label{eq:higgsinobenchmark}
\end{equation}
Unlike the generic pseudo-Dirac benchmark, where the interaction strength follows from the simplified relic relation, the Higgsino cross section is fixed by the SM weak interaction. The value $\delta\simeq377\,\keV$ is therefore selected almost entirely by the requirement that the predicted recoil rate be reduced to approximately one accepted LZ event.

The kinematics of this benchmark are unusually extreme. For scattering on $A=131$, the observed recoil requires
\begin{equation}
v_{\min}(248\,\keV)\simeq795\,\mathrm{km/s},
\end{equation}
which is only about $7\,\mathrm{km/s}$ below the mean detector-frame SHM cutoff,
\begin{equation}
v_{\rm esc}+\bar v_E\simeq794\,\mathrm{km/s}.
\end{equation}
The event therefore probes particles very close to the maximum speed available in the local halo. Equivalently, using Eq.~\eqref{eq:deltamax}, the largest splitting accessible for $m_\chi=1.1\,\TeV$ is approximately
\begin{equation}
\delta_{\max}^{\rm mean}\simeq385\,\keV
\end{equation}
when the mean Earth velocity is used, and increases to about
\begin{equation}
\delta_{\max}^{\rm summer}\simeq400\,\keV
\end{equation}
near the annual maximum of the detector-frame speed. The preferred value $\delta\simeq377\,\keV$ therefore lies only a few keV below the mean-speed kinematic ceiling.

This should be contrasted with the generic pseudo-Dirac benchmark, for which $\delta\simeq297\,\keV$ and $v_{\min}(248\,\keV)\simeq701\,\mathrm{km/s}$. The generic model probes the high-velocity tail but still has a substantial kinematic margin, whereas the Higgsino interpretation relies on a much narrower population of particles near the endpoint of the halo distribution. This difference explains both the narrower Higgsino recoil spectrum discussed above and its much stronger sensitivity to the astrophysical velocity distribution.

The location of the minimum of $v_{\min}(E_R)$ provides another useful way of understanding the spectrum. For the Higgsino benchmark,
\begin{equation}
E_R^\star=\delta\frac{\mu_A}{m_A}\simeq339\,\keV.
\end{equation}
This lies above the recoil energy of the observed event and close to, or beyond, the region where the LZ acceptance begins to fall rapidly. The $248\,\keV$ candidate therefore lies on the lower-energy side of the kinematically preferred recoil scale. The event is not centered exactly at $E_R^\star$ because the observable rate is determined by the product of the halo velocity integral, nuclear response, and detector efficiency, rather than by kinematics alone.

The importance of the high-velocity tail can also be seen directly by repeating the rate calculation without annual averaging. If the Earth speed is fixed to its mean value, the splitting required to obtain one event decreases by a few keV. Although this shift is small in $\delta$, it is significant because the benchmark lies close to the kinematic endpoint: a small change in the maximum available speed produces a large change in the number of halo particles capable of scattering.

The required splitting can finally be related to the heavy electroweak gaugino scale. If, for illustration, the bino and wino masses are equal,
\begin{equation}
M_1=M_2\equiv M,
\end{equation}
Eq.~\eqref{eq:higgsinosplit} reduces approximately to
\begin{equation}
\delta\simeq\frac{m_Z^2}{M},
\end{equation}
and therefore
\begin{equation}
M\simeq\frac{m_Z^2}{\delta}
\simeq2.2\times10^7\,\GeV.
\end{equation}
Thus a splitting near $377\,\keV$ corresponds to a characteristic gaugino scale of order $2\times10^7\,\GeV$, or several tens of PeV, if the bino and wino contributions are comparable and have the same sign. More general choices of $M_1$, $M_2$, their relative signs and phases, $\tan\beta$, and threshold corrections can modify this mapping substantially. The low-energy inelastic phenomenology can nevertheless remain similar as long as the resulting neutral-Higgsino splitting remains in the few-hundred-keV range.

\subsection{A dramatic annual-modulation prediction}

The proximity of the Higgsino benchmark to the kinematic endpoint has an important observational consequence: its recoil rate is expected to exhibit a much stronger annual modulation than an ordinary WIMP signal. The physical origin is simple. The detector-frame DM velocity distribution changes during the year because the Earth's orbital velocity adds to or subtracts from the motion of the Solar System through the Galactic halo. For a conventional elastic WIMP, this produces only a modest variation in the total rate. For the Higgsino benchmark, however, $v_{\min}$ lies so close to the maximum available DM speed that even a change of order $10$--$20\,\mathrm{km/s}$ can dramatically alter the number of particles able to cross the inelastic threshold.

To quantify this effect, we recompute the recoil rate as a function of the Earth-frame speed rather than imposing a sinusoidal modulation directly on the event rate. We adopt the nominal Standard Halo Model (SHM) parameters used throughout the recast: $v_0=238\,\mathrm{km/s}$, $v_{\rm esc}=544\,\mathrm{km/s}$, $\bar v_E=250.2\,\mathrm{km/s}$, $\Delta v_E=14.9\,\mathrm{km/s}$,
and approximate the annual variation of the magnitude of the Earth-frame velocity as
\begin{equation}
v_E(t)=\bar v_E+\Delta v_E
\cos\left[
\frac{2\pi(t-t_0)}{T}
\right],
\end{equation}
where $t_0\simeq 2~\mathrm{June}$ and $T=365.25~\mathrm{days}$.
For each value of $v_E(t)$, the velocity integral entering the differential recoil spectrum is evaluated using the truncated-Maxwellian SHM,
\begin{equation}
\eta(v_{\min},v_E)
=
\int_{v>v_{\min}}
\frac{f_{\rm SHM}(\mathbf v;v_E)}{v}\,d^3v,
\end{equation}
with the appropriate piecewise expression for the finite Galactic escape speed. The time dependence therefore enters the calculation through the available high-velocity phase space and is propagated nonlinearly into the recoil spectrum. In particular, when $v_{\min}$ approaches $v_{\rm esc}+v_E$, a small change in $v_E$ can produce a much larger fractional change in $\eta$ than would be obtained from a simple sinusoidal modulation of the total rate.

\begin{figure}[t]
\includegraphics[width=\columnwidth]{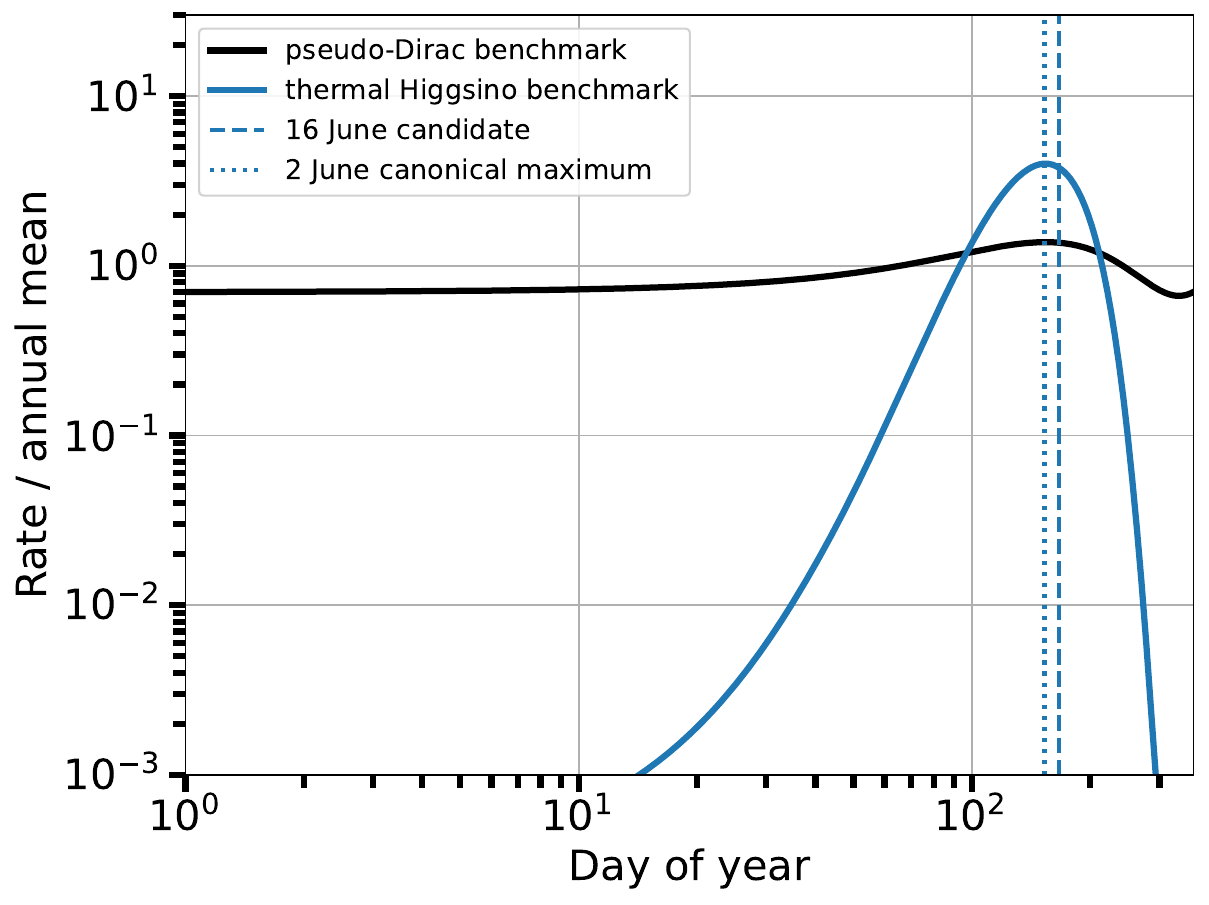}
\caption{Annual modulation of the high-energy LZ signal for the pseudo-Dirac and thermal-Higgsino benchmarks. The ordinate is the predicted event rate in the $215$--$300\,\keV$ recoil interval divided by its annual mean, so that unity corresponds to the time-averaged rate. The black curve shows the generic pseudo-Dirac benchmark, while the blue curve shows the thermal Higgsino. The vertical dotted line marks the canonical maximum of the detector-frame Earth speed near 2 June, and the vertical dashed line marks the date of the observed LZ candidate, 16 June. The Higgsino rate is strongly concentrated around the annual velocity maximum because the required $v_{\min}$ lies extremely close to the SHM kinematic cutoff, whereas the pseudo-Dirac benchmark exhibits a much milder modulation.}
\label{fig:tail}
\end{figure}

For each of 366 equally spaced times over one year, we then calculate the differential recoil spectrum for natural xenon, including the isotope abundances and the Helm nuclear form factor, and multiply it by the central LZ nuclear-recoil efficiency. The resulting accepted rate is integrated over the high-energy interval used in Fig.~\ref{fig:tail},
\begin{equation}
R(t)=
\int_{215\,\mathrm{keV}}^{300\,\mathrm{keV}}
dE_R\,
\epsilon_{\rm LZ}(E_R)
\frac{dR}{dE_R}(E_R,t).
\end{equation}
The curve shown in the figure is finally normalized to its annual mean,
\begin{equation}
\widehat R(t)=\frac{R(t)}
{\langle R\rangle_{\rm year}},
\,\,
\langle R\rangle_{\rm year}
=
\frac{1}{N_t}\sum_{i=1}^{N_t}R(t_i),
\,\,
N_t=366.
\end{equation}
Thus the calculation retains the nonlinear dependence of the scattering rate on the high-velocity tail, while the normalization removes the overall cross-section and exposure dependence. The only quantity displayed in Fig.~\ref{fig:tail} is therefore the predicted seasonal shape of the accepted high-energy signal.

Fig.~\ref{fig:tail} quantifies this effect for the two inelastic benchmarks. The horizontal axis gives the day of the year, while the vertical axis shows the predicted event rate in the $215$--$300\,\keV$ high-energy window divided by its annual mean. The curves therefore display the \emph{shape} of the seasonal modulation rather than the absolute number of events. A value equal to one corresponds to the annual-average rate; values larger or smaller than one indicate enhancement or suppression relative to that average.
The black curve in Fig.~\ref{fig:tail} corresponds to the generic pseudo-Dirac benchmark with $m_\chi=1\,\TeV$ and $\delta\simeq297\,\keV$. Its required velocity at the candidate energy, $v_{\min}\simeq701\,\mathrm{km/s}$, remains sufficiently below the SHM cutoff that a non-negligible population of particles can scatter throughout the year. The rate consequently varies only moderately around its annual mean. The modulation is nevertheless larger than for a conventional low-threshold elastic WIMP because this benchmark still probes the high-speed part of the halo.
The blue curve shows the qualitatively different behavior of the thermal Higgsino. Since $v_{\min}(248\,\keV)\simeq795\,\mathrm{km/s}$ lies at the mean detector-frame velocity cutoff, the accessible phase space changes dramatically during the Earth's orbit. Near the beginning of June, when the Earth and Solar velocities add approximately constructively, the high-speed tail extends sufficiently far to produce an enhanced scattering rate. During the opposite part of the year, the available maximum speed decreases and most of the velocity distribution falls below the inelastic threshold. The normalized rate therefore changes by orders of magnitude over the year and is strongly concentrated around the annual maximum.
The vertical dotted line in Fig.~\ref{fig:tail} marks the canonical maximum of the detector-frame speed, taken here to occur near 2 June, while the dashed line marks 16 June, the date of the LZ candidate. The event therefore occurred close in time to the part of the year when the Higgsino signal is predicted to be largest. This temporal coincidence is qualitatively interesting, but it should not be assigned an independent statistical significance. The model was examined after the event was observed, the detector live time is not necessarily uniform throughout the year, and the precise modulation depends on the full vector motion of the Earth rather than on the scalar sinusoidal approximation used here. A rigorous timing analysis would have to convolve the predicted time-dependent rate with the actual LZ live-time distribution and detector conditions.

The calculation presented in Fig.~\ref{fig:tail} should therefore be regarded as an astrophysical modulation diagnostic rather than as a timing likelihood. In particular, we do not include the detailed time dependence of the LZ detector efficiency, data-taking periods, detector conditions, or the exact three-dimensional Earth velocity. These effects are irrelevant for illustrating the extreme sensitivity of the Higgsino benchmark to the annual variation of the high-velocity tail, but they would be essential for assigning a quantitative probability to the observed event date.

For this reason, we do not combine the event date with the spectral information to obtain an additional significance. Instead, Fig.~\ref{fig:tail} should be interpreted as a prediction of the Higgsino hypothesis. If the $248\,\keV$ candidate originates from a Higgsino with a splitting close to $377\,\keV$, future high-energy nuclear-recoil events should be strongly concentrated around the part of the year when the detector-frame DM speed is maximal. In contrast, the generic pseudo-Dirac benchmark predicts a much broader seasonal distribution. The timing distribution of additional events could therefore provide an independent way to discriminate between these two inelastic interpretations.

\subsection{Indirect detection: thermal Higgsino and gamma-ray line searches}

The thermal Higgsino has a qualitatively different indirect-detection phenomenology from the generic pseudo-Dirac model. Although direct $Z$ exchange is inelastic, present-day Higgsino pairs still annihilate through electroweak interactions into gauge bosons. In addition to a continuum gamma-ray spectrum, electroweak loop and endpoint processes produce a characteristic line-like signal near the DM mass. Gamma-ray line searches therefore provide an important and largely independent test of the $m_\chi\simeq1.1\,\TeV$ thermal-Higgsino interpretation.

The strongest constraints around the TeV scale currently come from Galactic-center observations with imaging atmospheric Cherenkov telescopes. The latest H.E.S.S. Inner Galaxy Survey uses $546$ hours of observations and finds no significant line. For its reference Einasto Milky-Way profile, the reported $95\%$ limits reach
\begin{align}
\langle\sigma v\rangle_{\rm line}^{95\%}(1\,\TeV)
&\simeq2.3\times10^{-28}\,\mathrm{cm^3\,s^{-1}},\\
\langle\sigma v\rangle_{\rm line}^{95\%}(10\,\TeV)
&\simeq2.4\times10^{-27}\,\mathrm{cm^3\,s^{-1}},
\end{align}
and H.E.S.S. finds that the thermal-Higgsino parameter space is probed for some Milky-Way density profiles \cite{HESS2026}. MAGIC provides an independent Galactic-center line search based on $223$ hours of observations, with limits of approximately $5\times10^{-28}\,\mathrm{cm^3\,s^{-1}}$ at $1\,\TeV$ and $10^{-25}\,\mathrm{cm^3\,s^{-1}}$ at $100\,\TeV$ for its cuspy-profile interpretation \cite{MAGIC2023Lines}. The difference between the two experiments should not be interpreted solely as an instrumental effect, because the derived limits also depend on the ROI, exposure, event selection, energy response, and, most importantly, the assumed DM-density profile and corresponding $J$ factor.

\begin{figure}[t]
\includegraphics[width=\columnwidth]{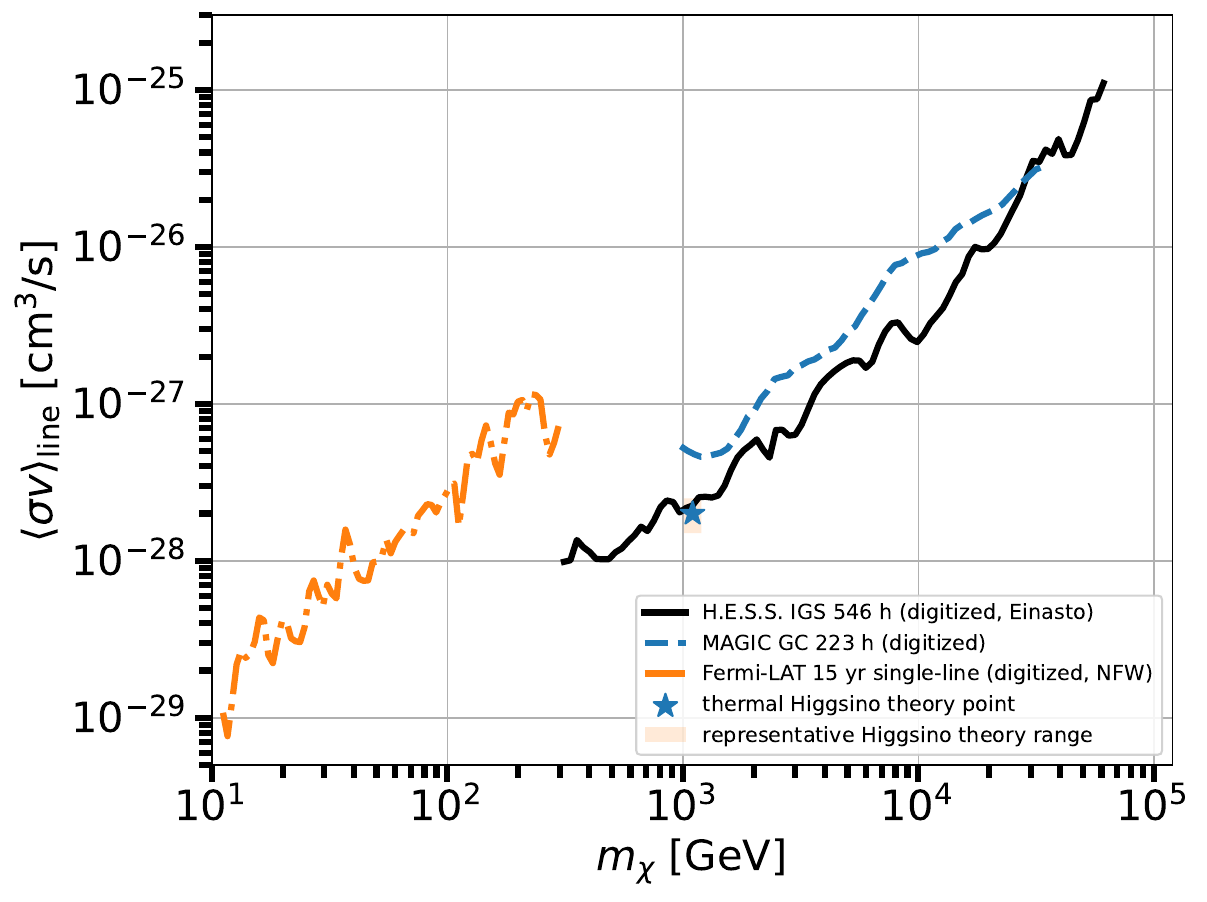}
\caption{Gamma-ray line constraints relevant to the thermal-Higgsino interpretation. The H.E.S.S. curve (black solid) shows the digitized 95\% C.L. observed upper limit for the baseline Einasto profile from the H.E.S.S. Inner Galaxy Survey analysis (546 h). The MAGIC curve (blue dashed) shows the digitized 95\% C.L. Galactic-Center limit for the 223 h MAGIC dataset, digitized from the published comparison figure. The Fermi-LAT curve (orange dash-dotted) shows the digitized 15-year single-line NFW limit from de la Torre Luque et al. The star at $m_\chi \simeq 1.1\,{\rm TeV}$ marks the representative thermal-Higgsino prediction adopted in this work, while the shaded band indicates the approximate theoretical range $(1.5$--$2.5)\times 10^{-28}\,{\rm cm}^3\,{\rm s}^{-1}$. Since the three searches use different regions of interest and halo assumptions, the figure should be interpreted as a comparison of characteristic sensitivities rather than as a statistically combined exclusion.}
\label{fig:gammalines}
\end{figure}

At lower masses, Fermi-LAT provides the most relevant gamma-ray line constraints. The official Pass-8 analysis searched $0.2$--$500\,\GeV$ and found no statistically significant line \cite{FermiLAT2015Lines}. A later analysis using $15$ years of Fermi-LAT data and the best-energy-resolution EDISP3 events searched approximately $10$--$300\,\GeV$, again finding no significant feature and improving the sensitivity near the upper end of the Fermi-LAT range \cite{FermiLAT15yrLines}. For an NFW profile, the published single-line limits span approximately $10^{-29}$--$10^{-27}\,\mathrm{cm^3\,s^{-1}}$ over this interval. Figure~\ref{fig:gammalines} shows the digitized single-line curve; it does not directly extend to the $1.1\,\TeV$ thermal-Higgsino mass.

At still higher masses, HAWC provides complementary line constraints using dwarf spheroidal galaxies and is particularly relevant in the multi-TeV regime \cite{HAWC2020Lines}. Because HAWC probes dwarf galaxies whereas H.E.S.S. and MAGIC use Galactic-center regions, the corresponding limits involve different astrophysical systematics and should not be interpreted as if they were derived for a common DM-density model.

For the thermal Higgsino itself, state-of-the-art calculations including electroweak corrections and endpoint photons predict a line-like annihilation rate of order
\begin{equation}
\langle\sigma v\rangle_{\rm line}
\sim(1.5\text{--}2.5)\times10^{-28}\,\mathrm{cm^3\,s^{-1}}
\end{equation}
near $m_\chi\simeq1.1\,\TeV$, depending on the precise experimental line window and theoretical treatment \cite{Beneke2020Higgsino,Rodd2024}. 
For a thermal Higgsino with $m_\chi\simeq1.08\,\TeV$, detailed calculations
including Sommerfeld enhancement predict a two-body line cross section of
order
\begin{equation}
\langle\sigma v\rangle_{\rm line}
\equiv
\langle\sigma v\rangle_{\gamma\gamma}
+
\frac{1}{2}
\langle\sigma v\rangle_{\gamma Z}
\sim
10^{-28}\,\mathrm{cm^3\,s^{-1}}
\end{equation}
\cite{Rodd2024}. The full line-like photon spectrum is larger and broader
than the strict two-body line because electroweak endpoint radiation provides
an additional $\mathcal{O}(1)$ contribution. Ref.~\cite{Rodd2024} accounts
for these effects using \textsc{DM}$\gamma$\textsc{Spec}. For the
illustrative sensitivity comparison below, we therefore adopt the
representative normalization
$\langle\sigma v\rangle_{\rm line\mbox{-}like}
=2\times10^{-28}\,\mathrm{cm^3\,s^{-1}}$, following the normalization used
for the approximate H.E.S.S. estimate in Ref.~\cite{Rodd2024}.

Fig.~\ref{fig:gammalines} summarizes the energy ranges relevant to the comparison. The Fermi-LAT curve occupies the sub-TeV region and therefore does not directly constrain the $1.1\,\TeV$ thermal Higgsino. Around the Higgsino mass, the relevant limits are instead those from H.E.S.S. and MAGIC. The representative Higgsino prediction lies below the MAGIC limit and very close to the current H.E.S.S. sensitivity for its reference Einasto profile. Log--log interpolation of the digitized H.E.S.S. curve gives an indicative sensitivity near $1.1\,\TeV$ of
\begin{equation}
\langle\sigma v\rangle_{\rm line}^{95\%}
\simeq2.3\times10^{-28}\,\mathrm{cm^3\,s^{-1}}.
\end{equation}
The representative Higgsino prediction, $2.0\times10^{-28}\,\mathrm{cm^3\,s^{-1}}$, is therefore slightly below, but close to, the corresponding H.E.S.S. reference sensitivity.

The proximity of the predicted signal to the H.E.S.S. limit makes gamma-ray searches particularly relevant for the LZ Higgsino interpretation. The conclusion is nevertheless strongly dependent on the Galactic DM profile because the line flux from the Galactic center scales with the line-of-sight integral of $\rho_{\rm DM}^2$. A more cuspy or higher-normalization profile increases the predicted flux and strengthens the constraint, while a shallower or lower-$J$ profile weakens it. The thermal Higgsino should therefore not be described as either generically excluded or generically unconstrained: it is already being probed for some halo profiles and remains viable for others.

Future gamma-ray observations can substantially sharpen this test. CTAO forecasts indicate that a $1.1\,\TeV$ thermal Higgsino can be detectable for a range of inner-Galaxy density profiles \cite{CTAO2026Higgsino,Rodd2024}. This provides a particularly useful complementarity with LZ. The direct-detection interpretation depends primarily on the local velocity distribution and, for the Higgsino benchmark, on its extreme high-speed tail, whereas the gamma-ray prediction depends mainly on the Galactic-center density profile and the associated $J$ factor. These two astrophysical uncertainties are physically distinct and should therefore be varied independently in any global interpretation.

\section{Combined interpretation and benchmark comparison}

The results above show that the LZ high-energy candidate can be accommodated by more than one microscopic realization of inelastic DM. The two benchmark scenarios studied in detail provide complementary examples of the same basic mechanism: the observed recoil is generated by an endothermic transition between two nearly degenerate Majorana states, with a mass splitting of a few hundred keV. Their main difference is the amount of particle-physics information that is fixed independently of the LZ event.

Table~\ref{tab:benchmarks} summarizes the central benchmark values. In the generic pseudo-Dirac model, the heavy-mediator, off-resonance thermal relic condition fixes the effective interaction strength, while the LZ rate determines the inelastic splitting. This gives
\begin{equation}
m_\chi\simeq1.0\,\TeV,\,\,
\delta\simeq297\,\keV,\,\,
\sigma_N\simeq6.5\times10^{-43}\,\mathrm{cm}^2.
\end{equation}
At the candidate energy, the required velocity is $v_{\min}(248\,\keV)\simeq701\,\mathrm{km/s}$ and the minimum of $v_{\min}(E_R)$ occurs at $E_R^\star\simeq265\,\keV$. The event therefore lies close to the characteristic recoil-energy scale of the benchmark while remaining sufficiently below the SHM kinematic cutoff that a non-negligible part of the high-velocity halo can contribute.

The thermal Higgsino is considerably more predictive. Its relic abundance fixes the DM mass close to $1.1\,\TeV$, while the leading $Z$-mediated inelastic scattering strength is determined by electroweak interactions. Matching the LZ rate therefore mainly selects the neutral-state splitting,
\begin{equation}
m_\chi\simeq1.1\,\TeV,\,\,
\delta\simeq377\,\keV,\,\,
\sigma_n^{\widetilde H}\simeq7.4\times10^{-39}\,\mathrm{cm}^2.
\end{equation}
This benchmark requires $v_{\min}(248\,\keV)\simeq795\,\mathrm{km/s}$ and has $E_R^\star\simeq339\,\keV$. It consequently lies at the mean maximum detector-frame speed of the local halo and is considerably more sensitive to the extreme velocity tail and to annual modulation.

\begin{table*}[t]
\caption{Central benchmark values from the public-information recast. Cross sections are quoted per nucleon. The generic pseudo-Dirac model uses an isoscalar normalization, whereas the Higgsino entry is neutron normalized and the isotope-dependent weak charge is implemented explicitly.}
\label{tab:benchmarks}
\begin{ruledtabular}
\begin{tabular}{lcc}
Quantity & Generic pseudo-Dirac & Thermal Higgsino\\
\hline
$m_\chi$ & $1.0\,\TeV$ & $1.1\,\TeV$\\
$\delta$ & $297\,\keV$ & $377\,\keV$\\
$\sigma_N$ & $6.5\times10^{-43}\,\mathrm{cm}^2$ & $7.4\times10^{-39}\,\mathrm{cm}^2$\\
$v_{\min}(248\,\keV)$ & $701\,\mathrm{km/s}$ & $795\,\mathrm{km/s}$\\
$E_R^\star$ & $265\,\keV$ & $339\,\keV$\\
Relic mechanism & $\chi_1\chi_2$ coannihilation & electroweak annihilation/coannihilation\\
Late-time indirect signal & suppressed if $\chi_2$ is depleted & electroweak annihilation remains active\\
Main astrophysical sensitivity & high-velocity tail & extreme high-velocity tail\\
Characteristic prediction & broad high-$E_R$ continuation & strong annual modulation\\
\end{tabular}
\end{ruledtabular}
\end{table*}

An important consequence of this comparison is that the LZ event does not determine a unique scattering cross section. In endothermic scattering, increasing $\delta$ reduces the phase-space fraction of halo particles capable of producing the recoil, so a larger cross section is required to maintain the same event yield. The same single event can therefore correspond to $\sigma_N\sim10^{-43}\,\mathrm{cm}^2$ for a splitting near $0.30\,\mathrm{MeV}$ or to an electroweak-scale cross section of order $10^{-39}\,\mathrm{cm}^2$ for a splitting near $0.37\,\mathrm{MeV}$. The recoil event alone cannot distinguish between these possibilities; the relic density, the particle-physics structure, indirect detection, and future spectral or timing information are required to break this degeneracy.

The two benchmarks illustrate different levels of model dependence. In the generic pseudo-Dirac construction, the direct--relic relation follows from the assumed heavy-mediator contact interaction and can be modified by finite-mediator effects, resonant annihilation, additional channels, or the detailed cosmological evolution of the excited state. The thermal Higgsino, by contrast, has substantially less freedom: its thermal mass and leading inelastic interaction are fixed by electroweak physics, so the LZ event primarily constrains the neutral-state splitting. Both scenarios nevertheless exploit the same hierarchy $\delta\ll T_f$, which allows a few-hundred-keV splitting to be negligible during freeze-out while strongly affecting present-day nuclear scattering.

Their late-time phenomenology is also qualitatively different. In the generic model, the dominant thermal process is the off-diagonal coannihilation $\chi_1\chi_2\to q\bar q$. If $\chi_2$ is efficiently depleted after freeze-out and no additional unsuppressed annihilation channels are present, the present-day indirect signal can be strongly suppressed. The viability of this scenario therefore depends on the late-time excited-state abundance and on the ultraviolet completion discussed in Sec.~\ref{sec:minpseudo}. The thermal Higgsino does not share this suppression: present-day Higgsino pairs continue to annihilate through electroweak interactions, producing continuum and line-like gamma-ray signals. Gamma-ray observations therefore provide an immediate and largely independent consistency test of the Higgsino interpretation.

The strong kinematic difference between the benchmarks provides another potential discriminator. The generic pseudo-Dirac point remains appreciably below the SHM kinematic ceiling and predicts a relatively broad distribution of high-energy recoils. The Higgsino lies very close to the maximum accessible velocity and therefore predicts a much stronger seasonal concentration of events. Its rate is also more sensitive to changes in the extreme local velocity distribution, including possible high-speed substructure or debris populations \cite{Graham2025}. Additional LZ events could therefore distinguish the two scenarios through both their recoil-energy distribution and their time dependence.

The LZ operator analysis also motivates a broader set of particle models beyond the two benchmarks considered explicitly here. Inelastic or even elastic SD interactions based on $\mathcal{O}_4^{s,v}$ remain phenomenologically interesting because LZ finds a non-negligible local preference for these operators. A pseudo-Dirac or Majorana fermion coupled through an axial-vector mediator provides a simple microscopic realization. Depending on the Lorentz structure of the annihilation current, the late-time annihilation rate can be $p$-wave or helicity suppressed, so the relic-density and indirect-detection phenomenology can differ substantially from both the vector pseudo-Dirac and Higgsino cases. A dedicated calculation is therefore required rather than extrapolating the $\mathcal{O}_1$ results.

Momentum-dependent covariant interactions such as $\mathcal{L}_{10}$ and $\mathcal{L}_{16}$ provide another possibility. These interactions can generate hard recoil spectra without relying entirely on an endothermic threshold. Transition dipole interactions are particularly interesting because they can combine momentum dependence with pseudo-Dirac inelasticity. Such models, however, require an independent matching between the relativistic interaction, the nuclear response, and the thermal annihilation processes; the analytic direct--relic relation derived for the universal-vector benchmark does not apply to them.

It is useful in this context to distinguish three levels of theoretical description. The nonrelativistic operators used by LZ characterize nuclear scattering but do not by themselves specify the particle model. A renormalizable simplified model introduces explicit mediator and DM fields and determines the leading laboratory and cosmological processes. A fully ultraviolet-complete theory must in addition provide the gauge structure, symmetry-breaking sector, anomaly cancellation, and any additional particles required for consistency.

The universal-vector pseudo-Dirac benchmark used here lies at the simplified-model level. A natural completion promotes $V_\mu$ to the gauge boson of a broken $U(1)'$, introduces a dark Higgs field that generates $m_V$, and generates the small Majorana masses through symmetry-breaking interactions. If SM quarks carry nontrivial $U(1)'$ charges, anomaly cancellation and gauge-invariant SM Yukawa interactions can require additional fermions or a more structured charge assignment. The universal quark coupling employed in the analytic benchmark should therefore be regarded as a useful low-energy realization until a complete gauge model is specified.

The Higgsino is much closer to a complete ultraviolet construction. A nearly pure Higgsino naturally arises in the MSSM or in split-supersymmetry-like spectra with heavy bino and wino states. Integrating out the Majorana gauginos generates the small neutral-Higgsino splitting, while the electroweak gauge interactions determine the leading scattering and annihilation processes. The inferred splitting points to a very high gaugino scale, but the low-energy Higgsino description remains an excellent EFT even when the detailed high-scale supersymmetric spectrum is unspecified.

Axial-vector completions of $\mathcal{O}_4$ face requirements similar to those of the vector benchmark: a massive axial $Z'$ must be embedded in a consistent gauge theory with an appropriate symmetry-breaking sector and anomaly cancellation. The covariant $\mathcal{L}_{10}$ and related dipole interactions are instead higher-dimensional at the particle level and must ultimately arise by integrating out heavier electrically or dark-charged states, often through loop processes. Appendix~\ref{app:l10} gives an illustrative mediator realization of the $\mathcal{L}_{10}$ interaction used in the LZ basis. These ultraviolet differences matter because relic density, collider bounds, and indirect detection depend on the complete particle content rather than only on the nuclear operator measured by LZ.

\begin{figure}[t]
\includegraphics[width=\columnwidth]{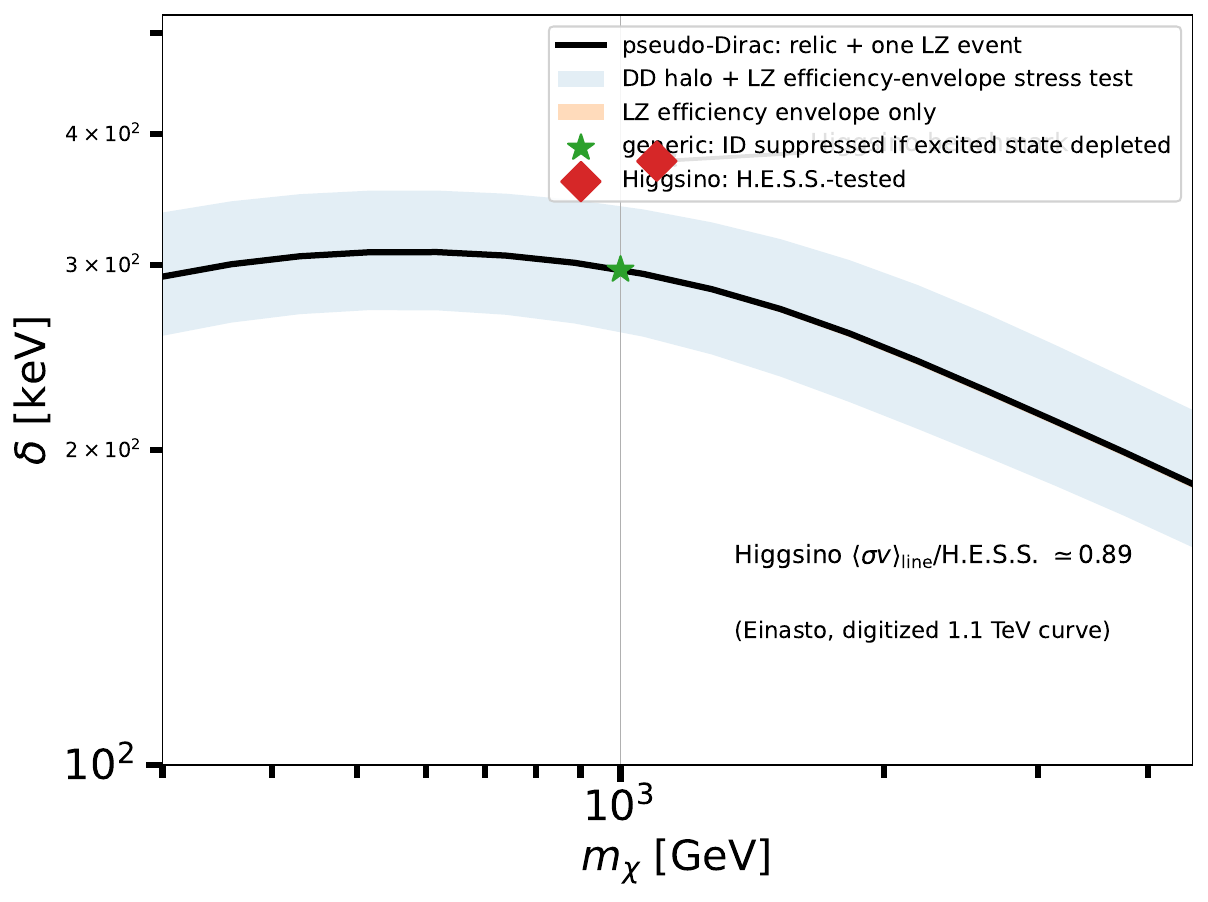}
\caption{Combined model-viability summary in the $m_\chi$--$\delta$ plane. The black curve gives the generic pseudo-Dirac locus for which the reference thermal relic abundance and the central LZ one-event normalization are simultaneously reproduced. The surrounding band shows the conservative halo-plus-efficiency stress test. Viability of this branch with respect to indirect detection additionally assumes efficient depletion of the excited state. The Higgsino marker shows the thermal $m_\chi\simeq1.1\,\TeV$ solution selected by the LZ rate. Its particle physics is more constrained, but it is directly testable with TeV gamma-ray line searches; the annotation compares a representative Higgsino line prediction with the log-interpolated H.E.S.S. Einasto sensitivity near $1.1\,\TeV$. The figure is intended as a qualitative comparison of logically distinct constraints rather than as a common global likelihood.}
\label{fig:modelviability}
\end{figure}

The combined status of the two benchmark scenarios is summarized in Fig.~\ref{fig:modelviability}. The black curve represents the generic pseudo-Dirac parameter combinations that simultaneously reproduce the reference thermal relic abundance and the central one-event LZ normalization. The surrounding band shows how this relation changes under the conservative halo and detector-efficiency stress test. This should be interpreted as a conditional viable region: direct detection and freeze-out are simultaneously reproduced along the curve, while indirect-detection safety additionally requires efficient depletion of $\chi_2$ and the absence of other unsuppressed late-time channels.
The isolated Higgsino point in Fig.~\ref{fig:modelviability} should be contrasted with the continuous pseudo-Dirac locus. In the generic model, the mediator parameters provide sufficient freedom for the thermal and LZ requirements to be satisfied over a broad range of $m_\chi$ and $\delta$, subject to the assumptions discussed above. In the Higgsino case, the thermal mass and scattering interaction are essentially fixed, leaving the splitting as the main quantity selected by LZ. The figure therefore summarizes a general lesson of the high-energy candidate: direct detection constrains the recoil rate and kinematics, but only the combination with cosmology, indirect searches, and the underlying particle theory determines whether a given interpretation is viable.

Additional high-energy nuclear-recoil events would substantially sharpen this comparison. Their recoil-energy distribution would test whether the signal resembles the broader pseudo-Dirac spectrum or the more kinematically compressed Higgsino spectrum, while their seasonal distribution would probe the extreme modulation predicted by the Higgsino benchmark. Improved TeV gamma-ray observations would independently test the Higgsino branch, whereas collider and cosmological studies of the mediator and excited-state evolution would primarily constrain the generic pseudo-Dirac scenario.

\subsection{Solar capture and the new Higgsino constraint}
\label{sec:solar_capture}

A particularly important new constraint on the Higgsino interpretation comes from dark-matter capture in the Sun. Very recently, Pospelov and Ramani studied the solar capture of quasi-Dirac electroweak dark matter and showed that the large gravitational potential of the Sun allows dark matter particles to reach velocities of order $1300$--$1400,\mathrm{km/s}$ in the solar interior. These velocities open inelastic scattering channels on heavy solar elements that are inaccessible to terrestrial direct-detection experiments. The captured Higgsinos can subsequently annihilate in the Sun, dominantly through $\chi\chi\rightarrow W^+W^-,ZZ$, producing high-energy neutrinos that are constrained by IceCube \cite{PospelovRamani2026}.

For the cosmologically preferred Higgsino mass $m_\chi\simeq1.08,\TeV$, Pospelov and Ramani find a robust lower bound on the neutral-Higgsino splitting of
\begin{equation}
\delta \gtrsim 566,\keV,
\end{equation}
when the predicted spin-dependent and loop-induced elastic interactions are included. Even considering tree-level inelastic $Z$ exchange alone, they obtain a bound of approximately $\delta\gtrsim506,\keV$. This is substantially stronger than the terrestrial kinematic limit and directly excludes the Higgsino interpretation of the LZ event. In particular, our benchmark value $\delta\simeq377,\keV$, as well as the conservative stress-test range $\delta\simeq337$--$427,\keV$, lies well below the solar-capture bound. At the splitting required by the LZ event, the corresponding annihilation rate in the Sun exceeds the IceCube limit by approximately three orders of magnitude \cite{PospelovRamani2026}. The solar-capture constraint therefore changes the status of the Higgsino interpretation from a viable direct-detection and thermal-relic solution to a model that is strongly disfavored once solar neutrino observations are included.

The same conclusion cannot, however, be directly transferred to the generic pseudo-Dirac vector model studied in this work. Although the two scenarios share the same basic endothermic transition mechanism, their microscopic interactions and solar phenomenology are different. In our benchmark, the nucleon scattering cross section required by the thermal relic relation is $\sigma_N\simeq6.5\times10^{-43},\mathrm{cm}^2$, compared with $\sigma_n\simeq7.4\times10^{-39},\mathrm{cm}^2$ for the Higgsino. Thus, at the level of the elementary scattering normalization, the generic model has a cross section approximately four orders of magnitude smaller, leading to a correspondingly less efficient solar capture rate for otherwise comparable kinematics. More importantly, the annihilation process is not fixed by electroweak symmetry to the $W^+W^-$ and $ZZ$ final states. In the minimal realization considered here, the relevant thermal process is $\chi_1\chi_2\rightarrow q\bar q$ through the vector mediator, while the late-time signal depends sensitively on the abundance and evolution of the excited state $\chi_2$.

Consequently, the solar-capture analysis of Ref.~\cite{PospelovRamani2026} should be regarded as a decisive constraint on the Higgsino benchmark, but not as a quantitative exclusion of the generic pseudo-Dirac model. Establishing the latter requires a dedicated treatment of capture and subsequent energy loss inside the Sun, including the solar velocity distribution, the composition of the solar medium, the inelastic transition rate, the evolution of the captured $\chi_1$ and $\chi_2$ populations, and their annihilation and neutrino yields. In particular, efficient depletion of $\chi_2$ in the Galactic halo does not by itself guarantee that the excited state remains absent inside the Sun, where inelastic up-scattering can regenerate it.

We therefore leave a complete calculation of solar capture and solar-neutrino constraints for both benchmark models to future work. Such an analysis will provide an important complementary test of the LZ interpretation and will determine whether the generic pseudo-Dirac solution remains viable once the full solar-capture phenomenology is consistently included.

\subsection{Predictions and near-term tests}

A decisive near-term test of the DM interpretation is simply the accumulation of additional exposure. The LZ experiment is continuing to collect data toward a total program exposure of approximately $1000$ live days \cite{LZWebsite2026}. If the candidate event is produced by a stationary DM population, and if the signal rate inferred from the present $220$ live-day dataset remains unchanged, the simplest expectation follows from linear exposure scaling. Using the same central estimator adopted in the rate recast, $\hat s=1-0.0106=0.9894$ signal events in $220$ live days, gives
\begin{align}
N_{\rm sig}(T)
&=
0.9894\,\frac{T}{220\,{\rm d}},
\\
N_{\rm sig}(1000\,{\rm d})
&\simeq
4.50,
\\
N_{\rm sig,add}
&\simeq
3.51,
\end{align}
where $N_{\rm sig,add}$ denotes the expected number of signal events accumulated after the presently analyzed dataset. This estimate assumes approximately unchanged fiducial mass, detector response, selection efficiency, and time-averaged DM rate.

For the central value above, the probability of observing at least one additional signal event during the remaining exposure is
\begin{equation}
P(N_{\rm add}\geq1)
=
1-\exp\left(-N_{\rm sig,add}\right)
\simeq 97\%.
\end{equation}
This number should not, however, be interpreted as a precision prediction. The normalization is inferred from a single event and consequently carries a very large statistical uncertainty. For one observed event, the exact $90\%$ Garwood confidence interval for the underlying Poisson mean is
\begin{equation}
0.051 < \mu < 4.744.
\end{equation}
Scaling this interval by the ratio $1000/220$ gives the broad range
\begin{equation}
0.23 \lesssim N_{\rm sig}(1000\,{\rm d}) \lesssim 21.6.
\end{equation}
This interval is only a counting-statistics illustration and does not replace the multidimensional LZ likelihood, which contains information on the location of the event in the $S1c$--$S2c$ plane, the different background populations, nuisance parameters, and veto sidebands. Nevertheless, it makes clear that the present event does not imply a precise prediction of approximately four additional events. Rather, if the central signal interpretation is correct, the signal should become considerably easier to test as the exposure increases. Conversely, observing no additional signal-like events by the end of the planned exposure would place substantial pressure on the central interpretation.

\begin{figure}[t]
\includegraphics[width=\columnwidth]{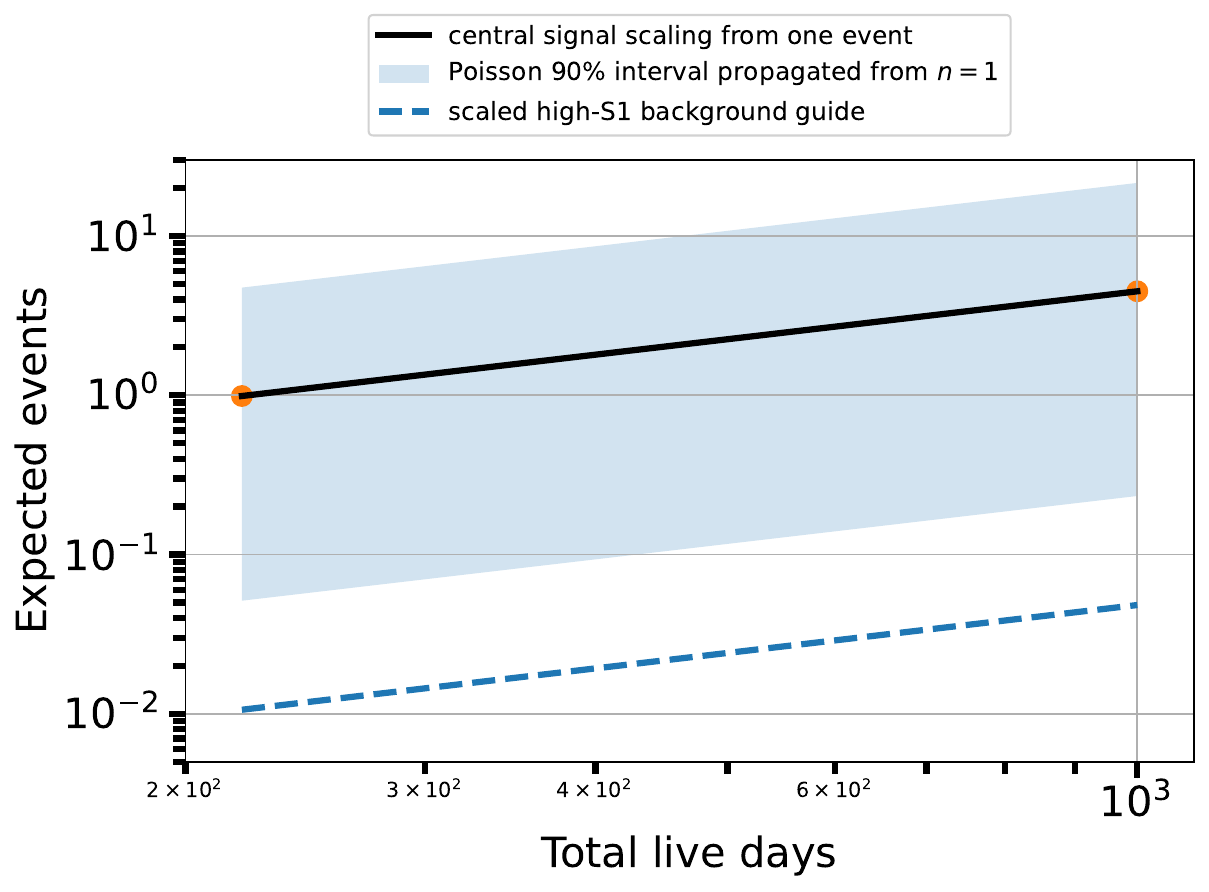}
\caption{Illustrative exposure forecast for the LZ candidate. The solid central curve scales the central signal estimator $\hat s=0.9894$ inferred from one observed event and the $0.0106$ high-$S1$ background guide linearly with exposure, while the shaded band is obtained by propagating the exact $90\%$ Poisson interval associated with a single observed count. The dashed curve has a different meaning: it scales the $0.0106$-event post-fit background quoted by LZ specifically for the $500<S1c<600$~phd interval, which contains the candidate at $S1c=540.1$~phd. It therefore illustrates the very small modeled background in the high-$S1$ region of the event and should not be interpreted as the total background expected in the full LZ search region. All curves assume simple linear exposure scaling and are intended as guides rather than substitutes for the full LZ likelihood.}
\label{fig:futurelz}
\end{figure}

Figure~\ref{fig:futurelz} illustrates these simple exposure scalings. The solid central curve corresponds to the central signal estimator used above,
\begin{equation}
N_{\rm sig}(T)=0.9894\,\frac{T}{220\,{\rm d}},
\end{equation}
and therefore passes through $N_{\rm sig}=0.9894$ at the exposure of the present analysis. The surrounding band is obtained by scaling the exact $90\%$ Poisson interval associated with this single observed event. Its large width is therefore not an uncertainty associated with detector efficiency or with a particular particle-physics model; it primarily reflects the unavoidable normalization uncertainty arising from having only one candidate event.

The dashed curve requires a more careful interpretation. LZ reports a total post-fit background expectation of only
\begin{equation}
N_{\rm bkg}^{500<S1c<600}
=
0.0106\pm0.0008
\end{equation}
events in the bottom panel of Fig.~5 of Ref.~\cite{LZ2026Extended}, corresponding specifically to the interval
\begin{equation}
500<S1c<600~{\rm phd}.
\end{equation}
This is the high-$S1$ slice containing the candidate event, which has $S1c=540.1$~phd. The very small background expectation reflects the strong separation between the electron-recoil and nuclear-recoil populations at large $S1c$: LZ explicitly notes that ER discrimination improves at high energies and that very little ER leakage is expected for $S1c>250$~phd \cite{LZ2026Extended}. The dashed curve in Fig.~\ref{fig:futurelz} simply scales this particular number linearly with live time,
\begin{equation}
N_{\rm bkg}^{500<S1c<600}(T)
=
0.0106
\frac{T}{220\,{\rm d}},
\end{equation}
giving
\begin{equation}
N_{\rm bkg}^{500<S1c<600}(1000\,{\rm d})
\simeq
0.048.
\end{equation}
It therefore illustrates how small the measured background model is in the high-$S1$ region containing the event. It is \emph{not} the expected total background of a future LZ analysis, nor is it the background integrated over the complete extended-energy ROI. In particular, a future analysis will require a new simultaneous fit of the science and veto samples, and the background rates, detector conditions, efficiencies, and nuisance parameters need not scale exactly with exposure. The dashed line should consequently be regarded only as a useful reference showing the unusually low background level in the immediate high-$S1$ region relevant for the candidate.

The comparison between the two curves is nevertheless instructive. Under the central DM interpretation, the signal expectation grows from approximately one event to about $4.5$ events over the full program, whereas a naive scaling of the presently fitted high-$S1$ background gives substantially less than one event. Therefore, several additional NR-like events in approximately the same high-$S1$ region would be difficult to attribute to a simple linear extrapolation of the presently modeled background. On the other hand, the broad signal band emphasizes that the absence of several new events would not by itself immediately exclude the DM interpretation; the quantitative conclusion must ultimately be obtained from the full LZ likelihood.

If the event is due to inelastic DM, however, several tests are more informative than the total number of events alone. Additional candidates would make it possible to determine whether the excess represents a generic population of high-energy nuclear recoils or follows the characteristic spectrum expected from a particular DM interaction. LZ finds that the event can be accommodated by several interactions, including inelastic $\mathcal{O}_1$ and $\mathcal{O}_4$ scattering and a number of covariant effective interactions, with local significances exceeding $3\sigma$ for a substantial subset of the models tested \cite{LZ2026Extended}. A single event primarily tests signal versus background, whereas multiple events would begin to probe the shape of the recoil spectrum and its distribution in the $S1c$--$S2c$ plane, thereby discriminating among the SI identity response, SD nuclear responses, and momentum-dependent interactions.

First, additional events should preferentially populate the high-energy part of the nuclear-recoil window. For endothermic scattering, the minimum incoming DM velocity is minimized around the characteristic recoil energy $\er^\star$ derived above. For the generic pseudo-Dirac benchmark,
\begin{equation}
\er^\star\simeq265\,\keV,
\end{equation}
which lies close to the observed recoil energy. Additional events should therefore preferentially occur in the same broad high-energy region rather than producing the steeply falling low-energy spectrum characteristic of ordinary elastic SI scattering. For the Higgsino benchmark, $\er^\star$ lies above the efficiently accessible LZ recoil window. The observable spectrum is consequently pushed against the upper-energy boundary: it tends to increase toward the high-energy part of the acceptance before eventually being suppressed by the detector efficiency and by the diminishing phase space of sufficiently fast halo particles. The detailed distribution of future recoil energies would therefore provide considerably more information than the total event count.

Second, annual modulation provides a particularly interesting discriminator. Endothermic scattering selects particles from the high-velocity tail of the Galactic DM distribution. The scattering rate is therefore considerably more sensitive to the Earth's velocity relative to the halo than for ordinary elastic WIMPs. LZ also emphasizes that inelastic models can exhibit substantial annual modulation, with the rate expected to peak approximately around June~2 \cite{LZ2026Extended}. The generic benchmark already requires comparatively fast DM particles, whereas the Higgsino benchmark lies especially close to the kinematic boundary. Its rate can consequently vary strongly during the year. A future LZ release providing the times of additional high-energy NR-like events together with the detector live-time distribution could therefore test whether such events preferentially occur around late May and June. This test must use the actual live-time mask, since a nonuniform exposure during the year can otherwise mimic or obscure modulation.

Third, the dependence on the target nucleus offers a complementary test. As follows from Eq.~\eqref{eq:estar}, the kinematics of endothermic scattering depend directly on the DM--nucleus reduced mass. Heavy target nuclei can access substantially larger mass splittings than light nuclei because they permit a larger transfer of kinetic energy into the excitation of the incoming DM state. Xenon is therefore especially well suited to probing the $\delta\sim0.3$--$0.4\,\mathrm{MeV}$ region considered here. Comparing xenon results with experiments employing different nuclear targets could help distinguish a genuine inelastic-scattering threshold from an unidentified detector-specific background. Conversely, sufficiently light nuclei rapidly lose kinematic sensitivity to these large splittings.

Fourth, the neutron hypothesis can be tested with increasing precision. LZ notes that a neutron must have an incident energy of at least approximately $8$ MeV to generate a $250\,$ keV elastic recoil in xenon. At such energies the neutron--xenon elastic cross section becomes increasingly forward peaked, making a single isolated recoil of this size without accompanying lower-energy interactions unlikely \cite{LZ2026Extended}. The collaboration also performed dedicated simulations of muon-induced cascades, corresponding to approximately $1200$ years of equivalent LZ exposure, without finding signal-like deposits; the resulting $90\%$ confidence-level upper limit on single-scatter events of any energy from muon-induced neutrons is $4.6\times 10^{-4}$ events. Additional exposure, together with the neutron-tagged veto samples and multiple-scatter sidebands, will therefore provide an increasingly stringent test of a neutron-background interpretation.

Finally, the two particle-physics interpretations considered here lead to qualitatively different predictions outside LZ. The generic pseudo-Dirac contact model requires an ultraviolet mediator completion and may therefore be probed independently through collider or mediator searches. The Higgsino realization is considerably more predictive because the same electroweak interactions responsible for its thermal history also generate indirect-detection signatures. In particular, gamma-ray line searches from the Galactic Center provide an important complementary probe, with future CTA and SWGO observations potentially improving the sensitivity. A confirmed population of high-energy nuclear recoils together with a null gamma-ray result would therefore constrain either the inner Milky-Way DM distribution or the canonical thermal Higgsino interpretation, while leaving a more generic model with a depleted excited state comparatively unconstrained by present-day annihilation searches.

\section{Systematic uncertainties and limitations}

The numerical benchmarks presented above should be regarded as representative central values rather than precision determinations. The dominant uncertainty is statistical: a single candidate event does not determine the signal normalization accurately. Whenever a cross section is inferred by requiring approximately one event in the present exposure, the corresponding Poisson uncertainty should therefore be kept separate from detector, nuclear, and astrophysical systematics. The benchmark cross sections quoted in this work are central values, while Fig.~\ref{fig:futurelz} illustrates the much broader allowed normalization. The numerical precision of the root-finding procedure should consequently not be interpreted as a physical uncertainty.

A second limitation is our simplified treatment of the detector response. Equation~\eqref{eq:nsig} works directly in recoil-energy space, whereas the LZ likelihood is constructed in the $\{S1c,\log_{10}S2c\}$ plane using calibrated NEST detector response, veto samples, and correlated nuisance parameters \cite{LZ2026Extended}. We use a vector digitization of the published final nuclear-recoil efficiency, including its uncertainty envelope, which improves upon a simple analytic approximation. Nevertheless, this remains a one-dimensional efficiency treatment and cannot reproduce the full mapping between recoil energy and the measured $S1c$--$S2c$ observables.

Nuclear-response uncertainties are also important because the candidate lies at large recoil energy. For xenon,
\begin{equation}
q=\sqrt{2m_AE_R}\simeq0.246\,\GeV
\end{equation}
at $E_R\simeq248\,\keV$, where nuclear diffraction effects are significant. A Helm form factor is sufficient for simple SI estimates, but precision predictions require isotope-dependent nuclear response functions. LZ uses DMFormfactor nuclear density matrices for its NREFT signal calculations \cite{LZ2026Extended,Fitzpatrick2013}, and rates near form-factor minima can be particularly sensitive to the nuclear modeling.

Astrophysical uncertainties are especially relevant for the inelastic interpretations considered here. The generic benchmark probes approximately
\begin{equation}
v_{\min}\sim700\,\mathrm{km/s},
\end{equation}
while the Higgsino benchmark reaches
\begin{equation}
v_{\min}\sim795\,\mathrm{km/s}.
\end{equation}
Both therefore depend strongly on the high-velocity tail of the Galactic DM distribution. Variations in the escape velocity, the local speed distribution, substructure, or contributions associated with the Large Magellanic Cloud can modify both the inferred mass splitting and the scattering rate \cite{Baxter2021,Graham2025}. This uncertainty is particularly important for the Higgsino benchmark, which lies close to the kinematic cutoff.

Finally, the relic-density relation derived for the generic model assumes a heavy mediator, approximately massless quarks, no additional annihilation channels, and a standard radiation-dominated cosmology. Resonant annihilation, a mediator close to threshold, additional dark-sector channels, or a nonstandard pre-BBN expansion history can modify the connection between relic density and direct detection encoded in Eq.~\eqref{eq:masterrelation}. The corresponding benchmark should therefore be interpreted as the simplest thermal realization rather than as a model-independent prediction.

Overall, the most robust conclusion is qualitative: interactions that suppress low-energy recoils and shift the spectrum toward the upper part of the LZ window provide a natural explanation of an isolated recoil near $250\,\keV$. The precise values of the cross section and mass splitting remain substantially more uncertain and will require additional LZ events and a full detector-level likelihood analysis.

\section{Conclusions}

The $248\,\keV$ LZ nuclear-recoil candidate is challenging to interpret in terms of ordinary elastic SI DM. When normalized to reproduce the high-energy event, the corresponding recoil spectrum is strongly concentrated at lower energies and would predict a much larger low-energy population. This conclusion does not extend to all standard elastic interactions: for example, LZ still finds a $2.7\sigma$ local preference for elastic $\mathcal{O}_4$ scattering at $m_\chi=1\,\TeV$. Nevertheless, endothermic scattering provides a particularly simple way to suppress the low-energy rate and shift the recoil spectrum toward the observed energy, naturally pointing to TeV-scale DM with a mass splitting of a few hundred keV.

A simple realization is a pseudo-Dirac fermion with an off-diagonal isoscalar vector interaction. For $m_\chi\simeq1\,\TeV$, imposing the observed relic abundance in the heavy-mediator limit gives
\begin{equation}
\sigma_N\simeq6.5\times10^{-43}\,\mathrm{cm}^2,
\end{equation}
while reproducing approximately one LZ event selects
\begin{equation}
\delta\simeq297\,\keV.
\end{equation}
This benchmark provides a direct connection between the relic-density requirement and the scattering rate suggested by the LZ event. Present-day indirect-detection constraints can remain weak if freeze-out is dominated by coannihilation of the two nearly degenerate states and the excited state is subsequently depleted.

A thermal Higgsino provides a more predictive alternative. Its relic abundance fixes the DM mass close to $1.1\,\TeV$, while off-diagonal $Z$ exchange gives
\begin{equation}
\sigma_N\simeq7.4\times10^{-39}\,\mathrm{cm}^2.
\end{equation}
Matching the LZ event then requires a larger splitting,
\begin{equation}
\delta\simeq0.37\,\mathrm{MeV},
\end{equation}
which places the scattering process close to the Galactic kinematic boundary. As a consequence, the Higgsino interpretation predicts a strong sensitivity to the high-velocity tail and potentially a large annual modulation. The fact that the candidate occurred on 16 June is qualitatively compatible with the expected early-June enhancement of an inelastic signal, but a single event obviously cannot establish a modulation signature.

The Higgsino scenario is also significantly more constrained by indirect detection. Gamma-ray line searches already probe part of the relevant parameter space, depending on the assumed Galactic DM distribution, and therefore provide an important complement to LZ. The two benchmark interpretations considered here consequently lead to rather different predictions: the generic pseudo-Dirac model can evade present-day annihilation bounds through depletion of the excited state, whereas the thermal Higgsino connects the direct-detection signal to comparatively well-defined gamma-ray signatures.

Future LZ data will be decisive. If the candidate is associated with DM, additional events should preferentially populate the high-energy nuclear-recoil region rather than the low-energy spectrum expected from conventional elastic SI scattering. Their recoil energies and arrival times could begin to test both the inelastic spectral shape and the expected annual modulation. Combined with gamma-ray searches, these measurements could distinguish a flexible pseudo-Dirac thermal relic from the more predictive Higgsino interpretation.

Finally, solar capture provides an important additional constraint on the Higgsino interpretation. The recent analysis of Ref.~\cite{PospelovRamani2026} finds that IceCube limits on neutrinos from DM annihilation in the Sun require $\delta\gtrsim506\,\keV$ even for tree-level inelastic $Z$ exchange, and $\delta\gtrsim566\,\keV$ when the full predicted Higgsino interactions are included. This seems to exclude the $\delta\simeq377\,\keV$ splitting required by the LZ candidate, making solar capture a particularly strong constraint on the Higgsino scenario. The corresponding bound does not directly apply to our generic pseudo-Dirac vector model, whose substantially smaller scattering cross section and different annihilation channels lead to a different solar-capture phenomenology; a dedicated study of solar capture, excited-state evolution, and neutrino production in both models will be presented in future work.

Several contemporaneous studies have explored complementary particle-physics realizations of the LZ candidate: neutral-current absorption of a $\simeq247\,\mathrm{MeV}$ fermion, although the required benchmark is in significant tension with KamLAND; a thermal $\sim1.1\,\mathrm{TeV}$ Higgsino; and a mixed axion--inelastic-electroweak-doublet scenario in which a $380$--$420\,\mathrm{GeV}$ doublet constitutes a subdominant DM component and accounts for the recoil through a splitting of $\delta\simeq323$--$330\,\mathrm{keV}$ \cite{LouLu2026,WuZhangZhu2026,Visinelli2026}.
For the thermal-Higgsino realization specifically, the solar-capture analysis of Ref.~\cite{PospelovRamani2026} finds an IceCube bound $\delta>566\,\mathrm{keV}$, excluding the splitting required to reproduce the LZ event under its assumptions.

A more precise quantitative interpretation will ultimately require the official LZ Data Release, including the full $S1c$--$S2c$ likelihood, NEST detector response and nuisance parameters, and the same isotope-dependent DMFormfactor nuclear responses used by the Collaboration. The calculation presented here should therefore be viewed as a transparent phenomenological baseline that identifies the particle-physics scenarios most naturally associated with the observed high-energy recoil and the complementary measurements capable of testing them.

\begin{acknowledgments}
MDM acknowledges support from the research grant {\sl TAsP (Theoretical Astroparticle Physics)} funded by Istituto Nazionale di Fisica Nucleare (INFN).
\end{acknowledgments}

\newpage

\appendix

\section{\texorpdfstring{Covariant interaction at maximum LZ significance: $\mathcal{L}_{10}^{s}$}{Covariant interaction at maximum LZ significance: L10s}}
\label{app:l10}

Supplemental Table S7 of Ref.~\cite{LZ2026Extended} shows that the largest local significance among the isoscalar covariant interactions at $m_\chi=1\,\TeV$ reaches $3.4\sigma$. This maximum is shared by $\mathcal{L}_{10}^{s}$ and $\mathcal{L}_{16}^{s}$. We discuss $\mathcal{L}_{10}^{s}$ explicitly because LZ also uses this interaction as one of the illustrative examples in the main text and emphasizes that it produces a broad, double-peaked recoil spectrum extending to high energies \cite{LZ2024NREFT,LZ2026Extended}.

For a spin-$1/2$ WIMP and a nucleon $N$, the relevant covariant interaction is
\begin{equation}
\mathcal{L}_{10}^{N}
=
\left(
\bar\chi i\sigma^{\mu\nu}
\frac{q_\nu}{m_M}
\chi
\right)
\left(
\bar N i\sigma_{\mu\alpha}
\frac{q^\alpha}{m_M}
N
\right),
\label{eq:L10covariant}
\end{equation}
with coefficient $d_{10}^{N}$. In the LZ convention, $m_M$ is taken to be the nucleon mass. The isoscalar and isovector interactions correspond to the sum and difference of the proton and neutron coefficients.

Its nonrelativistic reduction is
\begin{equation}
\mathcal{L}_{10}^{N}
\longrightarrow
4
\left[
\frac{\bm q^{2}}{m_M^{2}}\mathcal{O}_{4}
-
\frac{m_N^{2}}{m_M^{2}}\mathcal{O}_{6}
\right],
\label{eq:L10NR}
\end{equation}
where
\begin{align}
\mathcal{O}_{4}
&=
\bm S_\chi\cdot\bm S_N,
\\
\mathcal{O}_{6}
&=
\left(
\bm S_\chi\cdot\frac{\bm q}{m_N}
\right)
\left(
\bm S_N\cdot\frac{\bm q}{m_N}
\right).
\end{align}
The explicit momentum dependence explains why $\mathcal{L}_{10}$ can produce a hard recoil spectrum without requiring an inelastic threshold. In contrast with ordinary coherent SI scattering, the rate is controlled by momentum-dependent spin responses and can therefore remain sizable in the high-recoil region relevant for the LZ candidate.

Equation~\eqref{eq:L10covariant} should be regarded as an effective interaction rather than a complete particle model. A schematic realization can involve a heavy vector mediator with dipole-type interactions,
\begin{align}
\mathcal{L}_{\rm med}
\supset{}&
-\frac{1}{4}V_{\mu\nu}V^{\mu\nu}
+
\frac{1}{2}m_V^2V_\mu V^\mu
\nonumber\\
&+
\frac{\kappa_\chi}{2\Lambda_\chi}
\bar\chi\sigma^{\mu\nu}\chi V_{\mu\nu}
+
\frac{\kappa_N}{2\Lambda_N}
\bar N\sigma^{\mu\nu}N V_{\mu\nu}.
\label{eq:L10mediator}
\end{align}
After integrating out a sufficiently heavy mediator, the resulting interaction contains the same momentum-dependent tensor structure as Eq.~\eqref{eq:L10covariant}. A consistent relic-density calculation would, however, require specifying the ultraviolet completion, including the origin of the dipole interactions and their matching onto quark and gluon operators. We therefore do not attempt to connect the $\mathcal{L}_{10}$ interpretation directly to the relic-density relation derived for the pseudo-Dirac vector benchmark.

\section{Inelastic-scattering kinematics}
\label{app:inelastic_kinematics}

We summarize here the nonrelativistic kinematics of endothermic DM--nucleus scattering and derive the relations used throughout the main text. We consider the process
\begin{equation}
\chi_1 + A \rightarrow \chi_2 + A,
\end{equation}
where the incoming and outgoing DM states have masses
\begin{equation}
m_{\chi_1}=m_\chi,
\qquad
m_{\chi_2}=m_\chi+\delta,
\end{equation}
with
\begin{equation}
\delta>0.
\end{equation}
The scattering is therefore endothermic: part of the incoming kinetic energy is required to produce the heavier state $\chi_2$.

We work in the laboratory frame, in which the target nucleus is initially at rest. Let $\bm v$ and $\bm v'$ denote the initial and final DM velocities and let $\bm q$ be the momentum transferred to the nucleus. To leading order in the nonrelativistic expansion, momentum conservation gives
\begin{equation}
m_\chi \bm v
=
m_\chi \bm v'
+
\bm q,
\label{eq:momentum_cons_inelastic}
\end{equation}
where corrections of relative order $\delta/m_\chi$ have been neglected. The nuclear recoil energy is related to the momentum transfer through
\begin{equation}
E_R
=
\frac{\bm q^2}{2m_A}.
\label{eq:ERq}
\end{equation}

Energy conservation gives
\begin{equation}
\frac{1}{2}m_\chi v^2
=
\frac{1}{2}m_\chi v'^2
+
\frac{\bm q^2}{2m_A}
+
\delta.
\label{eq:energy_cons_inelastic}
\end{equation}
Using Eq.~\eqref{eq:momentum_cons_inelastic} to write
\begin{equation}
\bm v'
=
\bm v-\frac{\bm q}{m_\chi},
\end{equation}
and substituting into Eq.~\eqref{eq:energy_cons_inelastic}, one obtains
\begin{equation}
\bm v\cdot\bm q
=
\frac{\bm q^2}{2m_\chi}
+
\frac{\bm q^2}{2m_A}
+
\delta.
\end{equation}
Introducing the DM--nucleus reduced mass
\begin{equation}
\mu_A
=
\frac{m_\chi m_A}{m_\chi+m_A},
\label{eq:mureduced_appendix}
\end{equation}
this can be written as
\begin{equation}
\bm v\cdot\bm q
=
\frac{\bm q^2}{2\mu_A}
+
\delta.
\label{eq:vq_relation}
\end{equation}

For a fixed momentum transfer $q$, the smallest possible incoming speed occurs when the incoming velocity and momentum transfer are parallel,
\begin{equation}
\bm v\cdot\bm q=vq.
\end{equation}
Equation~\eqref{eq:vq_relation} therefore gives
\begin{equation}
v_{\min}
=
\frac{q}{2\mu_A}
+
\frac{\delta}{q}.
\label{eq:vmin_q}
\end{equation}
Using
\begin{equation}
q=\sqrt{2m_AE_R},
\end{equation}
we recover the standard expression
\begin{equation}
v_{\min}(E_R)
=
\frac{1}{\sqrt{2m_AE_R}}
\left[
\frac{m_AE_R}{\mu_A}
+
\delta
\right].
\label{eq:vmin_appendix}
\end{equation}

Equation~\eqref{eq:vmin_appendix} makes the main difference between elastic and endothermic scattering transparent. For $\delta=0$,
\begin{equation}
v_{\min}^{\rm elastic}
=
\sqrt{\frac{m_AE_R}{2\mu_A^2}},
\end{equation}
which increases monotonically with recoil energy. For $\delta>0$, instead, the additional term proportional to $\delta/\sqrt{E_R}$ strongly suppresses low-energy recoils. Very small recoil energies require very large incoming DM velocities because the particle must provide the excitation energy $\delta$ while transferring only a small momentum to the nucleus.

The function $v_{\min}(E_R)$ therefore has a minimum at a nonzero recoil energy. It is useful to rewrite Eq.~\eqref{eq:vmin_appendix} as
\begin{equation}
v_{\min}(E_R)
=
\sqrt{\frac{m_A}{2\mu_A^2}}\,\sqrt{E_R}
+
\frac{\delta}{\sqrt{2m_AE_R}}.
\end{equation}
Taking the derivative with respect to $E_R$,
\begin{equation}
\frac{d v_{\min}}{dE_R}
=
\frac{1}{2}
\sqrt{\frac{m_A}{2\mu_A^2E_R}}
-
\frac{\delta}
{2\sqrt{2m_A}E_R^{3/2}},
\end{equation}
and imposing
\begin{equation}
\frac{d v_{\min}}{dE_R}=0,
\end{equation}
gives
\begin{equation}
E_R^\star
=
\frac{\mu_A}{m_A}\delta.
\label{eq:estar_appendix}
\end{equation}
At this recoil energy, the required incoming speed reaches its absolute minimum,
\begin{equation}
v_{\min}^{\rm min}
=
v_{\min}(E_R^\star)
=
\sqrt{\frac{2\delta}{\mu_A}}.
\label{eq:vminminimum}
\end{equation}

The physical interpretation is particularly simple. At recoil energies below $E_R^\star$, increasing $E_R$ makes the scattering easier because a larger momentum transfer allows the excitation energy $\delta$ to be supplied more efficiently. Above $E_R^\star$, the ordinary cost of producing an increasingly energetic nuclear recoil dominates and $v_{\min}$ begins to increase again. Consequently, the halo integral entering the scattering rate is largest around this characteristic recoil scale.

It is important, however, not to identify $E_R^\star$ exactly with the peak of the experimentally observed recoil spectrum. Equation~\eqref{eq:estar_appendix} determines the recoil energy for which the kinematic requirement on the incoming velocity is weakest. The actual event spectrum also contains the nuclear response, isotope abundances, the velocity distribution, and the detector efficiency,
\begin{equation}
\frac{dR}{dE_R}
\propto
F_A^2(E_R)\,
\eta\left(v_{\min}(E_R)\right),
\end{equation}
before detector effects are included. Nuclear form-factor suppression at large momentum transfer can therefore move the maximum away from $E_R^\star$. Nevertheless, $E_R^\star$ provides a very useful estimate of the characteristic recoil scale selected by inelastic kinematics.

For a TeV-scale DM particle scattering on xenon,
\begin{equation}
m_A\simeq122\,\GeV,
\end{equation}
and
\begin{equation}
\mu_A\simeq109\,\GeV
\end{equation}
for $m_\chi=1\,\TeV$. Hence
\begin{equation}
\frac{\mu_A}{m_A}\simeq0.89,
\end{equation}
so that
\begin{equation}
E_R^\star\simeq0.89\,\delta.
\end{equation}
For the generic pseudo-Dirac benchmark,
\begin{equation}
\delta\simeq297\,\keV,
\end{equation}
this gives
\begin{equation}
E_R^\star\simeq265\,\keV,
\end{equation}
naturally close to the $248\,\keV$ LZ candidate. For the Higgsino benchmark with
\begin{equation}
\delta\simeq377\,\keV,
\end{equation}
one instead obtains
\begin{equation}
E_R^\star\simeq339\,\keV,
\end{equation}
above the upper part of the efficient LZ recoil window. The observable Higgsino spectrum is therefore sampled only on the lower-energy side of its preferred kinematic region and is driven toward the high-energy boundary of the LZ acceptance.

Another useful way to characterize the kinematics is to determine the range of recoil energies accessible to a DM particle with a fixed incoming speed $v$. Starting from Eq.~\eqref{eq:vmin_q}, scattering is possible whenever
\begin{equation}
v
\geq
\frac{q}{2\mu_A}
+
\frac{\delta}{q}.
\end{equation}
Multiplying by $q$ gives the quadratic inequality
\begin{equation}
q^2
-
2\mu_Avq
+
2\mu_A\delta
\leq0.
\end{equation}
The two limiting momentum transfers are therefore
\begin{equation}
q_{\pm}
=
\mu_Av
\left[
1
\pm
\sqrt{
1-
\frac{2\delta}
{\mu_Av^2}
}
\right],
\label{eq:qpm}
\end{equation}
which correspond to the recoil-energy interval
\begin{equation}
E_R^{-}(v)
\leq
E_R
\leq
E_R^{+}(v),
\end{equation}
with
\begin{equation}
E_R^{\pm}(v)
=
\frac{\mu_A^2v^2}{2m_A}
\left[
1
\pm
\sqrt{
1-
\frac{2\delta}
{\mu_Av^2}
}
\right]^2.
\label{eq:ERpm}
\end{equation}

A real solution exists only if
\begin{equation}
1-
\frac{2\delta}
{\mu_Av^2}
\geq0,
\end{equation}
or equivalently
\begin{equation}
v
\geq
v_\delta
\equiv
\sqrt{\frac{2\delta}{\mu_A}}.
\label{eq:vthreshold}
\end{equation}
This is identical to the minimum velocity obtained in Eq.~\eqref{eq:vminminimum}. At threshold,
\begin{equation}
v=v_\delta,
\end{equation}
the two recoil solutions merge,
\begin{equation}
E_R^{-}=E_R^{+}=E_R^\star.
\end{equation}
Thus, as the incoming DM speed approaches the minimum velocity necessary for upscattering, the available recoil-energy interval collapses around $E_R^\star$. This provides another way of understanding why endothermic scattering can produce a relatively narrow population of high-energy recoils.

Finally, the existence of a maximum DM speed in the laboratory frame imposes an upper limit on the mass splitting that can be probed. If the largest available speed is denoted by $v_{\max}$, scattering requires
\begin{equation}
v_\delta
=
\sqrt{\frac{2\delta}{\mu_A}}
\leq
v_{\max}.
\end{equation}
The largest kinematically accessible splitting is therefore
\begin{equation}
\delta_{\max}
=
\frac{1}{2}
\mu_Av_{\max}^2.
\label{eq:deltamax_appendix}
\end{equation}

As $\delta$ approaches $\delta_{\max}$, only particles in the extreme high-speed tail of the Galactic distribution can scatter. The event rate then becomes highly sensitive to the precise value of the Galactic escape speed, the Earth's motion, and departures from the Standard Halo Model. This also explains the enhanced annual modulation expected for strongly endothermic models: relatively small seasonal changes in the laboratory-frame velocity distribution can produce a large fractional change in the number of particles above the threshold in Eq.~\eqref{eq:vthreshold}. The Higgsino benchmark considered in this work lies particularly close to this regime, while the generic pseudo-Dirac benchmark is somewhat less extreme.

\section{Rate normalization and conversion to the LZ convention}
\label{app:rate}

For each xenon isotope, the differential nuclear scattering cross section used in the simplified recast is
\begin{equation}
\frac{d\sigma_A}{dE_R}
=
\frac{m_A\sigma_A}
{2\mu_A^2v^2}
F_A^2(E_R),
\end{equation}
with
\begin{equation}
\sigma_A
=
\sigma_N
\frac{\mu_A^2}{\mu_N^2}
Q_A^2.
\end{equation}
For an isoscalar SI interaction,
\begin{equation}
Q_A=A,
\end{equation}
whereas for the Higgsino interaction mediated by the $Z$ boson,
\begin{equation}
Q_A
=
N-
\left(
1-4\sin^2\theta_W
\right)Z.
\end{equation}

The differential rate is obtained by multiplying the scattering cross section by the local DM number density, the DM flux, and the number of target nuclei per detector mass. The implementation keeps the halo velocity distribution normalized in physical units of $\mathrm{km/s}$ and performs the appropriate conversion when evaluating the mean inverse speed.

For comparison with the LZ $\mathcal{O}_1^s$ result, the Supplemental Material gives
\begin{equation}
\left(
c_1^s m_v^2
\right)^2
=
\frac{
\sigma_N^{\rm SI}
\pi m_v^4
}{
\mu_N^2
},
\label{eq:lz_o1_conversion}
\end{equation}
where
\begin{equation}
m_v=246.2\,\GeV
\end{equation}
is the electroweak vacuum expectation value. Equation~\eqref{eq:lz_o1_conversion} corresponds to the scalar normalization used by LZ when presenting the inelastic $\mathcal{O}_1^s$ cross-section limits.

For a vector interaction, the coherent xenon charge differs from the scalar $A^2$ normalization. LZ notes that the corresponding xenon cross section can be obtained by multiplying the scalar-normalized result by approximately
\begin{equation}
\left[
\frac{A}
{
(A-Z)
-
\left(
1-4\sin^2\theta_W
\right)Z
}
\right]^2
\simeq
3.2.
\label{eq:vector_conversion}
\end{equation}
This distinction is important when comparing the pseudo-Dirac or Higgsino vector interaction directly with the scalar-normalized $\mathcal{O}_1$ limits shown by LZ.

\section{Digitized LZ efficiency}
\label{app:efficiency}

The signal efficiency used in the numerical recast is reconstructed from Supplemental Fig.~S2 of Ref.~\cite{LZ2026Extended}. The final black curve, which includes the trigger, S1 threshold, single-scatter and analysis selections, and the $S1c$--$S2c$ ROI requirement, is extracted from the vector graphics of the published figure and interpolated on a fine recoil-energy grid.

The digitization gives
\begin{equation}
\epsilon(248\,\keV)=0.9457,
\end{equation}
with a gray-envelope range
\begin{equation}
0.8835<\epsilon(248\,\keV)<0.9800,
\end{equation}
and
\begin{equation}
\epsilon(269.9\,\keV)=0.4984.
\end{equation}
These values are consistent with the LZ statement that the final efficiency is approximately $96\%$ on average between $14$ and $250\,\keV$ and falls to $50\%$ at $269.9\,\keV$ \cite{LZ2026Extended}. The published Supplemental figure also identifies the low-energy $50\%$ efficiency point at $5.4\,\keV$.

The gray uncertainty envelope is propagated by repeating the rate calculation using its lower and upper boundaries. Since LZ does not assign a statistical confidence level to this envelope, we interpret these variations as a detector systematic range rather than as a confidence interval. The digitized efficiency remains an approximation and does not replace the full two-dimensional NEST response in $S1c$ and $S2c$.

\section{One-event statistical guide}
\label{app:poisson}

A simple Poisson calculation provides intuition for the normalization uncertainty associated with one candidate event. Taking
\begin{equation}
n=1
\end{equation}
and using the high-$S1$ background guide
\begin{equation}
b=0.0106,
\end{equation}
the maximum-likelihood signal count is
\begin{equation}
\hat s=n-b=0.9894.
\end{equation}

For the Poisson likelihood
\begin{equation}
\mathcal{L}(s)
\propto
(s+b)^n
\exp[-(s+b)],
\end{equation}
the one-parameter profile-likelihood condition
\begin{equation}
\Delta
\left[
-2\ln\mathcal{L}
\right]
=
2.7055
\end{equation}
gives approximately
\begin{equation}
0.095
\lesssim
s
\lesssim
3.64
\end{equation}
at nominal $90\%$ confidence.

For a fixed signal spectrum, the same multiplicative uncertainty applies approximately to the inferred cross section. This interval is included only as an intuitive counting guide. It is not equivalent to the LZ confidence interval, because the experimental analysis uses the full event distributions in the $\{S1c,\log_{10}S2c\}$ plane, three event samples, nuisance constraints, and the profile likelihood constructed for each DM model.

\section{Derivation of the contact direct--relic relation}
\label{app:direct_relic}

For the generic pseudo-Dirac benchmark, we assume universal vector couplings to quarks. The coherent nonrelativistic nucleon matrix element then contains the vector charge
\begin{equation}
g_N=3g_q,
\end{equation}
which leads directly to the nucleon scattering cross section given in Eq.~\eqref{eq:sigmageneric}.

In the heavy-mediator limit and neglecting quark masses, the coannihilation cross section for one ordered pair is
\begin{equation}
\sigma_{12}v
\simeq
\sum_{q=1}^{6}
\frac{
N_c g_\chi^2g_q^2m_\chi^2
}{
\pi m_V^4
}
=
\frac{
18g_\chi^2g_q^2m_\chi^2
}{
\pi m_V^4
}.
\end{equation}
For nearly degenerate states with equal equilibrium abundances, the ordered pairs $12$ and $21$ each carry statistical weight $1/4$, giving
\begin{equation}
\sigma_{\rm eff}v
=
\frac{1}{2}\sigma_{12}v.
\end{equation}
Eliminating the combination
\begin{equation}
\frac{g_\chi g_q}{m_V^2}
\end{equation}
between the scattering and annihilation rates gives Eq.~\eqref{eq:masterrelation}.

The cancellation is specific to the minimal universal-vector contact benchmark. Nonuniversal quark charges, massive final states, resonant annihilation, additional coannihilation channels, axial interactions, or a mediator sufficiently light that the contact approximation fails modify this relation and should be treated in a complete ultraviolet model.

\section{Theoretical structure of the Higgsino dark-matter sector}
\label{sec:higgsinotheory}

A particularly predictive realization of inelastic DM is provided by a nearly pure Higgsino in the Minimal Supersymmetric Standard Model (MSSM), or more generally in a supersymmetric (SUSY) spectrum in which the electroweak gauginos and scalar superpartners are substantially heavier than the Higgsino. In contrast with the generic pseudo-Dirac model of Sec.~\ref{sec:minpseudo}, the particle content and the relevant couplings are largely fixed by electroweak gauge invariance. The Higgsino therefore provides a useful ultraviolet-motivated realization of the inelastic $\mathcal{O}_1$ phenomenology considered by LZ \cite{NagataShirai2015,KrallReece2018,Graham2025,LZ2026Extended}.

\subsection{Particle content and electroweakino Lagrangian}

The MSSM contains two Higgs doublets,
\begin{equation}
H_u=
\begin{pmatrix}
H_u^+\\
H_u^0
\end{pmatrix},
\qquad
H_d=
\begin{pmatrix}
H_d^0\\
H_d^-
\end{pmatrix},
\end{equation}
with hypercharges
\begin{equation}
Y_{H_u}=+\frac{1}{2},
\qquad
Y_{H_d}=-\frac{1}{2}.
\end{equation}
Their fermionic superpartners are the Higgsino doublets
\begin{equation}
\widetilde H_u=
\begin{pmatrix}
\widetilde H_u^+\\
\widetilde H_u^0
\end{pmatrix},
\qquad
\widetilde H_d=
\begin{pmatrix}
\widetilde H_d^0\\
\widetilde H_d^-
\end{pmatrix}.
\end{equation}
The supersymmetric $\mu$ term in the superpotential,
\begin{equation}
W_{\rm MSSM}\supset
\mu H_u\cdot H_d,
\end{equation}
generates the Higgsino mass term
\begin{equation}
\mathcal{L}_{\widetilde H,{\rm mass}}
=
-\mu\,
\epsilon_{\alpha\beta}
\widetilde H_u^\alpha
\widetilde H_d^\beta
+\mathrm{h.c.}
\label{eq:higgsino_mu_term}
\end{equation}
The parameter $\mu$ appearing here is the supersymmetric Higgsino mass parameter and should not be confused with the reduced masses $\mu_A$ and $\mu_N$ used in the direct-detection calculation.

The electroweak gaugino sector contains the bino $\widetilde B$ and the three winos $\widetilde W^a$, with soft Majorana masses
\begin{equation}
\mathcal{L}_{\rm soft}
\supset
-\frac{1}{2}M_1\widetilde B\widetilde B
-\frac{1}{2}M_2\widetilde W^a\widetilde W^a
+\mathrm{h.c.}
\label{eq:gaugino_masses}
\end{equation}
The Higgsino--gaugino interactions are fixed by the electroweak gauge couplings,
\begin{align}
\mathcal{L}_{\widetilde H\widetilde V H}
\supset{}&
-\sqrt{2}g
\left[
H_u^\dagger T^a\widetilde H_u
+
H_d^\dagger T^a\widetilde H_d
\right]
\widetilde W^a
\nonumber\\
&
-\sqrt{2}g'
\left[
Y_{H_u}H_u^\dagger\widetilde H_u
+
Y_{H_d}H_d^\dagger\widetilde H_d
\right]
\widetilde B
+\mathrm{h.c.}
\label{eq:higgsino_gaugino_interactions}
\end{align}
After electroweak symmetry breaking (EWSB),
\begin{equation}
\left\langle H_u^0\right\rangle
=
\frac{v_u}{\sqrt{2}},
\qquad
\left\langle H_d^0\right\rangle
=
\frac{v_d}{\sqrt{2}},
\end{equation}
with
\begin{equation}
v^2=v_u^2+v_d^2,
\qquad
v\simeq246\,\GeV,
\qquad
\tan\beta=\frac{v_u}{v_d},
\end{equation}
Eq.~\eqref{eq:higgsino_gaugino_interactions} generates mixing among the neutral Higgsinos, bino, and neutral wino.

Defining
\begin{equation}
s_W=\sin\theta_W,
\,\,
c_W=\cos\theta_W,
\,\,
s_\beta=\sin\beta,
\,\,
c_\beta=\cos\beta,
\end{equation}
the neutralino mass matrix in the basis
\begin{equation}
\psi^0=
\left(
-i\widetilde B,\,
-i\widetilde W^3,\,
\widetilde H_d^0,\,
\widetilde H_u^0
\right)
\end{equation}
is
\begin{equation}
\mathcal{M}_{\widetilde\chi^0}
=\!\!
\begin{pmatrix}
M_1
&
0
&
-m_Zs_Wc_\beta
&
m_Zs_Ws_\beta
\\
0
&
M_2
&
m_Zc_Wc_\beta
&
-m_Zc_Ws_\beta
\\
-m_Zs_Wc_\beta
&
m_Zc_Wc_\beta
&
0
&
-\mu
\\
m_Zs_Ws_\beta
&
-m_Zc_Ws_\beta
&
-\mu
&
0
\end{pmatrix}.
\label{eq:neutralino_mass_matrix}
\end{equation}
Similarly, in the charged basis
\begin{equation}
\psi^-=
\left(
-i\widetilde W^-,
\widetilde H_d^-
\right),
\qquad
\psi^+=
\left(
-i\widetilde W^+,
\widetilde H_u^+
\right),
\end{equation}
the chargino mass matrix is
\begin{equation}
\mathcal{M}_{\widetilde\chi^\pm}
=
\begin{pmatrix}
M_2
&
\sqrt{2}m_Ws_\beta
\\
\sqrt{2}m_Wc_\beta
&
\mu
\end{pmatrix}.
\label{eq:chargino_mass_matrix}
\end{equation}

The limit relevant for the present work is
\begin{equation}
|M_1|,|M_2|\gg|\mu|\gg m_Z.
\label{eq:pure_higgsino_limit}
\end{equation}
The light electroweakino spectrum then consists almost entirely of the two Higgsino doublets. It contains two neutral Majorana fermions,
\begin{equation}
\widetilde H_1^0,
\qquad
\widetilde H_2^0,
\end{equation}
and one charged Dirac fermion,
\begin{equation}
\widetilde H^\pm.
\end{equation}
For real positive $\mu$, a convenient phase convention for the neutral states is approximately
\begin{align}
\widetilde H_1^0
&\simeq
\frac{1}{\sqrt{2}}
\left(
\widetilde H_d^0-\widetilde H_u^0
\right),
\\
\widetilde H_2^0
&\simeq
\frac{i}{\sqrt{2}}
\left(
\widetilde H_d^0+\widetilde H_u^0
\right).
\label{eq:higgsino_mass_eigenstates}
\end{align}
Up to phase conventions, these are the two Majorana components of the neutral Dirac Higgsino that would exist if the gauginos were completely decoupled.

Assuming conserved $R$ parity, the lighter neutral state $\widetilde H_1^0$ is the lightest supersymmetric particle (LSP) and is stable. In the remainder of this work we identify
\begin{equation}
m_\chi=m_{\widetilde H_1^0}.
\end{equation}

\subsection{Origin of the neutral-state splitting}

In the formal limit
\begin{equation}
M_1,M_2\rightarrow\infty,
\end{equation}
the two neutral Higgsinos are exactly degenerate and combine into a Dirac fermion. This degeneracy can be understood as the consequence of an approximate Higgsino-number symmetry. Finite Majorana masses $M_1$ and $M_2$ break this symmetry. Mixing with the bino and wino after EWSB therefore generates small Majorana masses for the neutral Higgsinos and splits the Dirac state into $\widetilde H_1^0$ and $\widetilde H_2^0$.

Expanding Eq.~\eqref{eq:neutralino_mass_matrix} in inverse powers of the heavy gaugino masses gives, at leading order,
\begin{equation}
\delta
\equiv
m_{\widetilde H_2^0}
-
m_{\widetilde H_1^0}
\simeq
m_Z^2
\left(
\frac{s_W^2}{M_1}
+
\frac{c_W^2}{M_2}
\right)
+
\mathcal{O}
\left(
\frac{m_Z^2\mu}{M_{1,2}^2}
\right).
\label{eq:higgsino_split_theory}
\end{equation}
The leading contribution is independent of $\tan\beta$, while subleading terms contain additional dependence on $\mu$, $\tan\beta$, phases, and the relative signs of $M_1$ and $M_2$ \cite{NagataShirai2015,KrallReece2018}. Opposite signs of the bino and wino masses can also lead to partial cancellations in Eq.~\eqref{eq:higgsino_split_theory}.

The origin of the splitting can be viewed diagrammatically as two Higgsino--gaugino mixing insertions connected by a heavy Majorana bino or wino propagator. Since each mixing insertion is proportional to the Higgs vacuum expectation value, the induced Majorana mass scales parametrically as
\begin{equation}
\delta
\sim
\frac{g^2v^2}{M_{1,2}}
\sim
\frac{m_Z^2}{M_{1,2}}.
\end{equation}
This explains why a splitting in the sub-MeV range can coexist naturally with a TeV-scale Higgsino if the gauginos reside at a much larger mass scale.

For example, if
\begin{equation}
M_1=M_2=M,
\end{equation}
Eq.~\eqref{eq:higgsino_split_theory} simplifies to
\begin{equation}
\delta\simeq\frac{m_Z^2}{M}.
\label{eq:higgsino_equal_gaugino}
\end{equation}
The LZ-motivated value $\delta\simeq377\,\keV$ therefore corresponds to
\begin{equation}
M\simeq
\frac{m_Z^2}{\delta}
\simeq
2.2\times10^7\,\GeV,
\end{equation}
when the bino and wino contributions are comparable and have the same sign.

The charged Higgsino is also nearly degenerate with the neutral states. In the highly pure limit, electroweak radiative corrections produce a charged--neutral mass splitting of approximately
\begin{equation}
\Delta m_\pm
\equiv
m_{\widetilde H^\pm}
-
m_{\widetilde H_1^0}
\simeq
350\,\mathrm{MeV},
\label{eq:higgsino_charged_split}
\end{equation}
up to smaller tree-level contributions induced by the heavy gauginos \cite{KrallReece2018}. Thus, for the parameter region relevant here,
\begin{equation}
\delta
\ll
\Delta m_\pm
\ll
m_\chi.
\end{equation}
The low-energy spectrum therefore contains two extremely close neutral states separated by a few hundred keV and a charged state roughly a few hundred MeV above them.

\subsection{Gauge interactions and the origin of inelastic scattering}

The crucial direct-detection interaction follows directly from the Higgsino gauge kinetic terms,
\begin{equation}
\mathcal{L}_{\rm kin}
=
i\widetilde H_u^\dagger
\bar\sigma^\mu D_\mu\widetilde H_u
+
i\widetilde H_d^\dagger
\bar\sigma^\mu D_\mu\widetilde H_d.
\end{equation}
For the neutral components, the coupling to the $Z$ boson is
\begin{equation}
\mathcal{L}_{Z\widetilde H\widetilde H}
=
\frac{g}{2c_W}
Z_\mu
\left[
\widetilde H_d^{0\dagger}
\bar\sigma^\mu
\widetilde H_d^0
-
\widetilde H_u^{0\dagger}
\bar\sigma^\mu
\widetilde H_u^0
\right].
\label{eq:higgsino_Z_gauge}
\end{equation}
Substituting the mass eigenstates of Eq.~\eqref{eq:higgsino_mass_eigenstates} gives
\begin{equation}
\mathcal{L}_{Z\widetilde H_1\widetilde H_2}
=
\frac{ig}{2c_W}
Z_\mu
\left[
\widetilde H_2^{0\dagger}
\bar\sigma^\mu
\widetilde H_1^0
-
\widetilde H_1^{0\dagger}
\bar\sigma^\mu
\widetilde H_2^0
\right].
\label{eq:higgsino_Z_offdiagonal}
\end{equation}
The important point is that the leading neutral-current interaction is \emph{off diagonal}. The $Z$ boson converts the lighter state into the heavier one rather than coupling $\widetilde H_1^0$ to itself. Consequently, the tree-level process relevant for nuclear scattering is
\begin{equation}
\widetilde H_1^0+A
\rightarrow
\widetilde H_2^0+A,
\label{eq:higgsino_inelastic_process}
\end{equation}
which is precisely an endothermic transition.

This provides the microscopic origin of the inelastic kinematics discussed in Sec.~\ref{sec:inelastickinematics} and derived in more detail in Appendix~\ref{app:inelastic_kinematics}. The large $Z$-mediated interaction is not excluded by conventional low-energy direct-detection searches when the splitting is sufficiently large because the incoming DM particle must supply the additional energy $\delta$ needed to produce $\widetilde H_2^0$.

The vector coupling of the $Z$ boson to quarks is
\begin{equation}
\mathcal{L}_{Zq}
=
\frac{g}{2c_W}
Z_\mu
\overline q\gamma^\mu
\left[
T_3^q
-
2Q_qs_W^2
-
T_3^q\gamma^5
\right]q.
\end{equation}
At the nucleon level, its coherent vector component becomes
\begin{equation}
\mathcal{L}_{ZN}^{V}
=
\frac{g}{4c_W}
Z_\mu
\left[
\left(
1-4s_W^2
\right)
\overline p\gamma^\mu p
-
\overline n\gamma^\mu n
\right].
\label{eq:Z_nucleon_vector}
\end{equation}
Since
\begin{equation}
1-4s_W^2\simeq0.075,
\end{equation}
the proton vector coupling is strongly suppressed and the coherent interaction is predominantly with neutrons.

After integrating out the $Z$ boson, the leading nonrelativistic interaction is the identity operator $\mathcal{O}_1$. Neglecting the nuclear form factor and the inelastic phase-space suppression, the zero-momentum cross section for a nucleus with $Z$ protons and $N$ neutrons is
\begin{equation}
\sigma_A^{\widetilde H}
=
\frac{G_F^2\mu_A^2}{2\pi}
\left[
N-
\left(
1-4s_W^2
\right)Z
\right]^2.
\label{eq:higgsino_nucleus_cross_section}
\end{equation}
The coefficient in Eq.~\eqref{eq:higgsino_nucleus_cross_section} follows directly from the off-diagonal current in Eq.~\eqref{eq:higgsino_Z_offdiagonal}: combining it with Eq.~\eqref{eq:Z_nucleon_vector} gives the neutron contact coefficient $C_n=G_F/\sqrt{2}$ (up to an irrelevant sign), and hence $\sigma_n=\mu_N^2C_n^2/\pi$. The smaller $G_F^2\mu_N^2/(8\pi)$ result corresponds to assigning an additional factor $1/2$ to that transition current, i.e., $C_n=G_F/(2\sqrt{2})$; such a factor is not present after rotating the two neutral Higgsino Weyl fields to the Majorana mass basis. This normalization point is the origin of the factor-of-four discrepancy emphasized in Ref.~\cite{PospelovRamani2026}.
It is therefore useful to define the neutron-normalized cross section
\begin{equation}
\sigma_n^{\widetilde H}
=
\frac{G_F^2\mu_N^2}{2\pi}
\simeq
7.4\times10^{-39}\,\mathrm{cm}^2
\label{eq:higgsinosigma_theory}
\end{equation}
for a TeV-scale Higgsino. This interaction strength is fixed by the SM weak coupling and is not a free parameter of the Higgsino model. The relation between this weak-charge normalization and the $\mathcal{O}_1$ convention used by LZ is discussed further in Appendix~\ref{app:rate}.

It is important to distinguish this large inelastic interaction from ordinary elastic SI scattering. The diagonal $Z$ vector current vanishes for the neutral Majorana mass eigenstate, while the coupling of a pure Higgsino to the SM Higgs boson requires gaugino admixture and is suppressed in the limit of Eq.~\eqref{eq:pure_higgsino_limit}. Irreducible electroweak loop diagrams generate an elastic SI cross section below approximately $10^{-48}\,\mathrm{cm}^2$, with significant cancellations among the different contributions \cite{NagataShirai2015,KrallReece2018}. Thus the characteristic tree-level direct-detection signature of a very pure Higgsino is the much larger but kinematically restricted transition of Eq.~\eqref{eq:higgsino_inelastic_process}.

\subsection{Charged-current interactions and thermal freeze-out}

The charged Higgsino is not merely an additional state in the spectrum: it plays a central role in the thermal history. The charged-current interactions follow from the same electroweak kinetic terms,
\begin{align}
\mathcal{L}_{W\widetilde H\widetilde H}
\supset{}&
\frac{g}{\sqrt{2}}
W_\mu^+
\left[
\widetilde H_u^{+\dagger}
\bar\sigma^\mu
\widetilde H_u^0
+
\widetilde H_d^{0\dagger}
\bar\sigma^\mu
\widetilde H_d^-
\right]
\nonumber\\
&
+\mathrm{h.c.}
\label{eq:higgsino_W_coupling}
\end{align}
Together with the $Z$ and photon interactions, these couplings generate annihilation and coannihilation processes involving all members of the Higgsino multiplet.

At freeze-out,
\begin{equation}
T_f
\simeq
\frac{m_\chi}{20\text{--}25}
\simeq
40\text{--}50\,\GeV
\end{equation}
for $m_\chi\simeq1.1\,\TeV$. Both the neutral splitting,
\begin{equation}
\delta\simeq3.7\times10^{-4}\,\GeV,
\end{equation}
and the charged--neutral splitting,
\begin{equation}
\Delta m_\pm\simeq0.35\,\GeV,
\end{equation}
are therefore negligible compared with $T_f$. The two neutral Majorana states and the charged Dirac state are all thermally populated and must be included in the effective annihilation rate.

Representative channels include
\begin{align}
\widetilde H_i^0\widetilde H_j^0
&\rightarrow
W^+W^-,
\ ZZ,
\\
\widetilde H_i^0\widetilde H^\pm
&\rightarrow
W^\pm Z,
\ W^\pm\gamma,
\ W^\pm h,
\\
\widetilde H^+\widetilde H^-
&\rightarrow
W^+W^-,
\ ZZ,
\ Z\gamma,
\ \gamma\gamma,
\ f\overline f.
\end{align}
In the limit in which all Higgsino states are degenerate and electroweak gauge-boson masses are neglected, the thermally averaged effective cross section takes the approximate form
\begin{equation}
\left\langle
\sigma_{\rm eff}v
\right\rangle_{\widetilde H}
\simeq
\frac{g^4}
{512\pi m_\chi^2}
\left[
21
+
3\tan^2\theta_W
+
11\tan^4\theta_W
\right].
\label{eq:higgsino_sigma_eff}
\end{equation}
The characteristic $1/m_\chi^2$ scaling leads approximately to
\begin{equation}
\Omega_{\widetilde H}h^2
\simeq
0.12
\left(
\frac{m_\chi}{1.1\,\TeV}
\right)^2,
\label{eq:higgsino_relic_scaling}
\end{equation}
with the precise thermal target depending on electroweak corrections and the treatment of the nearly degenerate states \cite{KrallReece2018,Rodd2024,Graham2025}. This is the origin of the thermal Higgsino mass
\begin{equation}
m_\chi\simeq1.1\,\TeV
\end{equation}
used throughout this work.

The fact that
\begin{equation}
\delta\ll T_f
\end{equation}
is essential. A splitting of several hundred keV has essentially no effect on the early-Universe abundance, but it has a dramatic effect on present-day direct detection because Galactic kinetic energies are themselves only of order $100\,\keV$. The relic density and the high-energy LZ event therefore probe very different manifestations of the same electroweak multiplet.

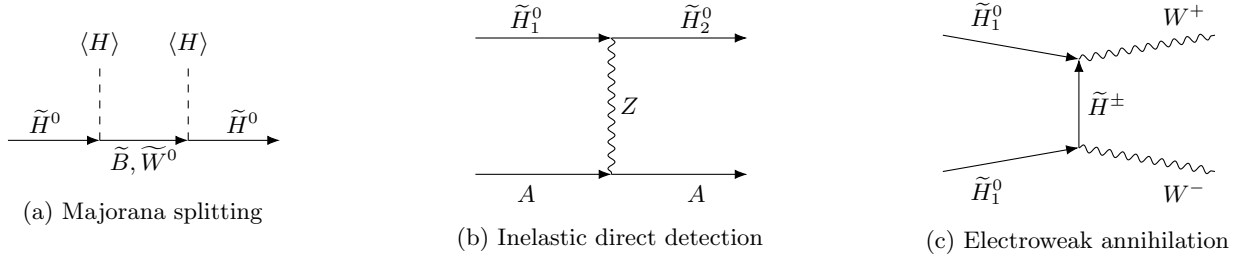
\begin{figure*}[t]
\centering

\begin{minipage}{0.31\textwidth}
\centering
\begin{tikzpicture}[>=Latex,scale=0.90]
\draw[->] (-2,0) -- (-0.65,0);
\draw[->] (0.65,0) -- (2,0);
\draw[->] (-0.65,0) -- (0.65,0);
\draw[dashed] (-0.65,0) -- (-0.65,1.15);
\draw[dashed] (0.65,0) -- (0.65,1.15);

\node at (-1.45,0.28) {$\widetilde H^0$};
\node at (1.45,0.28) {$\widetilde H^0$};
\node at (0,-0.30) {$\widetilde B,\widetilde W^0$};
\node at (-0.65,1.42) {$\langle H\rangle$};
\node at (0.65,1.42) {$\langle H\rangle$};
\end{tikzpicture}

\vspace{0.1cm}
{\small (a) Majorana splitting}
\end{minipage}
\hfill
\begin{minipage}{0.31\textwidth}
\centering
\begin{tikzpicture}[>=Latex,scale=0.90]
\draw[->] (-2,1) -- (0,1);
\draw[->] (0,1) -- (2,1);
\draw[->] (-2,-1) -- (0,-1);
\draw[->] (0,-1) -- (2,-1);

\draw[
decorate,
decoration={snake,amplitude=1.3pt,segment length=5pt}
]
(0,1) -- (0,-1);

\node at (-1.25,1.28) {$\widetilde H_1^0$};
\node at (1.25,1.28) {$\widetilde H_2^0$};
\node at (-1.25,-1.28) {$A$};
\node at (1.25,-1.28) {$A$};
\node[right] at (0,0) {$Z$};
\end{tikzpicture}

\vspace{0.1cm}
{\small (b) Inelastic direct detection}
\end{minipage}
\hfill
\begin{minipage}{0.31\textwidth}
\centering
\begin{tikzpicture}[>=Latex,scale=0.90]
\draw[->] (-2,1) -- (0,0.65);
\draw[->] (-2,-1) -- (0,-0.65);

\draw[->] (0,-0.65) -- (0,0.65);

\draw[
decorate,
decoration={snake,amplitude=1.3pt,segment length=5pt}
]
(0,0.65) -- (2,1);

\draw[
decorate,
decoration={snake,amplitude=1.3pt,segment length=5pt}
]
(0,-0.65) -- (2,-1);

\node at (-1.35,1.28) {$\widetilde H_1^0$};
\node at (-1.35,-1.28) {$\widetilde H_1^0$};
\node[right] at (0,0) {$\widetilde H^\pm$};
\node at (1.55,1.30) {$W^+$};
\node at (1.55,-1.30) {$W^-$};
\end{tikzpicture}

\vspace{0.1cm}
{\small (c) Electroweak annihilation}
\end{minipage}

\caption{Representative interactions governing the Higgsino phenomenology discussed in this work. Panel (a) illustrates the origin of the neutral Majorana splitting from two EWSB mixing insertions connected by a heavy bino or wino propagator, giving parametrically $\delta\sim m_Z^2/M_{1,2}$. Panel (b) shows the tree-level $Z$-mediated endothermic process $\widetilde H_1^0 A\rightarrow\widetilde H_2^0 A$ responsible for the LZ signal. Panel (c) shows a representative present-day annihilation process into electroweak gauge bosons through charged-Higgsino exchange; related annihilation and coannihilation channels determine the thermal relic abundance.}
\label{fig:higgsino_feynman}
\end{figure*}

Figure~\ref{fig:higgsino_feynman} summarizes the three ingredients that make the Higgsino interpretation unusually predictive. Heavy Majorana gauginos generate the small splitting, the SM $Z$ boson mediates the inelastic nuclear transition, and ordinary electroweak gauge interactions control the annihilation and coannihilation processes responsible for freeze-out. Unlike the generic pseudo-Dirac model, no independent mediator coupling is introduced to connect these phenomena.

\subsection{Present-day spectrum and indirect detection}

After freeze-out, the heavier electroweakino states disappear. The charged Higgsino rapidly decays to the lightest neutral state through weak transitions, while the heavier neutral state $\widetilde H_2^0$ can decay through
\begin{equation}
\widetilde H_2^0
\rightarrow
\widetilde H_1^0\gamma
\end{equation}
at one loop, as well as through suppressed three-body weak decays. For sub-GeV neutral splittings, the radiative channel is particularly important \cite{KrallReece2018}. Consequently, the Galactic DM population today consists essentially entirely of $\widetilde H_1^0$.

This differs qualitatively from the freeze-out epoch. Present-day annihilation therefore involves primarily
\begin{equation}
\widetilde H_1^0\widetilde H_1^0
\rightarrow
W^+W^-,
\ ZZ,
\end{equation}
with additional loop-induced monochromatic channels
\begin{equation}
\widetilde H_1^0\widetilde H_1^0
\rightarrow
\gamma\gamma,
\qquad
\widetilde H_1^0\widetilde H_1^0
\rightarrow
\gamma Z.
\end{equation}
The $W^+W^-$ and $ZZ$ final states generate broad continuum gamma-ray emission, whereas $\gamma\gamma$ and $\gamma Z$ produce line-like spectral features close to the DM mass. Electroweak radiative and Sommerfeld effects must be included for precision predictions of these signals \cite{Rodd2024}.

This point is particularly relevant for the comparison with the generic pseudo-Dirac model of Sec.~\ref{sec:minpseudo}. In that construction, freeze-out can be dominated by the off-diagonal process $\chi_1\chi_2\rightarrow q\overline q$, so depletion of $\chi_2$ can strongly reduce the present-day annihilation signal. For the Higgsino, instead, the light state $\widetilde H_1^0$ retains unsuppressed electroweak annihilation channels. The thermal relic-density requirement, the LZ inelastic-scattering rate, and gamma-ray searches are therefore tightly connected.

The Higgsino interpretation considered here can consequently be summarized by three largely independent theoretical inputs:
\begin{align}
m_{\widetilde H_1^0}
&\simeq
1.1\,\TeV
&&
\text{from thermal freeze-out},
\\
\sigma_n^{\widetilde H}
&\simeq
7.4\times10^{-39}\,\mathrm{cm}^2
&&
\text{from the SM weak interaction},
\\
\delta
&=
m_{\widetilde H_2^0}
-
m_{\widetilde H_1^0}
&&
\text{from heavy-gaugino mixing}.
\end{align}
The first two quantities are therefore essentially fixed once a thermal, nearly pure Higgsino is assumed. The LZ event predominantly measures the third. As shown in the following subsection, requiring approximately one accepted recoil in the present LZ exposure selects
\begin{equation}
\delta\simeq377\,\keV,
\end{equation}
placing the model remarkably close to the Galactic kinematic boundary.

\bibliography{LZ_248keV_DM_PRD_final}

\begin{thebibliography}{50}%
\makeatletter
\providecommand \@ifxundefined [1]{%
 \@ifx{#1\undefined}
}%
\providecommand \@ifnum [1]{%
 \ifnum #1\expandafter \@firstoftwo
 \else \expandafter \@secondoftwo
 \fi
}%
\providecommand \@ifx [1]{%
 \ifx #1\expandafter \@firstoftwo
 \else \expandafter \@secondoftwo
 \fi
}%
\providecommand \natexlab [1]{#1}%
\providecommand \enquote  [1]{``#1''}%
\providecommand \bibnamefont  [1]{#1}%
\providecommand \bibfnamefont [1]{#1}%
\providecommand \citenamefont [1]{#1}%
\providecommand \href@noop [0]{\@secondoftwo}%
\providecommand \href [0]{\begingroup \@sanitize@url \@href}%
\providecommand \@href[1]{\@@startlink{#1}\@@href}%
\providecommand \@@href[1]{\endgroup#1\@@endlink}%
\providecommand \@sanitize@url [0]{\catcode `\\12\catcode `\$12\catcode
  `\&12\catcode `\#12\catcode `\^12\catcode `\_12\catcode `\%12\relax}%
\providecommand \@@startlink[1]{}%
\providecommand \@@endlink[0]{}%
\providecommand \url  [0]{\begingroup\@sanitize@url \@url }%
\providecommand \@url [1]{\endgroup\@href {#1}{\urlprefix }}%
\providecommand \urlprefix  [0]{URL }%
\providecommand \Eprint [0]{\href }%
\providecommand \doibase [0]{https://doi.org/}%
\providecommand \selectlanguage [0]{\@gobble}%
\providecommand \bibinfo  [0]{\@secondoftwo}%
\providecommand \bibfield  [0]{\@secondoftwo}%
\providecommand \translation [1]{[#1]}%
\providecommand \BibitemOpen [0]{}%
\providecommand \bibitemStop [0]{}%
\providecommand \bibitemNoStop [0]{.\EOS\space}%
\providecommand \EOS [0]{\spacefactor3000\relax}%
\providecommand \BibitemShut  [1]{\csname bibitem#1\endcsname}%
\let\auto@bib@innerbib\@empty
\bibitem [{\citenamefont {Rubin}\ \emph {et~al.}(1980)\citenamefont {Rubin},
  \citenamefont {Ford},\ and\ \citenamefont {Thonnard}}]{Rubin1980}%
  \BibitemOpen
  \bibfield  {author} {\bibinfo {author} {\bibfnamefont {V.~C.}\ \bibnamefont
  {Rubin}}, \bibinfo {author} {\bibfnamefont {J.}~\bibnamefont {Ford},
  \bibfnamefont {W.~Kent}},\ and\ \bibinfo {author} {\bibfnamefont
  {N.}~\bibnamefont {Thonnard}},\ }\bibfield  {title} {\bibinfo {title}
  {Rotational properties of 21 sc galaxies with a large range of luminosities
  and radii, from ngc 4605 ($r=4$ kpc) to ugc 2885 ($r=122$ kpc)},\ }\href
  {https://doi.org/10.1086/158003} {\bibfield  {journal} {\bibinfo  {journal}
  {Astrophys. J.}\ }\textbf {\bibinfo {volume} {238}},\ \bibinfo {pages} {471}
  (\bibinfo {year} {1980})}\BibitemShut {NoStop}%
\bibitem [{\citenamefont {Clowe}\ \emph {et~al.}(2006)\citenamefont {Clowe},
  \citenamefont {Bradac}, \citenamefont {Gonzalez}, \citenamefont {Markevitch},
  \citenamefont {Randall}, \citenamefont {Jones},\ and\ \citenamefont
  {Zaritsky}}]{Clowe2006}%
  \BibitemOpen
  \bibfield  {author} {\bibinfo {author} {\bibfnamefont {D.}~\bibnamefont
  {Clowe}}, \bibinfo {author} {\bibfnamefont {M.}~\bibnamefont {Bradac}},
  \bibinfo {author} {\bibfnamefont {A.~H.}\ \bibnamefont {Gonzalez}}, \bibinfo
  {author} {\bibfnamefont {M.}~\bibnamefont {Markevitch}}, \bibinfo {author}
  {\bibfnamefont {S.~W.}\ \bibnamefont {Randall}}, \bibinfo {author}
  {\bibfnamefont {C.}~\bibnamefont {Jones}},\ and\ \bibinfo {author}
  {\bibfnamefont {D.}~\bibnamefont {Zaritsky}},\ }\bibfield  {title} {\bibinfo
  {title} {A direct empirical proof of the existence of dark matter},\ }\href
  {https://doi.org/10.1086/508162} {\bibfield  {journal} {\bibinfo  {journal}
  {Astrophys. J. Lett.}\ }\textbf {\bibinfo {volume} {648}},\ \bibinfo {pages}
  {L109} (\bibinfo {year} {2006})},\ \Eprint
  {https://arxiv.org/abs/astro-ph/0608407} {arXiv:astro-ph/0608407}
  \BibitemShut {NoStop}%
\bibitem [{\citenamefont {Aghanim}\ \emph {et~al.}(2020)\citenamefont {Aghanim}
  \emph {et~al.}}]{Planck2018}%
  \BibitemOpen
  \bibfield  {author} {\bibinfo {author} {\bibfnamefont {N.}~\bibnamefont
  {Aghanim}} \emph {et~al.} (\bibinfo {collaboration} {Planck Collaboration}),\
  }\bibfield  {title} {\bibinfo {title} {Planck 2018 results. vi. cosmological
  parameters},\ }\href {https://doi.org/10.1051/0004-6361/201833910} {\bibfield
   {journal} {\bibinfo  {journal} {Astron. Astrophys.}\ }\textbf {\bibinfo
  {volume} {641}},\ \bibinfo {pages} {A6} (\bibinfo {year} {2020})},\ \Eprint
  {https://arxiv.org/abs/1807.06209} {arXiv:1807.06209 [astro-ph.CO]}
  \BibitemShut {NoStop}%
\bibitem [{\citenamefont {Arcadi}\ \emph {et~al.}(2018)\citenamefont {Arcadi},
  \citenamefont {Dutra}, \citenamefont {Ghosh}, \citenamefont {Lindner},
  \citenamefont {Mambrini}, \citenamefont {Pierre}, \citenamefont {Profumo},\
  and\ \citenamefont {Queiroz}}]{Arcadi2018}%
  \BibitemOpen
  \bibfield  {author} {\bibinfo {author} {\bibfnamefont {G.}~\bibnamefont
  {Arcadi}}, \bibinfo {author} {\bibfnamefont {M.}~\bibnamefont {Dutra}},
  \bibinfo {author} {\bibfnamefont {P.}~\bibnamefont {Ghosh}}, \bibinfo
  {author} {\bibfnamefont {M.}~\bibnamefont {Lindner}}, \bibinfo {author}
  {\bibfnamefont {Y.}~\bibnamefont {Mambrini}}, \bibinfo {author}
  {\bibfnamefont {M.}~\bibnamefont {Pierre}}, \bibinfo {author} {\bibfnamefont
  {S.}~\bibnamefont {Profumo}},\ and\ \bibinfo {author} {\bibfnamefont {F.~S.}\
  \bibnamefont {Queiroz}},\ }\bibfield  {title} {\bibinfo {title} {The waning
  of the wimp? a review of models, searches, and constraints},\ }\href
  {https://doi.org/10.1140/epjc/s10052-018-5662-y} {\bibfield  {journal}
  {\bibinfo  {journal} {Eur. Phys. J. C}\ }\textbf {\bibinfo {volume} {78}},\
  \bibinfo {pages} {203} (\bibinfo {year} {2018})},\ \Eprint
  {https://arxiv.org/abs/1703.07364} {arXiv:1703.07364 [hep-ph]} \BibitemShut
  {NoStop}%
\bibitem [{\citenamefont {Aalbers}\ \emph {et~al.}(2025)\citenamefont {Aalbers}
  \emph {et~al.}}]{LZ2025WIMP}%
  \BibitemOpen
  \bibfield  {author} {\bibinfo {author} {\bibfnamefont {J.}~\bibnamefont
  {Aalbers}} \emph {et~al.} (\bibinfo {collaboration} {LZ Collaboration}),\
  }\bibfield  {title} {\bibinfo {title} {Dark matter search results from 4.2
  tonne-years of exposure of the lux-zeplin (lz) experiment},\ }\href
  {https://doi.org/10.1103/4dyc-z8zf} {\bibfield  {journal} {\bibinfo
  {journal} {Phys. Rev. Lett.}\ }\textbf {\bibinfo {volume} {135}},\ \bibinfo
  {pages} {011802} (\bibinfo {year} {2025})},\ \Eprint
  {https://arxiv.org/abs/2410.17036} {arXiv:2410.17036 [hep-ex]} \BibitemShut
  {NoStop}%
\bibitem [{\citenamefont {Aprile}\ \emph {et~al.}(2025)\citenamefont {Aprile}
  \emph {et~al.}}]{XENONnT2025}%
  \BibitemOpen
  \bibfield  {author} {\bibinfo {author} {\bibfnamefont {E.}~\bibnamefont
  {Aprile}} \emph {et~al.} (\bibinfo {collaboration} {XENON Collaboration}),\
  }\bibfield  {title} {\bibinfo {title} {Wimp dark matter search using a 3.1
  tonne-year exposure of the xenonnt experiment},\ }\href
  {https://doi.org/10.1103/msw4-t342} {\bibfield  {journal} {\bibinfo
  {journal} {Phys. Rev. Lett.}\ }\textbf {\bibinfo {volume} {135}},\ \bibinfo
  {pages} {221003} (\bibinfo {year} {2025})},\ \Eprint
  {https://arxiv.org/abs/2502.18005} {arXiv:2502.18005 [hep-ex]} \BibitemShut
  {NoStop}%
\bibitem [{\citenamefont {Di~Mauro}\ and\ \citenamefont
  {Xie}(2026)}]{DiMauroXie2026}%
  \BibitemOpen
  \bibfield  {author} {\bibinfo {author} {\bibfnamefont {M.}~\bibnamefont
  {Di~Mauro}}\ and\ \bibinfo {author} {\bibfnamefont {B.}~\bibnamefont {Xie}},\
  }\bibfield  {title} {\bibinfo {title} {$s$-channel dark matter simplified
  models in the resonance region},\ }\href {https://doi.org/10.1103/pdd9-jkl6}
  {\bibfield  {journal} {\bibinfo  {journal} {Phys. Rev. D}\ }\textbf {\bibinfo
  {volume} {113}},\ \bibinfo {pages} {015034} (\bibinfo {year}
  {2026})}\BibitemShut {NoStop}%
\bibitem [{\citenamefont {Kong}\ and\ \citenamefont
  {Di~Mauro}(2026)}]{KongDiMauro2026}%
  \BibitemOpen
  \bibfield  {author} {\bibinfo {author} {\bibfnamefont {C.}~\bibnamefont
  {Kong}}\ and\ \bibinfo {author} {\bibfnamefont {M.}~\bibnamefont
  {Di~Mauro}},\ }\bibfield  {title} {\bibinfo {title} {Comprehensive study of
  wimp models explaining the fermi-lat galactic center excess},\ }\href
  {https://doi.org/10.1103/v9vr-9skd} {\bibfield  {journal} {\bibinfo
  {journal} {Phys. Rev. D}\ }\textbf {\bibinfo {volume} {113}},\ \bibinfo
  {pages} {043031} (\bibinfo {year} {2026})},\ \Eprint
  {https://arxiv.org/abs/2511.21808} {arXiv:2511.21808 [hep-ph]} \BibitemShut
  {NoStop}%
\bibitem [{\citenamefont {Koechler}\ and\ \citenamefont
  {Di~Mauro}(2025)}]{KoechlerDiMauro2025}%
  \BibitemOpen
  \bibfield  {author} {\bibinfo {author} {\bibfnamefont {J.}~\bibnamefont
  {Koechler}}\ and\ \bibinfo {author} {\bibfnamefont {M.}~\bibnamefont
  {Di~Mauro}},\ }\bibfield  {title} {\bibinfo {title} {Leptophilic dark matter
  in $u(1)_{L_i-L_j}$ models: A solution to the fermi-lat galactic center
  excess consistent with cosmological and laboratory observations},\ }\href
  {https://doi.org/10.1103/nw98-38vr} {\bibfield  {journal} {\bibinfo
  {journal} {Phys. Rev. D}\ }\textbf {\bibinfo {volume} {112}},\ \bibinfo
  {pages} {115016} (\bibinfo {year} {2025})},\ \Eprint
  {https://arxiv.org/abs/2508.02775} {arXiv:2508.02775 [hep-ph]} \BibitemShut
  {NoStop}%
\bibitem [{\citenamefont {Pospelov}\ \emph {et~al.}(2008)\citenamefont
  {Pospelov}, \citenamefont {Ritz},\ and\ \citenamefont
  {Voloshin}}]{PospelovRitzVoloshin2008}%
  \BibitemOpen
  \bibfield  {author} {\bibinfo {author} {\bibfnamefont {M.}~\bibnamefont
  {Pospelov}}, \bibinfo {author} {\bibfnamefont {A.}~\bibnamefont {Ritz}},\
  and\ \bibinfo {author} {\bibfnamefont {M.~B.}\ \bibnamefont {Voloshin}},\
  }\bibfield  {title} {\bibinfo {title} {Secluded wimp dark matter},\ }\href
  {https://doi.org/10.1016/j.physletb.2008.02.052} {\bibfield  {journal}
  {\bibinfo  {journal} {Phys. Lett. B}\ }\textbf {\bibinfo {volume} {662}},\
  \bibinfo {pages} {53} (\bibinfo {year} {2008})},\ \Eprint
  {https://arxiv.org/abs/0711.4866} {arXiv:0711.4866 [hep-ph]} \BibitemShut
  {NoStop}%
\bibitem [{\citenamefont {Di~Mauro}\ and\ \citenamefont
  {Wang}(2026)}]{DiMauroWang2026}%
  \BibitemOpen
  \bibfield  {author} {\bibinfo {author} {\bibfnamefont {M.}~\bibnamefont
  {Di~Mauro}}\ and\ \bibinfo {author} {\bibfnamefont {Y.}~\bibnamefont
  {Wang}},\ }\bibfield  {title} {\bibinfo {title} {Phenomenology of secluded
  dark matter in three minimal bsm scenarios},\ }\href
  {https://doi.org/10.1103/44ls-ss78} {\bibfield  {journal} {\bibinfo
  {journal} {Phys. Rev. D}\ }\textbf {\bibinfo {volume} {113}},\ \bibinfo
  {pages} {075003} (\bibinfo {year} {2026})},\ \Eprint
  {https://arxiv.org/abs/2510.23771} {arXiv:2510.23771 [hep-ph]} \BibitemShut
  {NoStop}%
\bibitem [{\citenamefont {Di~Mauro}(2026)}]{DiMauroNeutrino2026}%
  \BibitemOpen
  \bibfield  {author} {\bibinfo {author} {\bibfnamefont {M.}~\bibnamefont
  {Di~Mauro}},\ }\bibfield  {title} {\bibinfo {title} {Two puzzles, one
  solution: Neutrino mass and secluded dark matter},\ }\href@noop {} {\bibfield
   {journal} {\bibinfo  {journal} {arXiv e-prints}\ } (\bibinfo {year}
  {2026})},\ \Eprint {https://arxiv.org/abs/2511.19622} {arXiv:2511.19622
  [hep-ph]} \BibitemShut {NoStop}%
\bibitem [{\citenamefont {{LUX-ZEPLIN
  Collaboration}}(2026{\natexlab{a}})}]{LZ2026Extended}%
  \BibitemOpen
  \bibfield  {author} {\bibinfo {author} {\bibnamefont {{LUX-ZEPLIN
  Collaboration}}},\ }\href
  {https://lz.lbl.gov/wp-content/uploads/sites/6/2026/08/LZ_Preprint_260901_Dark_Matter_EFT_Nuclear_Recoil_Search_at_Higher_Energies.pdf}
  {\bibinfo {title} {Search for dark matter particle interactions in an
  extended nuclear recoil energy window with the lux-zeplin (lz) experiment}}
  (\bibinfo {year} {2026}{\natexlab{a}}),\ \bibinfo {note} {lZ preprint, 1
  September 2026}\BibitemShut {NoStop}%
\bibitem [{\citenamefont {Fan}\ \emph {et~al.}(2010)\citenamefont {Fan},
  \citenamefont {Reece},\ and\ \citenamefont {Wang}}]{Fan2010}%
  \BibitemOpen
  \bibfield  {author} {\bibinfo {author} {\bibfnamefont {J.}~\bibnamefont
  {Fan}}, \bibinfo {author} {\bibfnamefont {M.}~\bibnamefont {Reece}},\ and\
  \bibinfo {author} {\bibfnamefont {L.-T.}\ \bibnamefont {Wang}},\ }\bibfield
  {title} {\bibinfo {title} {Non-relativistic effective theory of dark matter
  direct detection},\ }\href@noop {} {\bibfield  {journal} {\bibinfo  {journal}
  {JCAP}\ }\textbf {\bibinfo {volume} {11}},\ \bibinfo {pages} {042}},\ \Eprint
  {https://arxiv.org/abs/1008.1591} {arXiv:1008.1591 [hep-ph]} \BibitemShut
  {NoStop}%
\bibitem [{\citenamefont {Fitzpatrick}\ \emph {et~al.}(2013)\citenamefont
  {Fitzpatrick}, \citenamefont {Haxton}, \citenamefont {Katz}, \citenamefont
  {Lubbers},\ and\ \citenamefont {Xu}}]{Fitzpatrick2013}%
  \BibitemOpen
  \bibfield  {author} {\bibinfo {author} {\bibfnamefont {A.~L.}\ \bibnamefont
  {Fitzpatrick}}, \bibinfo {author} {\bibfnamefont {W.}~\bibnamefont {Haxton}},
  \bibinfo {author} {\bibfnamefont {E.}~\bibnamefont {Katz}}, \bibinfo {author}
  {\bibfnamefont {N.}~\bibnamefont {Lubbers}},\ and\ \bibinfo {author}
  {\bibfnamefont {Y.}~\bibnamefont {Xu}},\ }\bibfield  {title} {\bibinfo
  {title} {The effective field theory of dark matter direct detection},\
  }\href@noop {} {\bibfield  {journal} {\bibinfo  {journal} {JCAP}\ }\textbf
  {\bibinfo {volume} {02}},\ \bibinfo {pages} {004}},\ \Eprint
  {https://arxiv.org/abs/1203.3542} {arXiv:1203.3542 [hep-ph]} \BibitemShut
  {NoStop}%
\bibitem [{\citenamefont {Anand}\ \emph {et~al.}(2014)\citenamefont {Anand},
  \citenamefont {Fitzpatrick},\ and\ \citenamefont {Haxton}}]{Anand2014}%
  \BibitemOpen
  \bibfield  {author} {\bibinfo {author} {\bibfnamefont {N.}~\bibnamefont
  {Anand}}, \bibinfo {author} {\bibfnamefont {A.~L.}\ \bibnamefont
  {Fitzpatrick}},\ and\ \bibinfo {author} {\bibfnamefont {W.~C.}\ \bibnamefont
  {Haxton}},\ }\bibfield  {title} {\bibinfo {title} {Weakly interacting massive
  particle-nucleus elastic scattering response},\ }\href
  {https://doi.org/10.1103/PhysRevC.89.065501} {\bibfield  {journal} {\bibinfo
  {journal} {Phys. Rev. C}\ }\textbf {\bibinfo {volume} {89}},\ \bibinfo
  {pages} {065501} (\bibinfo {year} {2014})},\ \Eprint
  {https://arxiv.org/abs/1308.6288} {arXiv:1308.6288 [hep-ph]} \BibitemShut
  {NoStop}%
\bibitem [{\citenamefont {Graham}\ \emph {et~al.}(2025)\citenamefont {Graham},
  \citenamefont {Ramani},\ and\ \citenamefont {Wong}}]{Graham2025}%
  \BibitemOpen
  \bibfield  {author} {\bibinfo {author} {\bibfnamefont {P.~W.}\ \bibnamefont
  {Graham}}, \bibinfo {author} {\bibfnamefont {H.}~\bibnamefont {Ramani}},\
  and\ \bibinfo {author} {\bibfnamefont {S.~S.~Y.}\ \bibnamefont {Wong}},\
  }\bibfield  {title} {\bibinfo {title} {Enhancing direct detection of higgsino
  dark matter},\ }\href@noop {} {\bibfield  {journal} {\bibinfo  {journal}
  {Phys. Rev. D}\ }\textbf {\bibinfo {volume} {111}},\ \bibinfo {pages}
  {055030} (\bibinfo {year} {2025})},\ \Eprint
  {https://arxiv.org/abs/2409.07768} {arXiv:2409.07768 [hep-ph]} \BibitemShut
  {NoStop}%
\bibitem [{\citenamefont {Tucker-Smith}\ and\ \citenamefont
  {Weiner}(2001)}]{TuckerSmith2001}%
  \BibitemOpen
  \bibfield  {author} {\bibinfo {author} {\bibfnamefont {D.}~\bibnamefont
  {Tucker-Smith}}\ and\ \bibinfo {author} {\bibfnamefont {N.}~\bibnamefont
  {Weiner}},\ }\bibfield  {title} {\bibinfo {title} {{Inelastic dark matter}},\
  }\href {https://doi.org/10.1103/PhysRevD.64.043502} {\bibfield  {journal}
  {\bibinfo  {journal} {Phys. Rev. D}\ }\textbf {\bibinfo {volume} {64}},\
  \bibinfo {pages} {043502} (\bibinfo {year} {2001})},\ \Eprint
  {https://arxiv.org/abs/hep-ph/0101138} {arXiv:hep-ph/0101138} \BibitemShut
  {NoStop}%
\bibitem [{\citenamefont {Tucker-Smith}\ and\ \citenamefont
  {Weiner}(2005)}]{TuckerSmith2005}%
  \BibitemOpen
  \bibfield  {author} {\bibinfo {author} {\bibfnamefont {D.}~\bibnamefont
  {Tucker-Smith}}\ and\ \bibinfo {author} {\bibfnamefont {N.}~\bibnamefont
  {Weiner}},\ }\bibfield  {title} {\bibinfo {title} {The status of inelastic
  dark matter},\ }\href {https://doi.org/10.1103/PhysRevD.72.063509} {\bibfield
   {journal} {\bibinfo  {journal} {Phys. Rev. D}\ }\textbf {\bibinfo {volume}
  {72}},\ \bibinfo {pages} {063509} (\bibinfo {year} {2005})},\ \Eprint
  {https://arxiv.org/abs/hep-ph/0402065} {arXiv:hep-ph/0402065} \BibitemShut
  {NoStop}%
\bibitem [{\citenamefont {Barello}\ \emph {et~al.}(2014)\citenamefont
  {Barello}, \citenamefont {Chang},\ and\ \citenamefont {Newby}}]{Barello2014}%
  \BibitemOpen
  \bibfield  {author} {\bibinfo {author} {\bibfnamefont {G.}~\bibnamefont
  {Barello}}, \bibinfo {author} {\bibfnamefont {S.}~\bibnamefont {Chang}},\
  and\ \bibinfo {author} {\bibfnamefont {C.~A.}\ \bibnamefont {Newby}},\
  }\bibfield  {title} {\bibinfo {title} {A model independent approach to
  inelastic dark matter scattering},\ }\href
  {https://doi.org/10.1103/PhysRevD.90.094027} {\bibfield  {journal} {\bibinfo
  {journal} {Phys. Rev. D}\ }\textbf {\bibinfo {volume} {90}},\ \bibinfo
  {pages} {094027} (\bibinfo {year} {2014})},\ \Eprint
  {https://arxiv.org/abs/1409.0536} {arXiv:1409.0536 [hep-ph]} \BibitemShut
  {NoStop}%
\bibitem [{\citenamefont {Bramante}\ \emph {et~al.}(2016)\citenamefont
  {Bramante}, \citenamefont {Fox}, \citenamefont {Kribs},\ and\ \citenamefont
  {Martin}}]{Bramante2016}%
  \BibitemOpen
  \bibfield  {author} {\bibinfo {author} {\bibfnamefont {J.}~\bibnamefont
  {Bramante}}, \bibinfo {author} {\bibfnamefont {P.~J.}\ \bibnamefont {Fox}},
  \bibinfo {author} {\bibfnamefont {G.~D.}\ \bibnamefont {Kribs}},\ and\
  \bibinfo {author} {\bibfnamefont {A.}~\bibnamefont {Martin}},\ }\bibfield
  {title} {\bibinfo {title} {Inelastic frontier: Discovering dark matter at
  high recoil energy},\ }\href {https://doi.org/10.1103/PhysRevD.94.115026}
  {\bibfield  {journal} {\bibinfo  {journal} {Phys. Rev. D}\ }\textbf {\bibinfo
  {volume} {94}},\ \bibinfo {pages} {115026} (\bibinfo {year} {2016})},\
  \Eprint {https://arxiv.org/abs/1608.02662} {arXiv:1608.02662 [hep-ph]}
  \BibitemShut {NoStop}%
\bibitem [{\citenamefont {Krall}\ and\ \citenamefont
  {Reece}(2018)}]{KrallReece2018}%
  \BibitemOpen
  \bibfield  {author} {\bibinfo {author} {\bibfnamefont {R.}~\bibnamefont
  {Krall}}\ and\ \bibinfo {author} {\bibfnamefont {M.}~\bibnamefont {Reece}},\
  }\bibfield  {title} {\bibinfo {title} {{Last electroweak WIMP standing:
  pseudo-Dirac Higgsino status and compact stars as future probes}},\ }\href
  {https://doi.org/10.1088/1674-1137/42/4/043105} {\bibfield  {journal}
  {\bibinfo  {journal} {Chin. Phys. C}\ }\textbf {\bibinfo {volume} {42}},\
  \bibinfo {pages} {043105} (\bibinfo {year} {2018})},\ \Eprint
  {https://arxiv.org/abs/1705.04843} {arXiv:1705.04843 [hep-ph]} \BibitemShut
  {NoStop}%
\bibitem [{\citenamefont {Lewin}\ and\ \citenamefont
  {Smith}(1996)}]{LewinSmith1996}%
  \BibitemOpen
  \bibfield  {author} {\bibinfo {author} {\bibfnamefont {J.~D.}\ \bibnamefont
  {Lewin}}\ and\ \bibinfo {author} {\bibfnamefont {P.~F.}\ \bibnamefont
  {Smith}},\ }\bibfield  {title} {\bibinfo {title} {Review of mathematics,
  numerical factors, and corrections for dark matter experiments based on
  elastic nuclear recoil},\ }\href
  {https://doi.org/10.1016/S0927-6505(96)00047-3} {\bibfield  {journal}
  {\bibinfo  {journal} {Astropart. Phys.}\ }\textbf {\bibinfo {volume} {6}},\
  \bibinfo {pages} {87} (\bibinfo {year} {1996})}\BibitemShut {NoStop}%
\bibitem [{\citenamefont {Baxter}\ \emph {et~al.}(2021)\citenamefont {Baxter}
  \emph {et~al.}}]{Baxter2021}%
  \BibitemOpen
  \bibfield  {author} {\bibinfo {author} {\bibfnamefont {D.}~\bibnamefont
  {Baxter}} \emph {et~al.},\ }\bibfield  {title} {\bibinfo {title} {Recommended
  conventions for reporting results from direct dark matter searches},\ }\href
  {https://doi.org/10.1140/epjc/s10052-021-09655-y} {\bibfield  {journal}
  {\bibinfo  {journal} {Eur. Phys. J. C}\ }\textbf {\bibinfo {volume} {81}},\
  \bibinfo {pages} {907} (\bibinfo {year} {2021})},\ \Eprint
  {https://arxiv.org/abs/2105.00599} {arXiv:2105.00599 [hep-ex]} \BibitemShut
  {NoStop}%
\bibitem [{\citenamefont {De~Simone}\ \emph {et~al.}(2010)\citenamefont
  {De~Simone}, \citenamefont {Sanz},\ and\ \citenamefont
  {Sato}}]{DeSimone2010}%
  \BibitemOpen
  \bibfield  {author} {\bibinfo {author} {\bibfnamefont {A.}~\bibnamefont
  {De~Simone}}, \bibinfo {author} {\bibfnamefont {V.}~\bibnamefont {Sanz}},\
  and\ \bibinfo {author} {\bibfnamefont {H.~P.}\ \bibnamefont {Sato}},\
  }\bibfield  {title} {\bibinfo {title} {{Pseudo-Dirac Dark Matter Leaves a
  Trace}},\ }\href {https://doi.org/10.1103/PhysRevLett.105.121802} {\bibfield
  {journal} {\bibinfo  {journal} {Phys. Rev. Lett.}\ }\textbf {\bibinfo
  {volume} {105}},\ \bibinfo {pages} {121802} (\bibinfo {year} {2010})},\
  \Eprint {https://arxiv.org/abs/1004.1567} {arXiv:1004.1567 [hep-ph]}
  \BibitemShut {NoStop}%
\bibitem [{\citenamefont {Kahlhoefer}\ \emph {et~al.}(2016)\citenamefont
  {Kahlhoefer}, \citenamefont {Schmidt-Hoberg}, \citenamefont {Schwetz},\ and\
  \citenamefont {Vogl}}]{Kahlhoefer2016}%
  \BibitemOpen
  \bibfield  {author} {\bibinfo {author} {\bibfnamefont {F.}~\bibnamefont
  {Kahlhoefer}}, \bibinfo {author} {\bibfnamefont {K.}~\bibnamefont
  {Schmidt-Hoberg}}, \bibinfo {author} {\bibfnamefont {T.}~\bibnamefont
  {Schwetz}},\ and\ \bibinfo {author} {\bibfnamefont {S.}~\bibnamefont
  {Vogl}},\ }\bibfield  {title} {\bibinfo {title} {{Implications of unitarity
  and gauge invariance for simplified dark matter models}},\ }\href
  {https://doi.org/10.1007/JHEP02(2016)016} {\bibfield  {journal} {\bibinfo
  {journal} {JHEP}\ }\textbf {\bibinfo {volume} {02}},\ \bibinfo {pages}
  {016}},\ \Eprint {https://arxiv.org/abs/1510.02110} {arXiv:1510.02110
  [hep-ph]} \BibitemShut {NoStop}%
\bibitem [{\citenamefont {Fileviez~Perez}\ and\ \citenamefont
  {Wise}(2010)}]{FileviezPerez2010}%
  \BibitemOpen
  \bibfield  {author} {\bibinfo {author} {\bibfnamefont {P.}~\bibnamefont
  {Fileviez~Perez}}\ and\ \bibinfo {author} {\bibfnamefont {M.~B.}\
  \bibnamefont {Wise}},\ }\bibfield  {title} {\bibinfo {title} {{Baryon and
  lepton number as local gauge symmetries}},\ }\href
  {https://doi.org/10.1103/PhysRevD.82.011901} {\bibfield  {journal} {\bibinfo
  {journal} {Phys. Rev. D}\ }\textbf {\bibinfo {volume} {82}},\ \bibinfo
  {pages} {011901} (\bibinfo {year} {2010})},\ \Eprint
  {https://arxiv.org/abs/1002.1754} {arXiv:1002.1754 [hep-ph]} \BibitemShut
  {NoStop}%
\bibitem [{\citenamefont {Duerr}\ and\ \citenamefont
  {Fileviez~Perez}(2014)}]{Duerr2014}%
  \BibitemOpen
  \bibfield  {author} {\bibinfo {author} {\bibfnamefont {M.}~\bibnamefont
  {Duerr}}\ and\ \bibinfo {author} {\bibfnamefont {P.}~\bibnamefont
  {Fileviez~Perez}},\ }\bibfield  {title} {\bibinfo {title} {{Baryonic Dark
  Matter}},\ }\href@noop {} {\bibfield  {journal} {\bibinfo  {journal} {Phys.
  Lett. B}\ }\textbf {\bibinfo {volume} {732}},\ \bibinfo {pages} {101}
  (\bibinfo {year} {2014})},\ \Eprint {https://arxiv.org/abs/1309.3970}
  {arXiv:1309.3970 [hep-ph]} \BibitemShut {NoStop}%
\bibitem [{\citenamefont {Ekstedt}\ \emph {et~al.}(2016)\citenamefont
  {Ekstedt}, \citenamefont {Enberg}, \citenamefont {Ingelman}, \citenamefont
  {Lofgren},\ and\ \citenamefont {Mandal}}]{Ekstedt2016}%
  \BibitemOpen
  \bibfield  {author} {\bibinfo {author} {\bibfnamefont {A.}~\bibnamefont
  {Ekstedt}}, \bibinfo {author} {\bibfnamefont {R.}~\bibnamefont {Enberg}},
  \bibinfo {author} {\bibfnamefont {G.}~\bibnamefont {Ingelman}}, \bibinfo
  {author} {\bibfnamefont {J.}~\bibnamefont {Lofgren}},\ and\ \bibinfo {author}
  {\bibfnamefont {T.}~\bibnamefont {Mandal}},\ }\bibfield  {title} {\bibinfo
  {title} {{Constraining minimal anomaly free U(1) extensions of the Standard
  Model}},\ }\href@noop {} {\bibfield  {journal} {\bibinfo  {journal} {JHEP}\
  }\textbf {\bibinfo {volume} {11}},\ \bibinfo {pages} {071}},\ \Eprint
  {https://arxiv.org/abs/1605.04855} {arXiv:1605.04855 [hep-ph]} \BibitemShut
  {NoStop}%
\bibitem [{\citenamefont {Carrillo~Gonz{\\'a}lez}\ and\ \citenamefont
  {Toro}(2021)}]{CarrilloGonzalez2021}%
  \BibitemOpen
  \bibfield  {author} {\bibinfo {author} {\bibfnamefont {M.}~\bibnamefont
  {Carrillo~Gonz{\\'a}lez}}\ and\ \bibinfo {author} {\bibfnamefont
  {N.}~\bibnamefont {Toro}},\ }\href@noop {} {\bibinfo {title} {{Cosmology and
  Signals of Light Pseudo-Dirac Dark Matter}}} (\bibinfo {year} {2021}),\
  \Eprint {https://arxiv.org/abs/2108.13422} {arXiv:2108.13422 [hep-ph]}
  \BibitemShut {NoStop}%
\bibitem [{\citenamefont {Griest}\ and\ \citenamefont
  {Seckel}(1991)}]{GriestSeckel1991}%
  \BibitemOpen
  \bibfield  {author} {\bibinfo {author} {\bibfnamefont {K.}~\bibnamefont
  {Griest}}\ and\ \bibinfo {author} {\bibfnamefont {D.}~\bibnamefont
  {Seckel}},\ }\bibfield  {title} {\bibinfo {title} {Three exceptions in the
  calculation of relic abundances},\ }\href
  {https://doi.org/10.1103/PhysRevD.43.3191} {\bibfield  {journal} {\bibinfo
  {journal} {Phys. Rev. D}\ }\textbf {\bibinfo {volume} {43}},\ \bibinfo
  {pages} {3191} (\bibinfo {year} {1991})}\BibitemShut {NoStop}%
\bibitem [{\citenamefont {Steigman}\ \emph {et~al.}(2012)\citenamefont
  {Steigman}, \citenamefont {Dasgupta},\ and\ \citenamefont
  {Beacom}}]{Bringmann2012Thermal}%
  \BibitemOpen
  \bibfield  {author} {\bibinfo {author} {\bibfnamefont {G.}~\bibnamefont
  {Steigman}}, \bibinfo {author} {\bibfnamefont {B.}~\bibnamefont {Dasgupta}},\
  and\ \bibinfo {author} {\bibfnamefont {J.~F.}\ \bibnamefont {Beacom}},\
  }\bibfield  {title} {\bibinfo {title} {Precise relic wimp abundance and its
  impact on searches for dark matter annihilation},\ }\href
  {https://doi.org/10.1103/PhysRevD.86.023506} {\bibfield  {journal} {\bibinfo
  {journal} {Phys. Rev. D}\ }\textbf {\bibinfo {volume} {86}},\ \bibinfo
  {pages} {023506} (\bibinfo {year} {2012})},\ \Eprint
  {https://arxiv.org/abs/1204.3622} {arXiv:1204.3622 [hep-ph]} \BibitemShut
  {NoStop}%
\bibitem [{\citenamefont {Brahma}\ \emph {et~al.}(2024)\citenamefont {Brahma},
  \citenamefont {Heeba},\ and\ \citenamefont {Schutz}}]{Brahma2023}%
  \BibitemOpen
  \bibfield  {author} {\bibinfo {author} {\bibfnamefont {N.}~\bibnamefont
  {Brahma}}, \bibinfo {author} {\bibfnamefont {S.}~\bibnamefont {Heeba}},\ and\
  \bibinfo {author} {\bibfnamefont {K.}~\bibnamefont {Schutz}},\ }\bibfield
  {title} {\bibinfo {title} {Resonant pseudo-dirac dark matter as a sub-gev
  thermal target},\ }\href {https://doi.org/10.1103/PhysRevD.109.035006}
  {\bibfield  {journal} {\bibinfo  {journal} {Phys. Rev. D}\ }\textbf {\bibinfo
  {volume} {109}},\ \bibinfo {pages} {035006} (\bibinfo {year} {2024})},\
  \Eprint {https://arxiv.org/abs/2308.01960} {arXiv:2308.01960 [hep-ph]}
  \BibitemShut {NoStop}%
\bibitem [{\citenamefont {Chao}\ \emph {et~al.}(2023)\citenamefont {Chao},
  \citenamefont {Feng},\ and\ \citenamefont {Jin}}]{Chao2023Excited}%
  \BibitemOpen
  \bibfield  {author} {\bibinfo {author} {\bibfnamefont {W.}~\bibnamefont
  {Chao}}, \bibinfo {author} {\bibfnamefont {J.-J.}\ \bibnamefont {Feng}},\
  and\ \bibinfo {author} {\bibfnamefont {M.}~\bibnamefont {Jin}},\ }\bibfield
  {title} {\bibinfo {title} {$n_{\rm eff}$ from an excited dark matter state},\
  }\href {https://doi.org/10.1103/PhysRevD.107.015022} {\bibfield  {journal}
  {\bibinfo  {journal} {Phys. Rev. D}\ }\textbf {\bibinfo {volume} {107}},\
  \bibinfo {pages} {015022} (\bibinfo {year} {2023})},\ \Eprint
  {https://arxiv.org/abs/2202.12673} {arXiv:2202.12673 [hep-ph]} \BibitemShut
  {NoStop}%
\bibitem [{\citenamefont {Slatyer}(2016)}]{Slatyer2016}%
  \BibitemOpen
  \bibfield  {author} {\bibinfo {author} {\bibfnamefont {T.~R.}\ \bibnamefont
  {Slatyer}},\ }\bibfield  {title} {\bibinfo {title} {Energy injection and
  absorption in the cosmic dark ages},\ }\href@noop {} {\bibfield  {journal}
  {\bibinfo  {journal} {Phys. Rev. D}\ }\textbf {\bibinfo {volume} {93}},\
  \bibinfo {pages} {023521} (\bibinfo {year} {2016})},\ \Eprint
  {https://arxiv.org/abs/1506.03811} {arXiv:1506.03811 [astro-ph.CO]}
  \BibitemShut {NoStop}%
\bibitem [{\citenamefont {Nagata}\ and\ \citenamefont
  {Shirai}(2015)}]{NagataShirai2015}%
  \BibitemOpen
  \bibfield  {author} {\bibinfo {author} {\bibfnamefont {N.}~\bibnamefont
  {Nagata}}\ and\ \bibinfo {author} {\bibfnamefont {S.}~\bibnamefont
  {Shirai}},\ }\bibfield  {title} {\bibinfo {title} {{Higgsino Dark Matter in
  High-Scale Supersymmetry}},\ }\href {https://doi.org/10.1007/JHEP01(2015)029}
  {\bibfield  {journal} {\bibinfo  {journal} {JHEP}\ }\textbf {\bibinfo
  {volume} {01}},\ \bibinfo {pages} {029}},\ \Eprint
  {https://arxiv.org/abs/1410.4549} {arXiv:1410.4549 [hep-ph]} \BibitemShut
  {NoStop}%
\bibitem [{\citenamefont {Pospelov}\ and\ \citenamefont
  {Ramani}(2026)}]{PospelovRamani2026}%
  \BibitemOpen
  \bibfield  {author} {\bibinfo {author} {\bibfnamefont {M.}~\bibnamefont
  {Pospelov}}\ and\ \bibinfo {author} {\bibfnamefont {H.}~\bibnamefont
  {Ramani}},\ }\bibfield  {title} {\bibinfo {title} {Strong constraints on
  higgsino dark matter from solar capture},\ }\href@noop {} {\  (\bibinfo
  {year} {2026})},\ \Eprint {https://arxiv.org/abs/2609.02775}
  {arXiv:2609.02775 [hep-ph]} \BibitemShut {NoStop}%
\bibitem [{\citenamefont {Rodd}\ \emph {et~al.}(2024)\citenamefont {Rodd},
  \citenamefont {Safdi},\ and\ \citenamefont {Xu}}]{Rodd2024}%
  \BibitemOpen
  \bibfield  {author} {\bibinfo {author} {\bibfnamefont {N.~L.}\ \bibnamefont
  {Rodd}}, \bibinfo {author} {\bibfnamefont {B.~R.}\ \bibnamefont {Safdi}},\
  and\ \bibinfo {author} {\bibfnamefont {W.~L.}\ \bibnamefont {Xu}},\
  }\bibfield  {title} {\bibinfo {title} {{CTA and SWGO can discover Higgsino
  dark matter annihilation}},\ }\href
  {https://doi.org/10.1103/PhysRevD.110.043003} {\bibfield  {journal} {\bibinfo
   {journal} {Phys. Rev. D}\ }\textbf {\bibinfo {volume} {110}},\ \bibinfo
  {pages} {043003} (\bibinfo {year} {2024})}\BibitemShut {NoStop}%
\bibitem [{\citenamefont {{H.E.S.S. Collaboration}}(2026)}]{HESS2026}%
  \BibitemOpen
  \bibfield  {author} {\bibinfo {author} {\bibnamefont {{H.E.S.S.
  Collaboration}}},\ }\bibfield  {title} {\bibinfo {title} {Search for
  gamma-ray spectral lines from dark matter annihilation with the h.e.s.s.
  inner galaxy survey},\ }\href {https://doi.org/10.1103/d8tj-55kc} {\bibfield
  {journal} {\bibinfo  {journal} {Phys. Rev. Lett.}\ }\textbf {\bibinfo
  {volume} {137}},\ \bibinfo {pages} {091002} (\bibinfo {year} {2026})},\
  \Eprint {https://arxiv.org/abs/2608.07234} {arXiv:2608.07234 [astro-ph.HE]}
  \BibitemShut {NoStop}%
\bibitem [{\citenamefont {{MAGIC Collaboration}}(2023)}]{MAGIC2023Lines}%
  \BibitemOpen
  \bibfield  {author} {\bibinfo {author} {\bibnamefont {{MAGIC
  Collaboration}}},\ }\bibfield  {title} {\bibinfo {title} {Search for
  gamma-ray spectral lines from dark matter annihilation up to 100 tev toward
  the galactic center with magic},\ }\href
  {https://doi.org/10.1103/PhysRevLett.130.061002} {\bibfield  {journal}
  {\bibinfo  {journal} {Phys. Rev. Lett.}\ }\textbf {\bibinfo {volume} {130}},\
  \bibinfo {pages} {061002} (\bibinfo {year} {2023})},\ \Eprint
  {https://arxiv.org/abs/2212.10527} {arXiv:2212.10527 [astro-ph.HE]}
  \BibitemShut {NoStop}%
\bibitem [{\citenamefont {{Fermi-LAT
  Collaboration}}(2015)}]{FermiLAT2015Lines}%
  \BibitemOpen
  \bibfield  {author} {\bibinfo {author} {\bibnamefont {{Fermi-LAT
  Collaboration}}},\ }\bibfield  {title} {\bibinfo {title} {Updated search for
  spectral lines from galactic dark matter interactions with pass 8 data from
  the fermi large area telescope},\ }\href
  {https://doi.org/10.1103/PhysRevD.91.122002} {\bibfield  {journal} {\bibinfo
  {journal} {Phys. Rev. D}\ }\textbf {\bibinfo {volume} {91}},\ \bibinfo
  {pages} {122002} (\bibinfo {year} {2015})},\ \Eprint
  {https://arxiv.org/abs/1506.00013} {arXiv:1506.00013 [astro-ph.HE]}
  \BibitemShut {NoStop}%
\bibitem [{\citenamefont {De~La Torre~Luque}\ \emph {et~al.}(2024)\citenamefont
  {De~La Torre~Luque}, \citenamefont {Smirnov},\ and\ \citenamefont
  {Linden}}]{FermiLAT15yrLines}%
  \BibitemOpen
  \bibfield  {author} {\bibinfo {author} {\bibfnamefont {P.}~\bibnamefont
  {De~La Torre~Luque}}, \bibinfo {author} {\bibfnamefont {J.}~\bibnamefont
  {Smirnov}},\ and\ \bibinfo {author} {\bibfnamefont {T.}~\bibnamefont
  {Linden}},\ }\bibfield  {title} {\bibinfo {title} {Gamma-ray lines in 15
  years of fermi-lat data: New constraints on higgs portal dark matter},\
  }\href {https://doi.org/10.1103/PhysRevD.109.L041301} {\bibfield  {journal}
  {\bibinfo  {journal} {Phys. Rev. D}\ }\textbf {\bibinfo {volume} {109}},\
  \bibinfo {pages} {L041301} (\bibinfo {year} {2024})},\ \Eprint
  {https://arxiv.org/abs/2309.03281} {arXiv:2309.03281 [hep-ph]} \BibitemShut
  {NoStop}%
\bibitem [{\citenamefont {{HAWC Collaboration}}(2020)}]{HAWC2020Lines}%
  \BibitemOpen
  \bibfield  {author} {\bibinfo {author} {\bibnamefont {{HAWC
  Collaboration}}},\ }\bibfield  {title} {\bibinfo {title} {Search for
  gamma-ray spectral lines from dark matter annihilation in dwarf galaxies with
  the high-altitude water cherenkov observatory},\ }\href
  {https://doi.org/10.1103/PhysRevD.101.103001} {\bibfield  {journal} {\bibinfo
   {journal} {Phys. Rev. D}\ }\textbf {\bibinfo {volume} {101}},\ \bibinfo
  {pages} {103001} (\bibinfo {year} {2020})},\ \Eprint
  {https://arxiv.org/abs/1912.05632} {arXiv:1912.05632 [astro-ph.HE]}
  \BibitemShut {NoStop}%
\bibitem [{\citenamefont {Beneke}\ \emph {et~al.}(2020)\citenamefont {Beneke},
  \citenamefont {Hasner}, \citenamefont {Urban},\ and\ \citenamefont
  {Vollmann}}]{Beneke2020Higgsino}%
  \BibitemOpen
  \bibfield  {author} {\bibinfo {author} {\bibfnamefont {M.}~\bibnamefont
  {Beneke}}, \bibinfo {author} {\bibfnamefont {C.}~\bibnamefont {Hasner}},
  \bibinfo {author} {\bibfnamefont {K.}~\bibnamefont {Urban}},\ and\ \bibinfo
  {author} {\bibfnamefont {M.}~\bibnamefont {Vollmann}},\ }\bibfield  {title}
  {\bibinfo {title} {Precise yield of high-energy photons from higgsino dark
  matter annihilation},\ }\href {https://doi.org/10.1007/JHEP03(2020)030}
  {\bibfield  {journal} {\bibinfo  {journal} {JHEP}\ }\textbf {\bibinfo
  {volume} {03}},\ \bibinfo {pages} {030}},\ \Eprint
  {https://arxiv.org/abs/1912.02034} {arXiv:1912.02034 [hep-ph]} \BibitemShut
  {NoStop}%
\bibitem [{\citenamefont {Abe}\ \emph {et~al.}(2026)\citenamefont {Abe},
  \citenamefont {Inada}, \citenamefont {Moulin}, \citenamefont {Rodd},
  \citenamefont {Safdi},\ and\ \citenamefont {Xu}}]{CTAO2026Higgsino}%
  \BibitemOpen
  \bibfield  {author} {\bibinfo {author} {\bibfnamefont {S.}~\bibnamefont
  {Abe}}, \bibinfo {author} {\bibfnamefont {T.}~\bibnamefont {Inada}}, \bibinfo
  {author} {\bibfnamefont {E.}~\bibnamefont {Moulin}}, \bibinfo {author}
  {\bibfnamefont {N.~L.}\ \bibnamefont {Rodd}}, \bibinfo {author}
  {\bibfnamefont {B.~R.}\ \bibnamefont {Safdi}},\ and\ \bibinfo {author}
  {\bibfnamefont {W.~L.}\ \bibnamefont {Xu}},\ }\bibfield  {title} {\bibinfo
  {title} {Discovery prospects for the higgsino at the cherenkov telescope
  array observatory's northern site within the decade},\ }\href
  {https://doi.org/10.1103/5qjj-l4b5} {\bibfield  {journal} {\bibinfo
  {journal} {Phys. Rev. D}\ }\textbf {\bibinfo {volume} {114}},\ \bibinfo
  {pages} {023052} (\bibinfo {year} {2026})}\BibitemShut {NoStop}%
\bibitem [{\citenamefont {{LUX-ZEPLIN
  Collaboration}}(2026{\natexlab{b}})}]{LZWebsite2026}%
  \BibitemOpen
  \bibfield  {author} {\bibinfo {author} {\bibnamefont {{LUX-ZEPLIN
  Collaboration}}},\ }\href {https://lz.lbl.gov/} {\bibinfo {title} {The lz
  dark matter experiment}} (\bibinfo {year} {2026}{\natexlab{b}}),\ \bibinfo
  {note} {experiment status page, accessed 1 September 2026}\BibitemShut
  {NoStop}%
\bibitem [{\citenamefont {Lou}\ and\ \citenamefont {Lu}(2026)}]{LouLu2026}%
  \BibitemOpen
  \bibfield  {author} {\bibinfo {author} {\bibfnamefont {Y.}~\bibnamefont
  {Lou}}\ and\ \bibinfo {author} {\bibfnamefont {H.-T.}\ \bibnamefont {Lu}},\
  }\bibfield  {title} {\bibinfo {title} {Fermionic dark matter absorption and
  the high-energy event in {LUX-ZEPLIN}},\ }\href@noop {} {\  (\bibinfo {year}
  {2026})},\ \Eprint {https://arxiv.org/abs/2609.01592} {arXiv:2609.01592
  [hep-ph]} \BibitemShut {NoStop}%
\bibitem [{\citenamefont {Wu}\ \emph {et~al.}(2026)\citenamefont {Wu},
  \citenamefont {Zhang},\ and\ \citenamefont {Zhu}}]{WuZhangZhu2026}%
  \BibitemOpen
  \bibfield  {author} {\bibinfo {author} {\bibfnamefont {L.}~\bibnamefont
  {Wu}}, \bibinfo {author} {\bibfnamefont {Y.}~\bibnamefont {Zhang}},\ and\
  \bibinfo {author} {\bibfnamefont {B.}~\bibnamefont {Zhu}},\ }\bibfield
  {title} {\bibinfo {title} {{TeV} higgsino dark matter from {LZ} nuclear
  recoil to {Fermi-LAT} gamma rays},\ }\href@noop {} {\  (\bibinfo {year}
  {2026})},\ \Eprint {https://arxiv.org/abs/2609.01590} {arXiv:2609.01590
  [hep-ph]} \BibitemShut {NoStop}%
\bibitem [{\citenamefont {Visinelli}(2026)}]{Visinelli2026}%
  \BibitemOpen
  \bibfield  {author} {\bibinfo {author} {\bibfnamefont {L.}~\bibnamefont
  {Visinelli}},\ }\bibfield  {title} {\bibinfo {title} {A {Peccei--Quinn}
  origin for inelastic electroweak dark matter after {LUX-ZEPLIN}},\
  }\href@noop {} {\  (\bibinfo {year} {2026})},\ \Eprint
  {https://arxiv.org/abs/2609.02807} {arXiv:2609.02807 [hep-ph]} \BibitemShut
  {NoStop}%
\bibitem [{\citenamefont {{LUX-ZEPLIN Collaboration}}(2024)}]{LZ2024NREFT}%
  \BibitemOpen
  \bibfield  {author} {\bibinfo {author} {\bibnamefont {{LUX-ZEPLIN
  Collaboration}}},\ }\bibfield  {title} {\bibinfo {title} {Constraints on
  covariant wimp-nucleon effective field theory interactions from the first
  science run of the lux-zeplin experiment},\ }\href@noop {} {\bibfield
  {journal} {\bibinfo  {journal} {Phys. Rev. Lett.}\ }\textbf {\bibinfo
  {volume} {133}},\ \bibinfo {pages} {221801} (\bibinfo {year} {2024})},\
  \Eprint {https://arxiv.org/abs/2404.17666} {arXiv:2404.17666 [hep-ex]}
  \BibitemShut {NoStop}%
\end{thebibliography}%

\end{document}